\documentclass[journal,doublecolumn]{IEEEtran}
\usepackage{epsfig,color,amsmath,cite}
\usepackage{amsthm} %defined already in ieeeconf
\usepackage{amsmath}    %For theorems
\usepackage[T1]{fontenc}
\usepackage[utf8]{inputenc}
\usepackage{bm}
\usepackage{epstopdf}
\usepackage{amssymb}
\usepackage{url}
\usepackage{enumitem} %defined already in ieeeconf
\usepackage{multirow}
\usepackage{hhline}
\usepackage{booktabs}
\usepackage{mathtools}
\usepackage{makecell}

\usepackage[linesnumbered,boxed,commentsnumbered,ruled,vlined,longend]{algorithm2e}
\makeatletter

\SetKwProg{Init}{Initialization}{}{Proceed}
\SetKwProg{St}{Start}{}{Proceed}
\makeatother
\usepackage{comment}

\makeatother
\DeclareMathAlphabet\mathbfcal{OMS}{cmsy}{b}{n}

\usepackage{stackengine}

\def\comment{\textcolor{red}}

\newcommand{\m}{\boldsymbol}
\allowdisplaybreaks[4]
\usepackage[colorlinks = true,
linkcolor = blue,
urlcolor  = blue,
citecolor = blue,
anchorcolor = blue]{hyperref}
\usepackage[framemethod=TikZ]{mdframed}
\mdfdefinestyle{MyFrame}{%
	linecolor=black,
	outerlinewidth=1.25pt,
	roundcorner=1.25pt,
	innerrightmargin=5pt,
	innerleftmargin=5pt,}
	
\usepackage{tikz}
\usetikzlibrary{matrix,positioning,decorations.pathreplacing}
\usetikzlibrary{shapes.geometric, arrows}
\usetikzlibrary{backgrounds}
\usetikzlibrary{shapes}
\usetikzlibrary{tikzmark}
\usetikzlibrary{calc}
\usetikzlibrary{arrows,shapes,positioning,shadows,trees,mindmap}
\usepackage[edges]{forest}
\usetikzlibrary{arrows.meta}
\colorlet{linecol}{black!75}
\usepackage{xkcdcolors} % xkcd colors

\usepackage[noabbrev]{cleveref}

\usepackage{mathtools}

\DeclarePairedDelimiter\abs{\lvert}{\rvert}%
\DeclarePairedDelimiter\norm{\lVert}{\rVert}%

\makeatletter
\let\oldabs\abs
\def\abs{\@ifstar{\oldabs}{\oldabs*}}
\let\oldnorm\norm
\def\norm{\@ifstar{\oldnorm}{\oldnorm*}}
\makeatother

\usepackage[english]{babel}
\usepackage[utf8]{inputenc}
\usepackage[super]{nth}

\usepackage{graphicx}
\usepackage{float}
\usepackage[caption = false]{subfig}

\usepackage{array}
\usepackage{threeparttable}

\usepackage[english]{babel}
\usepackage[utf8]{inputenc}
\usepackage[super]{nth}

\SetKwRepeat{Do}{do}{while}%

\everymath{\displaystyle}

\newcolumntype{E}{>{\centering\arraybackslash}m{0.5in}}
\newcolumntype{Q}{>{\centering\arraybackslash}m{3in}}
\newcolumntype{K}{>{\centering\arraybackslash}m{1.6in}}

\definecolor{airforceblue}{rgb}{0.36, 0.54, 0.66}
\usepackage[most]{tcolorbox}
\newtcbox{\colorboxouline}[1][]{boxsep=0pt,top=2.5pt,bottom=1pt,left=2pt,right=1pt,colframe=airforceblue,colback=airforceblue!10,boxrule=0pt,toprule=0pt,leftrule=3pt,sharp corners, enhanced jigsaw,,#1}

\usepackage[export]{adjustbox}
\renewcommand{\arraystretch}{1.5}
\usepackage{threeparttable}
\usepackage{multirow}
\usepackage{annotate-equations}

\usepackage{colortbl} 

\usepackage{mdframed}

\mdfdefinestyle{theoremstyle}{% 
	linecolor=blue,linewidth=1pt,%
	frametitlerule=true,% 
	frametitlebackgroundcolor=gray!20, innertopmargin=\topskip,}

\usepackage[framemethod=TikZ]{mdframed}
\mdfsetup{skipabove=1mm,skipbelow=1mm}
\newcounter{theo}[section]
\newenvironment{theo}[1][]{%
	\stepcounter{theo}%
	\ifstrempty{#1}%
	{\mdfsetup{%
			frametitle={%
				\tikz[baseline=(current bounding box.east),outer sep=0pt]
				\node[anchor=east,rectangle,fill=white!80!blue]
				{\strut };}}
	}%
	{\mdfsetup{%
			frametitle={%
				\tikz[baseline=(current bounding box.east),outer sep=0pt]
				\node[anchor=east,rectangle,fill=white!80!blue]
				{\strut \textcolor{black!90}{#1}};}}%
	}%
	\mdfsetup{innertopmargin=10pt,linecolor=white!80!blue,%
		linewidth=3pt,topline=true,
		frametitleaboveskip=\dimexpr-\ht\strutbox\relax,}
	\begin{mdframed}[]\relax%
	}{\end{mdframed}}

\definecolor{pansypurple}{rgb}{0.47, 0.09, 0.29}
\definecolor{darkcerulean}{rgb}{0.15, 0.38, 0.61}
\mdfdefinestyle{RepStyle}{%
	skipabove=\topskip,
	skipbelow=\topskip,
	innerleftmargin=0ex,
	innerrightmargin=0ex,
	innertopmargin=0ex,
	linewidth=1.5pt,
	leftline=false,
	rightline=false,
	frametitlefont={\color{darkcerulean}\normalfont\bfseries},
	frametitlerule=false,
	frametitlebackgroundcolor=white,
	backgroundcolor=white,
	linecolor=darkcerulean,
	theoremseparator={},
	ntheorem=false % Disables numbering
}

\definecolor{darkblueilike}{rgb}{0.1843 0.4471 0.6196}

\definecolor{deepjunglegreen}{rgb}{0.0, 0.29, 0.29}
\newmdenv[  
topline=false,  
rightline=false,  
bottomline=false,  
leftline=true,  
linecolor=darkblueilike!90,  
linewidth=3pt,
innerleftmargin=1.5ex,  
backgroundcolor=darkblueilike!10,  
frametitle=CBSP,
frametitlefont={\color{darkblueilike!10}\normalfont\bfseries},
frametitlebackgroundcolor=darkblueilike!90
]{BStP}

\newmdenv[style=RepStyle]{Rep}

\definecolor{pinegreen}{rgb}{0.0, 0.47, 0.44}
\definecolor{upmaroon}{rgb}{0.48, 0.07, 0.07}
\definecolor{ruddybrown}{rgb}{0.73, 0.4, 0.16}
\definecolor{burntorange}{rgb}{0.8, 0.33, 0.0}

\definecolor{coolblack}{rgb}{0.0, 0.18, 0.39}
\definecolor{midnightgreen}{rgb}{0.0, 0.29, 0.33}
\definecolor{smokeytopaz}{rgb}{0.61, 0.77, 0.89}
\definecolor{copper}{rgb}{0.72, 0.45, 0.2}

\def\comment{\textcolor{red}}

\usepackage{tabularx} % For tables with adjustable-width columns
\usepackage{booktabs, multirow} % Required for table formatting and multirow
\usepackage{graphicx} % Required for text rotation
\usepackage{caption}  % For customizing table captions

\usepackage[most]{tcolorbox}
\newtcolorbox{shadedcvbox}[1][]{enhanced jigsaw,
	colback=white!93!blue,
	coltext={black},
	boxrule=0pt,
	arc=3mm,
	auto outer arc,
	boxsep=3pt,
	left=3pt,
	right=3pt,
	bottom=3pt,
	top=4pt,
	#1}

\newtcolorbox{custombox}[1]{
	colback=#1!10,    % Background color
	colframe=#1!80,   % Frame color
	boxrule=0pt,      % No border
	arc=3mm,          % Rounded corners
	auto outer arc,
	boxsep=3pt,       % Box separation
	left=3pt,
	right=3pt,
	bottom=3pt,
	top=4pt,
	enhanced jigsaw,  % Enhanced options
}

\newtcolorbox{custombox1}[1]{
	colback=#1!0,    % Background color
	colframe=#1!80,   % Frame color
	boxrule=3pt,      % No border
	arc=1mm,          % Rounded corners
	auto outer arc,
	boxsep=3pt,       % Box separation
	left=3pt,
	right=3pt,
	bottom=3pt,
	top=4pt,
	enhanced jigsaw,  % Enhanced options
}

\usepackage{xcolor}
\definecolor{indigodye}{rgb}{0.0, 0.25, 0.42}
\definecolor{sapphire}{rgb}{0.03, 0.15, 0.4}
\definecolor{goldmetallic}{rgb}{0.83, 0.69, 0.22}
\definecolor{darkcoral}{rgb}{0.8, 0.36, 0.27}
\definecolor{cordovan}{rgb}{0.54, 0.25, 0.27}
\definecolor{oxfordblue}{rgb}{0.0, 0.13, 0.28}
\definecolor{navyblue}{rgb}{0.0, 0.0, 0.5}
\definecolor{prussianblue}{rgb}{0.0, 0.19, 0.33}

\definecolor{atmFill}{HTML}{DDE6F1}   % powder blue-gray (Atmosphere)
\definecolor{atmDraw}{HTML}{A6B7CE}
\definecolor{atmText}{HTML}{243447}

\definecolor{oceFill}{HTML}{DDEDEA}   % soft greenish aqua (Hydrosphere)
\definecolor{oceDraw}{HTML}{A4C3B8}
\definecolor{oceText}{HTML}{26332E}

\definecolor{cryFill}{HTML}{E7E9EE}   % soft cool gray (Cryosphere)
\definecolor{cryDraw}{HTML}{B5BAC3}
\definecolor{cryText}{HTML}{2A2E35}

\definecolor{landFill}{HTML}{F1E7DD}  % warm beige (Land)
\definecolor{landDraw}{HTML}{D3BCA5}
\definecolor{landText}{HTML}{3E342B}

\definecolor{bioFill}{HTML}{E7EFE5}   % pale sage (Biosphere)
\definecolor{bioDraw}{HTML}{B7C8B2}
\definecolor{bioText}{HTML}{2E3B2B}

 \usepackage{lettrine}

\usepackage{tikz}
\usetikzlibrary{positioning, shapes, arrows.meta, shadows.blur}

\usepackage[subtle]{savetrees} 

\begin{document}
\title{\LARGE \textsc{Climate Science and Control Engineering: \\ Insights, Parallels, and Connections} \vspace{0.5cm}}

\author{Salma M. Elsherif$^{\dagger, \P}$
        and~Ahmad F. Taha$^{\dagger,*}$
\thanks{$^\dagger$Department of Civil and Environmental Engineering, Vanderbilt University, Nashville, TN, USA. Emails: salma.m.elsherif@vanderbilt.edu, ahmad.taha@vanderbilt.edu.} \thanks{$^*$Corresponding author.}
\thanks{This research was \textit{not} supported by any funding agency. The authors pursued this work in their free time. }
\thanks{$\P$Secondary appointment: Department of Irrigation and Hydraulics Engineering, Faculty of Engineering, Cairo University.}}

\maketitle

\begin{abstract}
Climate science is the multidisciplinary field that studies the Earth's climate and its evolution---past, present, and future. This simultaneously includes studying the natural and human-caused factors that meaningfully influence and often adversely impact it. At the very core of climate science are indispensable climate models that predict future climate scenarios, inform policy decisions, and dictate how a country's economy should change in light of the changing climate. Climate models capture a wide range of interacting dynamic processes via extremely complex ordinary and partial differential equations---processes that span varying spatiotemporal scales. To model these large‑scale complex processes, state-of-the-art climate science leverages supercomputers, advanced simulations, statistical methods, and machine learning techniques to predict future climate. %In nations where governments adhere to what conclusive scientific evidence dictates, these predictions are used to define energy policies, investment portfolios, and urbanization plans. 

An area of applied math and engineering that is rarely studied or applied in climate science is control engineering---the analytical study of dynamic systems. Given that climate systems are inherently dynamic, it is intuitive to analyze them within the framework of dynamic system science. This perspective has been underexplored in the literature. Motivated by recent control engineering initiatives,  in this manuscript, we provide a tutorial that: \textit{(i)} introduces the control engineering community to climate dynamics and modeling, including spatiotemporal scales and challenges in climate modeling; \textit{(ii)} offers a fresh perspective on climate models from a control systems viewpoint; and \textit{(iii)} explores the relevance and applicability of various advanced graph and network control-based approaches in building a physics-informed framework for learning, control and estimation in climate systems. We also present simple and then more complex climate models, depicting fundamental ideas and processes that are instrumental in building climate change projections. 

This tutorial also builds parallels and observes connections between various contemporary problems at the forefront of climate science and their control theoretic counterparts. We specifically observe that an abundance of climate science problems can be linguistically reworded and mathematically framed as control theoretic ones. From state estimation, reachability analysis, and feedback control to model order reduction and system identification, the parallels between the two disciplines are delineated by describing the climate science equivalent of these controls concepts.  This perhaps merits further investigation of some of these problems through a control lens.  Grounded in control engineering, this tutorial in particular offers a new and robust foundation to potentially enhance strategic decision-making and support effective mitigation efforts. Finally, and as a disclaimer, we make no claims regarding whether this framework will be exceptionally useful to solve some climate science problems---the tutorial merely presents a fresh outlook on the problem while building reasonable analogies. 
\end{abstract}

\newpage

{\small     \tableofcontents}

\newpage

\section*{Acronyms}

\vspace{0.1cm}

\begin{table}[h]
    \centering
    \normalsize
    \renewcommand{\arraystretch}{1} 
        \begin{tabular}{r r}
        \textbf{CMIP} & \textit{Coupled Model Intercomparison Project} \\
\textbf{EBM} & \textit{Energy Balance Model}                   \\
\textbf{ESM} & \textit{Earth System Model}                     \\
\textbf{GCM} & \textit{General Circulation Model}              \\
\textbf{GHGs} & \textit{Greenhouse Gases}                      \\
\textbf{IAM} & \textit{Integrated Assessment Model}            \\
\textbf{IPCC} & \textit{Intergovernmental Panel on Climate Change} \\
\textbf{MPC} & \textit{Model Predictive Control}     \\
\textbf{NOAA} & \textit{National Oceanic and Atmospheric Administration} \\
\textbf{NWP} & \textit{Numerical Weather Prediction}   \\
\textbf{ODE} & \textit{Ordinary Differential Equation         }\\
\textbf{PDE} & \textit{Partial Differential Equation}          \\
\textbf{RCM} & \textit{Regional Climate Model}     \\
\textbf{SAI} & \textit{Stratospheric Aeorosal Injection}     \\
\textbf{SRM} & \textit{Solar Radiation Management}     

\end{tabular}
\end{table}

\vspace{0.5cm}

\begin{IEEEkeywords}
Climate modeling, control engineering, dynamic systems, climate-driven decision making, climate mitigation strategies.
\end{IEEEkeywords}

\section{Introduction and Tutorial Objectives}
% \IEEEPARstart{C}{Climate} 
\lettrine[lraise=0.1, nindent=0em, slope=-.5em]{C}{limate}  change is one of the most pressing challenges of our era, influencing ecosystems, economies, and societies on a global scale. Accurate climate modeling is essential for predicting future scenarios and informing strategic policies aimed at mitigation and adaptation. These models simulate the complex interactions between the Earth's atmosphere, oceans, land, ice, and biosphere---capturing feedback mechanisms, nonlinear behaviors, and the influence of human activities and their impacts. 

In fact, climate scientists and policymakers focus on projecting future temperature changes and understanding the role of greenhouse gases (GHGs), particularly carbon dioxide (CO$_2$), in driving these changes. The elevated concentrations of GHGs increase the greenhouse effect, leading to global warming and associated climate disruptions. Accurate projections of temperature and GHGs levels are vital for assessing impacts of climate change such as sea-level rise, extreme weather events, and ecosystem alterations, thereby guiding effective policy decisions.

\begin{figure}[t]
    \centering
    \includegraphics[width=1\linewidth]{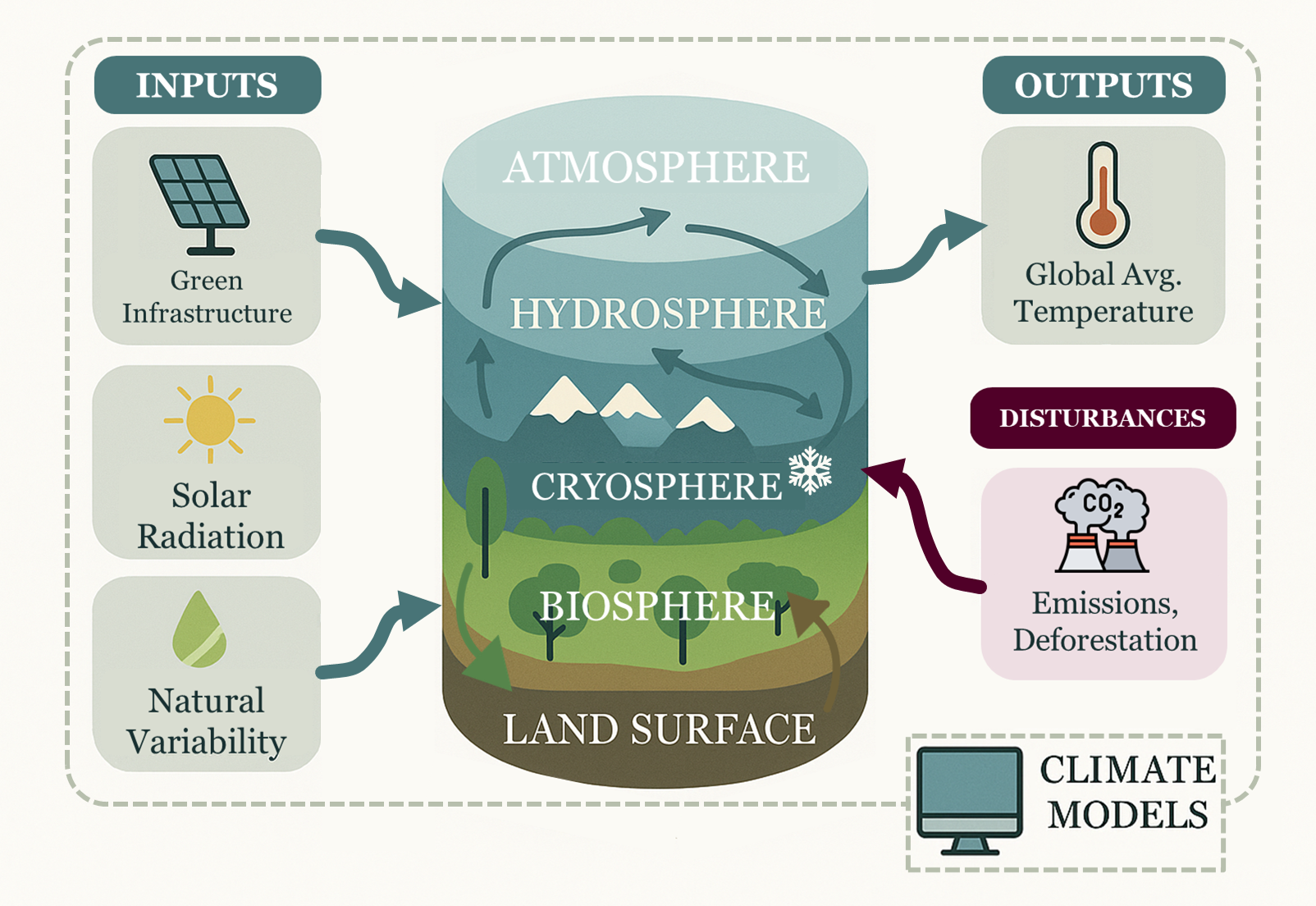}
    \caption{\textit{Climate as a dynamic system.} In this tutorial, we explain how the Earth’s climate can be viewed as a complex dynamical system with inputs, outputs, internal states, and external disturbances. We describe how the key components of the climate system---atmosphere, hydrosphere, cryosphere, biosphere, and land surface---interact and evolve over time. We cover the roles of natural and anthropogenic inputs (e.g., solar radiation, green infrastructure, and natural variability) and disturbances (e.g., emissions and deforestation), and how they influence the system’s trajectory. We delineate the different types and levels of climate models and highlight how those models map the climate inputs and disturbances to evolving states and observable outputs such as global average temperature, in the form of scenario-based simulations and projections. Throughout the tutorial, we highlight and examine these dynamic processes through a control-theoretic lens and discuss how control tools can provide structure for understanding, simulating, and potentially influencing climate dynamics. {We note here, as a much-needed disclaimer, that \textit{jointly} modeling all of the dynamic system components and inputs/outputs in this figure is virtually impossible as such models would not be practical computationally. Climate models wisely stick to subsets of these interactions rather than the whole kitchen sink, and we, too, respect that wisdom (mostly). This tutorial offers only some perspectives that are neither complete nor extraordinary. It just makes climate science more accessible to control system researchers and hopes for more cross-disciplinary learning and developments.  }}
    \label{fig:ClimDynm}
    % \vspace{-0.4cm}
\end{figure}

Despite significant advances in computational power, statistical and stochastic methods, and machine learning integration, control engineering principles are rarely presented as an alternative to study climate-related problems. Climate systems are at their core dynamic in nature, governed by underlying physical laws and exhibiting complex behaviors that evolve over time (Fig. \ref{fig:ClimDynm}). However, the application of control engineering principles (which are fundamentally suited to analyzing and managing such dynamic systems) remains relatively underexplored in the field of climate science.  The objectives of this tutorial are three-fold: \textit{(i)} introduce control engineering researchers to climate science and modeling; \textit{(ii)} produce a fresh outlook on climate problems; and \textit{(iii)} build parallels between various known and well-studied climate science problems and their control theoretic counterparts. 

\vspace{0.1cm}
\noindent \textbf{Why Control Theory for Climate Science and the Tutorial's Audience.}  
% The opportunities for control researchers and scientists in this field are vast, and the possibilities are boundless. 
This tutorial is introduced to begin addressing this challenge from our perspective as control researchers. Over the past decades, control theory has taken a proactive role in tackling real-life problems in dynamic systems, including infrastructure systems (e.g., energy systems~\cite{10684337} and transportation networks~\cite{10858622}), aerospace engineering~\cite{10479554}, epidemiology~\cite{VATTIATO2022100657}, and medicine~\cite{doyle2014closed}.  Control engineering offers a structured framework to analyze dynamic systems, optimize their performance, and design robust solutions for managing uncertainties while ensuring performance guarantee. By leveraging this perspective, we can expand our understanding of climate models and uncover new pathways for their utility and implications. Furthermore, the integration of advanced control-based approaches, such as graph theory and network control, holds significant promise for building physics-informed frameworks that can bridge the gap between modeling, control, and decision-making. 

The readership of this tutorial are either control systems research who are looking for applications for some of their theoretic techniques, control engineers interested in climate modeling and mitigation strategies, or climate scientists who are interested in applying new techniques to solve either vintage climate problems or more stressing and modern ones.  
 
\vspace{0.1cm}
\noindent \textbf{Some Efforts in Controls for Climate.} While some efforts have been made to apply optimization and control techniques to climate-related challenges, they have largely focused on specific goals or methodologies distinct from the approach presented here. For example, a group of researchers has examined weather and climate systems as dynamic systems influenced by human activity, with a particular emphasis on aerosol injections. In the studies \cite{soldatenkoWeatherClimateManipulation2017,soldatenkoOptimalControlPerspective2021}, the authors formulated optimal control problems to explore deliberate weather and climate modification. The authors modeled the atmosphere as a closed-loop dynamical system, where state variables included atmospheric parameters such as temperature and pressure, while control variables represented human interventions like aerosol injections. Their objective was to modify natural atmospheric processes to achieve desired weather outcomes, framed as a dynamic optimization problem. In their most recent work \cite{soldatenkoDesigningPrioriScenarios2025}, they applied optimal control theory to develop strategies for stratospheric aerosol injections, aiming to stabilize global climate temperatures by adjusting aerosol emission rates. {Another recent paper \cite{brody2025using} formally defines control inputs within climate models and demonstrates how scenario generation for aerosol injection can be treated as a control-guided dynamic process. In parallel, the work of \cite{kravitz2016geoengineering} frames geoengineering as a multivariable feedback control problem, emphasizing the design of climate intervention strategies that explicitly account for system uncertainty and feedback latency via a simple control methodologies. This design-driven view builds on physical and spatial insights from \cite{rasch2008overview} and \cite{ban2010geoengineering}, who explore the spatial heterogeneity and hemispheric responses to solar radiation management.  }

%\cite{ban2010geoengineering,brody2025using,rasch2008overview,kravitz2016geoengineering}

In addition, the study \cite{atoliaOptimalControlGlobal2023} has regarded the global climate system as a controlled dynamic system, with controls corresponding to economic activities causing emissions of GHGs. These efforts have applied nonlinear optimization techniques to solve dynamic control problems, aiming to maximize economic outcomes while adhering to environmental constraints. Similarly, the work in \cite{kellettFeedbackDynamicsOptimal2019} provides a tutorial on the \textit{dynamic integrated model of climate and economy model}, which employs discrete-time control to estimate optimal pathways for balancing GHGs emissions and economic development. 

The study \cite{caetanoReductionCO2Emission2009} takes a related approach by modeling the dynamic relationship between CO$_2$ emissions and investments in reforestation and clean technology, highlighting the importance of efficiently allocating resources to achieve emission reduction targets such as those outlined in the Kyoto Protocol. Another earlier noteworthy effort in \cite{filarApplicationOptimizationProblem1996}, has employed a two-stage optimization process to develop optimal CO$_2$ emission reduction strategies, demonstrating sensitivity to regional and global environmental targets. %Similarly, recent efforts in \cite{soldatenkoDesigningPrioriScenarios2025} have applied optimal control theory to develop strategies for stratospheric aerosol injections, aiming to stabilize global climate temperatures by adjusting aerosol emission rates. 

\vspace{0.1cm}
\noindent \textbf{Recent Control Engineering Initiatives.} The work in \cite{liangClimateModificationDirected2008} explores the application of control theory to systematically guide climate modification efforts. The study demonstrates how control analysis can inform the design of climate regulators through simple model applications and acknowledges the potential of applying control theory to climate regulation pathways. The magazine paper~\cite{khargonekarClimateChangeMitigation2024} has highlighted the potential for the control systems community to contribute to tackling challenges in climate change for specific applications such as transportation infrastructure, electric power grids, and agriculture. For instance, they outline, at a conceptual level, how control principles could support the integration of renewable energy into power systems, improve transportation system efficiency, and enhance decision-making frameworks in climate-aware agriculture, but without presenting any mathematical models, formulations, or methodological details. That is, this study primarily serves as a call to action by identifying broad opportunity areas rather than developing specific actual methods or techniques for implementation. This tutorial takes motivation from~\cite{khargonekarClimateChangeMitigation2024}. 

These aforementioned studies either focus on specific optimization techniques for particular problems or abstractly discuss opportunities without detailing practical implications or actionable methods. In contrast, this manuscript aims to address this interdisciplinary gap by providing a comprehensive tutorial that explores the intersection of climate science and control engineering. This tutorial covers the climate dynamics and the practical control-theoretic techniques applicable to such dynamics. By doing so, we hope to foster a deeper understanding of the potential contributions that control theory can make to tackling climate-driven challenges and inspire innovative research in this critical area. 

\begin{figure}[t]
    \centering
    \includegraphics[width=0.95\linewidth]{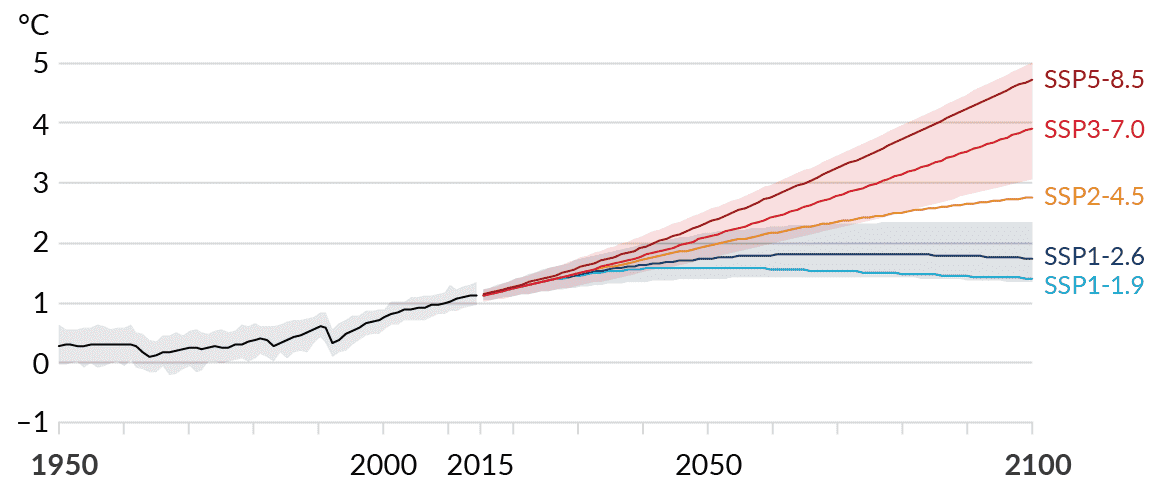}
    \caption{Projected global mean surface temperature changes relative to 1850--1900 levels under five emissions scenarios: very low emissions (\textit{SSP1-1.9}), low emissions (\textit{SSP1-2.6}), midlevel emissions (\textit{SSP2-4.5}), high emissions  (\textit{SSP3-7.0}), and very high emissions (\textit{SSP5-8.5}), illustrating the impact of different policy and socioeconomic pathways on future warming. Shading indicates 5\% to 95\% confidence intervals. Figure adapted from the IPCC Sixth Assessment Report (AR6) \cite{calvinIPCC2023Climate2023}. This tutorial shows how these (now culturally mainstream) figures are generated via dynamic models in climate science.}
    \label{fig:TempScenarios}
    \vspace{-0.4cm}
\end{figure}

To offer a conceptual preview, Fig.~\ref{fig:TempScenarios} illustrates how global temperature trajectories are influenced by alternative policy and emission scenarios as published in a report aimed at policymakers \cite{calvinIPCC2023Climate2023}---more details on this report are included in the following sections. These projections, developed through complex dynamical models of the climate system, exemplify the types of problems addressed in this tutorial: how to model, predict, and control the evolution of climate variables. Throughout the following sections, we build the foundations necessary to understand how such projections are constructed, and how control engineering perspectives can contribute to designing effective climate interventions. We now proceed to delineate the main objectives and broader implications of this tutorial.

\noindent 
\textbf{Tutorial Objectives and Broader Implications.} This tutorial aims to build a foundational framework that connects the principles of control engineering with the complexities of climate science. 

\begin{figure}
    % \centering
    % \includegraphics[width=0.5\linewidth]{}
    \begin{tikzpicture}[
    % font=\sffamily\small,
    node distance=0.9cm and 1.0cm,
    box/.style={rectangle, rounded corners=3pt, draw=sapphire!90, line width=1pt, align=left,
        fill=sapphire!80, text width=8.7cm, minimum height=1.0cm
        },
    objective/.style={rectangle, rounded corners=3pt, draw=sapphire!90, line width=1pt,
        fill=indigodye!10, text width=8.7cm, minimum height=1.2cm, align=left},
    open/.style={rectangle, rounded corners=3pt, draw=indigodye!70, line width=1pt,
        fill=indigodye!70, text width=8.7cm, minimum height=1.2cm, align=left},
    title/.style={font=\bfseries\large, align=center, text width=8cm},
    arrow/.style={
    line width=2.5pt, % Changed to a thicker line width
    -{Latex[length=3mm, width=3.5mm]},
    color=sapphire!90
  }
]

% === Goal ===
\node[box] (goal) {%
\textcolor{indigodye!10}{\textbf{Tutorial Goal:} {\textbf{Bridge Control and Climate Science}}\\
$\bullet$ Establish a framework linking control and climate research.\\
$\bullet$ Produce brief intro to various dynamic climate models.\\
$\bullet$ Summarize control efforts in climate literature.\\
$\bullet$ Build parallels between climate and control problems.  \\
$\bullet$ Formulate novel high-level climate control problems.  }};

% === Objective 1 ===
\node[objective, below=0.5cm of goal] (obj1) {%
\textcolor{sapphire!90}{\textbf{(Sec. \ref{sec:ClmtModels}--\ref{sec:ClimMdlEvol}) Introduce Climate Components}\\
$\bullet$ Discuss history of climate models and explain key physical processes \& dynamic nature of climate systems.\\
$\bullet$ Present  the applicability and goal of climate models.\\
$\bullet$ Review model hierarchies, spatiotemporal scales, and limitations.
}};

% === Objective 2 ===
\node[objective, below=0.5cm of obj1] (obj2) {%
\textcolor{sapphire!90}{\textbf{(Sec. \ref{sec:ClmvsWth}--\ref{sec:MoreComplxModel}) Control Modeling for Climate Problems}\\
$\bullet$ Recast climate models as MIMO state-space systems.\\
$\bullet$ Describe rudimentary and more complex climate models. \\
$\bullet$ Identify states, inputs, uncertainty, and measurable outputs.\\
$\bullet$ Facilitate estimation, control, uncertainty quantification. }};

% === Objective 3 ===
\node[objective, below=0.5cm of obj2] (obj3) {%
\textcolor{sapphire!90}{\textbf{(Sec. \ref{sec:ClmvsWth}--\ref{sec:MoreComplxModel}) Control-Theoretic Insights \& Applications}\\
$\bullet$ Demonstrate role of control for climate regulation.\\
$\bullet$ Describe observability and controllability-related problems in climate dynamics.\\
$\bullet$ Explore control-based, geoengineering design frameworks for mitigation and adaptation strategies. 
}};

% === Objective 4 ===
\node[objective, below=0.5cm of obj3] (obj4) {%
\textcolor{sapphire!90}{\textbf{(Sec.~\ref{sec:Clmt&CntrlEng}) Parallels Between Climate \& Control Sciences}\\
$\bullet$ Observability analysis and state estimation $\leftrightarrow$ state inference from sparse data and missing data problem.\\
$\bullet$ System identification $\leftrightarrow$ climate model calibration.\\
% $\bullet$ Robust control $\leftrightarrow$ uncertainty quantification and m.\\
$\bullet$ Optimal control $\leftrightarrow$ mitigation and adaptation policies.
}};

% === Open Problems ===
\node[open, below=0.5cm of obj4] (open) {%
\textcolor{sapphire!10}{\textbf{\textbf{(Sec.~\ref{sec:Clmt&CntrlEng}) Open Problems in Climate Science:}}\\
$\bullet$ Inferring hidden states from sparse/noisy observations.\\
$\bullet$ Quantifying uncertainty in nonlinear, large-scale models.\\
$\bullet$ Designing robust and optimal mitigation/control policies.\\
$\bullet$ Integrating physics-based and data-driven learning frameworks. 
}};

% === Broader Implications ===
\node[box, below=0.5cm of open, draw=prussianblue!80, fill=prussianblue!80] (imp) {%
\textcolor{sapphire!10}{\textbf{(Sec.~\ref{sec:Conc}) Tutorial Summary and Concrete Problems:}\\
$\bullet$ Climate models as testbeds for advancing control theory.\\
$\bullet$ Reachability \& observability of large-scale climate models.\\
$\bullet$ Optimized sensor placement and orbital designs.\\
$\bullet$ Scalable model order reduction and system identification. 
}};

% % === Footer ===
% \node[below=0.5cm of imp, align=center, font=\footnotesize\itshape, text width=9.5cm] (footer) {%
% \textbf{Tutorial Flow:} (II) Climate Basics $\rightarrow$ (III) Model Evolution $\rightarrow$ (IV--V) Control Representations $\rightarrow$ (VI--VIII) Insights \& Open Problems $\rightarrow$ (IX) Summary \& Outlook
% };

% === Clean vertical arrows ===
\foreach \i/\j in {goal/obj1, obj1/obj2, obj2/obj3, obj3/obj4, obj4/open, open/imp}
    \draw[arrow] (\i.south) -- (\j.north);

\end{tikzpicture}
    \caption{Tutorial summary and paper organization.}
    \label{fig:outline}
\end{figure}

\begin{itemize}
\item \textit{Introducing Climate Dynamics and Models to the Control Community:} 
We aim to provide control engineers and scientists with a comprehensive understanding of climate systems, highlighting their dynamic nature and the physical processes they encompass. We shed the light on how these dynamics can be conceptualized as controllable systems with multiple inputs, states, and outputs. We also survey the literature on the state-of-the-art in climate models, discussing their various scales included processes, limitations, and challenges. ({{Sections \ref{sec:ClmtModels} and \ref{sec:ClimMdlEvol}}})

\item \textit{Control-Oriented Representations of Climate Models:} We introduce and present how climate models can be interpreted as control-oriented dynamical representations with the temporal evolution of variables/parameters as states, and system control inputs (e.g., in some cases, policies or human influence), and measured outputs of the systems corresponding to observable climate indicators. This structured, analytical approach enables the application of control and network-based techniques to climate systems. Such methodologies can enhance accurate modeling and prediction, determine optimal mitigation strategies, facilitate systematic analysis, and integrate and quantify uncertainties. ({{Sections \ref{sec:ClmvsWth}, \ref{sec:MathModels}, and \ref{sec:MoreComplxModel}}}) 

\item \textit{Control-Theoretic Insights and Applications:} We survey and demonstrate the relevance, potential, and applicability of control-theoretic and network-based techniques that can be applied to the control-oriented representations of climate dynamics. These techniques are further explored for their implications and outcomes in developing physics-informed frameworks for climate learning, control, and estimation. Leveraging such frameworks supports decision-making processes and drives mitigation efforts. Additionally, we identify potential research directions and practical applications that can benefit from this interdisciplinary approach. Some control engineering efforts have taken place under the umbrella of climate geoengineering. Such efforts are briefly summarized too.  ({{Sections \ref{sec:MathModels}, \ref{sec:MoreComplxModel},  \ref{sec:GeoEngCntProb}, and \ref{sec:Clmt&CntrlEng}}})

\item \textit{Drawing Parallels Between Climate Science Challenges and Control Engineering Concepts:} We highlight conceptual and structural similarities between problems in climate science and control engineering. These include, for instance, the role of observability in inferring unmeasured climate states from sparse data. We also discuss the resemblance between climate model calibration and system identification, the parallels between uncertainty quantification in climate projections and robust control under model uncertainty, and the interpretation of mitigation policies as constrained optimal control problems. We note that although we make these observations in this manuscript, the aforementioned problems have \textit{not} been studied through the lens of control engineering. By outlining these connections, we aim to foster a shared vocabulary and problem structure that can bridge the two fields and promote new interdisciplinary strategies. ({{Sections \ref{sec:MathModels}, \ref{sec:MoreComplxModel}, and \ref{sec:Clmt&CntrlEng}}})

\end{itemize}

It is important to emphasize that the purpose of this tutorial is not to teach climate scientists control theory, nor to suggest that existing climate science lacks methodological depth. Rather, this tutorial is intended for the control engineering community, aiming to introduce climate systems as a rich, complex, and practically relevant domain where control-theoretic tools can be meaningfully applied. Climate models provide natural testbeds for developing and evaluating methods in large-scale dynamical systems, estimation under uncertainty, and robust decision-making---all areas where control theory has much to contribute. We simply describe what control systems theory can potentially deliver for climate research.  The structure of the tutorial is designed to support this goal and the aforementioned objectives. The tutorial organization is provided in Fig.~\ref{fig:outline}.
% it begins with the basics and foundations of climate science and dynamics (Section~\ref{sec:ClmtModels}). Section \ref{sec:ClimMdlEvol} explains the evolution and classification of climate models. Section \ref{sec:ClmvsWth} discusses the distinction between weather and climate and how these concepts can be interpreted within the control-oriented framework introduced in this tutorial. Sections \ref{sec:MathModels} and~\ref{sec:MoreComplxModel} highlight the dynamic nature of climate models from a control-theoretic perspective and poses questions that can be tackled using control methods. Section~\ref{sec:GeoEngCntProb} presents high-level description of some geoengineering as control problems, from the recent literature. This is followed by Section \ref{sec:Clmt&CntrlEng}, which examines climate dynamics through the lens of control engineering, offering interpretations and insights. In particular, this section draws explicit parallels between challenges in climate science and their counterparts in control theory. Section \ref{sec:Conc} concludes the tutorial with its summary and potential next steps and recommendations for the interested researchers. 

\begin{table*}[t!]
	\caption{Climate Processes and Modeling List of References.}~\label{tab:ClimRef}
	\centering
	\renewcommand{\arraystretch}{1.2} % Adjusts row height
	\begin{tabularx}{\textwidth}{>{\bfseries}l X p{2cm}}
		\toprule
		Topics & \textbf{Overview} & \textbf{References} \\
		\midrule
%		\rowcolor{cyan!10}	
		Atmospheric Processes & Foundations of atmospheric physics and their mathematical representations & \cite{houghtonPhysicsAtmospheres2002,andrewsIntroductionAtmosphericPhysics2010,holtonIntroductionDynamicMeteorology2013, achatzAtmosphericDynamics2022} \\  
%		\rowcolor{cyan!5}	
		Ocean Processes & Physical mechanisms and processes governing the ocean &  \cite{OceanCirculationClimate2001, kanthaNumericalModelsOceans2000, talleyDescriptivePhysicalOceanography2011, olbersOceanDynamics2012, apelPrinciplesOceanPhysics2013, gangopadhyayIntroductionOceanCirculation2022} \\
%		\rowcolor{cyan!10}	
		Ocean Processes Modeling & Introduction to the physical, mathematical, and numerical foundations of computer models used to understand and predict the global ocean climate system  &  \cite{griffiesFundamentalsOceanClimate2018} \\
%		\rowcolor{cyan!5}	
		Atmosphere and Ocean & Atmospheric and ocean processes and interactions &  \cite{nihoulCoupledOceanAtmosphereModels1985, krausAtmosphereOceanInteraction1994, wellsAtmosphereOceanPhysical2011,  gillAtmosphereOceanDynamics2016, vallisAtmosphericOceanicFluid2017, vonstorchConceptLargeScaleConditioning2018, xieCoupledAtmosphereOceanDynamics2022, staniforthGlobalAtmosphericOceanic2022, zardiAtmosphereOceanInteractions2024} \\
%		\rowcolor{cyan!10}	
		Ocean and Climate & Interactions between oceanic processes and the climate system, highlighting the oceans' integral role in climate regulation & \cite{biggOceansClimate2003} \\
%		\rowcolor{cyan!5}	
		Atmosphere and Climate & Analysis and prediction of atmosphere processes and their feedbacks with the other systems spheres  & \cite{moteNumericalModelingGlobal2000,pytharoulisClimateAtmosphericDynamics2021}\\
%		\rowcolor{cyan!10}	
		Atmosphere/Temperature Variability & Examines atmospheric temperature variability from a dynamical perspective and explores how it responds to climate change & \cite{tamarin-brodskyDynamicalPerspectiveAtmospheric2019} \\ 
%		\rowcolor{cyan!5} 
		Cryosphere Processes & Fundamentals and processes of ice sheets, glacier, and frozen water parts & \cite{walshCrysophereHydrology2006,colbeckDynamicsSnowIce2012,marshallCryosphere2012,veenFundamentalsGlacierDynamics2013,fowlerGlaciersIceSheets2021,barryGlobalCryosphereSecond2022} \\
%		\rowcolor{cyan!10} 
		Hydrosphere & Focus on the physics and dynamics of all the water on earth --- some references include the glaciers in this sphere & \cite{arayaHydrosphere2005,walshCrysophereHydrology2006,hayesEarthsHydrosphere2018} \\ 
%		\rowcolor{cyan!5} 
		Biosphere & Overview of the evolution of the living organisms on Earth and their interactions with their environment and other spheres & \cite{bengtssonGeospherebiosphereInteractionsClimate2001,cowellEarthsBiosphereEvolution2003,smilEarthsBiosphereEvolution2003,dobretsovBiosphereOriginEvolution2008,vernadskyBiosphere2012,ramblerGlobalEcologyScience2013} \\
%		\rowcolor{cyan!10} 
		Land Surface / Lithosphere & Components and geodynamic processes of the lithosphere & \cite{stuweGeodynamicsLithosphereIntroduction2003,artemevaLithosphereInterdisciplinaryApproach2011,hosaniLithosphereDynamicsSedimentary2012} \\
		\midrule
%		\rowcolor{green!10}	
		Climate Dynamics and Processes & Climate components, process, interactions& \cite{sellersGlobalClimaticModel1969,trenberthClimateSystemModeling1992,chenGeodynamicalInterconnectionsAtmosphere1998,dijkstraNonlinearClimateDynamics2013,goosseClimateSystemDynamics2015,cookClimateDynamics2nd2025} \\
%		\rowcolor{green!5} 
		Climate Models & Climate models evolution, hierarchy, processes, comparisons, and limitations & \cite{mcguffieClimateModellingPrimer1998,mcguffieFortyYearsNumerical2001,washingtonIntroductionThreeDimensionalClimate2005,randallClimateModelsTheir2007,glecklerPerformanceMetricsClimate2008,heldGapSimulationUnderstanding2005,provenzaleClimateModels2014,stockerIntroductionClimateModelling2011,kawamlehConfirmingClimateChange2022,northEnergyBalanceClimate2017,eyringTakingClimateModel2019,imkellerStochasticClimateModels2001, alloydClimateModellingPhilosophical2018, baderClimateModelsAssessment,jeevanjeePerspectiveClimateModel2017} \\
%		\rowcolor{green!10} 
		Regional Climate Models & Focus on evolution and development of regional climate models& \cite{dickinsonRegionalClimateModel1989,giorgiIntroductionSpecialSection1999,rummukainenStateoftheartRegionalClimate2010,collinsChallengesOpportunitiesImproved2018,foleyUncertaintyRegionalClimate2010,feserRegionalClimateModels2011,giorgiThirtyYearsRegional2019,tapiadorRegionalClimateModels2020} \\
		\midrule
%		\rowcolor{yellow!10} 
		Climate Change & Encompassing topics such as the physical science basis, observed impacts, mitigation strategies, policy frameworks, and adaptation measures & \cite{houghtonClimateChange19951995,jonesRegionalHydrologicalResponse1996,flemingHistoricalPerspectivesClimate1998,houghtonClimateChange20012001,dawsonCompleteGuideClimate2009,bulkeleyGoverningClimateChange2010,neelinClimateChangeClimate2010,councilAdvancingScienceClimate2011,pierceSelectingGlobalClimate2009,eggletonShortIntroductionClimate2012,tannerClimateChangeDevelopment2014,franchitoStudiesClimateChange2015,metzControllingClimateChange2010,burroughsClimateChangeMultidisciplinary2001,obrienAdaptiveChallengeClimate2015,forsterLatestClimateModels2020,massonUrbanClimatesClimate2020,desslerIntroductionModernClimate2021,thunbergClimateBookFacts2024} \\
		\midrule
%		\rowcolor{red!7} 
		IPCC Assessment Reports & IPCC has released six assessment reports on climate change, its causes, potential impacts, and response options & \cite{changeIPCCFirstAssessment1990,changeIPCCSecondAssessment1995,fosterIPCCThirdAssessment2001,pachauriIPCCFourthAssessment2007,ipccIPCCFifthAssessment2014,calvinIPCC2023Climate2023} \\
		\bottomrule
	\end{tabularx}
\end{table*}

 The information presented in this manuscript on climate modeling and associated processes is based on the references listed in Tab. \ref{tab:ClimRef}. This compilation serves as a resource for researchers seeking foundational and advanced knowledge in the field. The table encompasses a range of topics, including atmospheric dynamics, ocean circulation, coupling climate dynamics, climate modeling, and climate change risks, providing a comprehensive overview of the literature that has shaped our current understanding of climate systems.

\begin{table*}[t]
	\centering
	\caption{Summary of Climate System Components, Examples of their Governing Processes, and Modeling Characteristics.}~\label{tab:ClimCompCharct}
\renewcommand{\arraystretch}{1.2} % Adjusts row height
%	\rowcolors{2}{gray!20}{gray!10} % Start coloring from the first data row
\begin{tabularx}{\textwidth}{>{\bfseries}l X l >{\hsize=.6\hsize}X >{\hsize=1.4\hsize}X}
	\toprule
		\textbf{Component} & \textbf{Processes Examples \& Modeled States} & \textbf{Time Scale} & \textbf{Spatial Scale} & \textbf{Numerical Representation} \\ \midrule
		Atmosphere & Large-scale circulation, states: wind velocity components, pressure, temperature, specific humidity & Days to weeks & 100–1000 km & Nonlinear PDEs: momentum, mass, and energy equations for a rotating, stratified atmosphere; vertical motion simplified using hydrostatic balance
		 \\ \cline{2-5}
		& Convection and turbulence, states: vertical velocity, heat and moisture fluxes & Minutes to hours & 1–10 km & High-resolution nonlinear PDEs (Navier-Stokes + thermodynamics)---often parameterized \\ \cline{2-5}
		& Radiative transfer, states: radiative fluxes (shortwave and longwave) & Seconds to hours & 1D vertical (column-wise), applied across global grid & Radiative transfer  integro-differential equations, often with semi-empirical approximations to account for absorption, scattering, and emission \\ \midrule
		Hydrosphere & Ocean circulation (thermohaline, wind-driven), states: ocean currents, ocean temperature, salinity, sea surface height & Months to millennia & 10–1000 km & Nonlinear PDEs: conservation of momentum, mass, temperature, and salinity under Earth's rotation; solved using layered 3D ocean models \\ \cline{2-5}
		& Surface wave dynamics, states: wave height, frequency spectrum, direction & Seconds to hours & 1-100 m & Nonlinear PDEs (wave action balance equation) with algebraic source/sink terms (e.g., wind input, dissipation) \\ \midrule
		Cryosphere & Ice sheet flow, states: ice thickness, velocity, basal temperature & Years to millennia & 10–1000 km & Nonlinear PDEs for slow viscous flow, using shallow ice and shallow shelf approximations to simulate spreading, sliding, and thinning of ice
		 \\ \cline{2-5}
		& Snow cover and melt dynamics, states: snow depth, snow temperature, liquid water content & Days to seasons & 1D vertical (column wise) over a horizontal grid of scale 1–100 km & 1D ODEs/PDEs describing snow temperature and melt based on energy balance; includes radiation, conduction, and phase change processes
		 \\ \midrule
		Land surface & Soil moisture and heat transport, states: soil moisture content, soil temperature profile & Hours to seasons & 1D vertical (column wise) over a horizontal grid of scale 0.1–10 km & Coupled 1D PDEs for water and heat transport in the soil column (Richards' equation and heat conduction) \\ \cline{2-5}
		& Surface-atmosphere exchange (energy, momentum), states: surface temperature, latent and sensible heat fluxes, momentum flux & Minutes to hours & 1-10 m when modeled locally \& grid-cell scale of the model when parameterized & Algebraic flux formulations and empirical parameterizations for heat, moisture, and momentum exchange; sometimes coupled with simple ODEs (e.g., surface temperature evolution)
		 \\ \midrule
		Biosphere & Vegetation dynamics and photosynthesis, states: biomass, carbon assimilation rate & Seasonal to decades & 1–1000 km & Coupled ODEs representing carbon uptake, biomass allocation, and plant response to environment \\ \cline{2-5}
		& Biogeochemical cycling (carbon, nitrogen), states: carbon and nitrogen pools (e.g., soil organic carbon, plant carbon), nutrient fluxes & Years to centuries & Regional to global & Coupled ODEs for elemental stocks and flows (e.g., carbon pools), with algebraic constraints for conservation and stoichiometry \\ 
		\bottomrule
\end{tabularx}
\end{table*}

\section{Birds' Eye View of Climate Models: From Simple Physics to the Discipline of Earth Science}~\label{sec:ClmtModels}
This section provides a high-level overview of the climate system and its core components, setting the stage for how simple physical principles scale up to form the basis of Earth system science defined as {the field that studies the Earth as one integrated system, where the interactions between its various components are examined and studied}. These components are comprised of five main ones of the climate system (Fig. \ref{fig:ClimComp}). A brief introduction to each is given next, while Tab.~\ref{tab:ClimCompCharct} summarizes the key physical and modeling characteristics of each. The table outlines the typical time scales (ranging from minutes to millennia), spatial scales (from meters to thousands of kilometers), and types of mathematical representations used in numerical climate models. As observed, each of these components is governed by fundamental conservation laws (e.g., of mass, momentum, and energy), with majority typically expressed through ordinary differential equations (ODEs) or partial differential equations (PDEs). These range from 1D vertical ODEs in snow and soil models, to high-dimensional nonlinear PDEs in atmosphere and ocean dynamics. This framing not only highlights the physics-based structure of climate models but also underscores their direct relevance to the principles of dynamic systems modeling and analysis.

% \comment{add more dynamic systems context in this section and in the following bullet points: explain briefly what kind of basic conservation models are used, how they're typically modeled (ODEs vs PDEs), describe the typical time and spatial scales, and say ``oh this demonstrates relevance to dynamic system principles".} 
% \comment{add your ugly figure here}
% \comment{add the phrase five components and put them in a box in addition to the figure (the ugly one).} 

\begin{custombox1}{white!60!blue}
\textbf{The five main climate components:}
\begin{enumerate}
	\item \textit{Atmosphere:} The layer of gases surrounding Earth, responsible for weather patterns, climate regulation, and the greenhouse effect.
	\item \textit{Hydrosphere:} All water on Earth, including oceans, lakes, rivers, and groundwater, play an important role in heat storage and transport.
	\item \textit{Cryosphere:} Frozen water on Earth, such as glaciers, ice sheets, and permafrost, affecting sea levels and Earth's energy balance. Note that, some scientific disciplines and studies consider the cryosphere as part of the hydrosphere.
	\item \textit{Land Surface:} The solid Earth, including landforms, soils, and surface processes, which influence climate through heat exchange, water storage, and interactions with vegetation.
	\item \textit{Biosphere:} All living organisms, from microbes to forests, that interact with the climate system through carbon, water, and energy cycles.
\end{enumerate}
\end{custombox1}

\begin{figure}[t]
	\centering
    \begin{tikzpicture}[every node/.style={align=center}]
\def\radius{2.2} 
\node[rectangle, draw=blue!20, fill=blue!20, rounded corners, minimum width=2.0cm, yshift=1.2cm, minimum height=1.2cm] 
(atmo) at ({90}:\radius) {{\textbf{Atmosphere}}\\[0.6mm] {\smaller Gases \& Winds}};

% \node[rectangle, draw=cyan!20, fill=cyan!20, rounded corners, minimum width=2.0cm, minimum height=1.2cm, xshift=-1.2cm] 
% (ocean) at ({162}:\radius) {\textbf{Hydrosphere}\\[0.6mm] {\smaller Currents \& Salinity}};

% \node[rectangle, draw=gray!20, fill=gray!20, rounded corners, minimum width=2.0cm, minimum height=1.2cm] 
% (cryo) at ({234}:\radius) {\textbf{Cryosphere}\\[0.6mm] {\smaller Ice \& Snow}};

% \node[rectangle, draw=green!20, fill=green!20, rounded corners, minimum width=2.0cm, minimum height=1.2cm, xshift=0.7cm] 
% (land) at ({306}:\radius) {\textbf{Land}\\[0.6mm] {\smaller Soil \& Vegetation}};

% \node[rectangle, draw=orange!20, fill=orange!20, rounded corners, minimum width=2.0cm, minimum height=1.2cm, xshift=1.2cm] 
% (bio) at ({18}:\radius) {\textbf{Biosphere}\\[0.6mm] {\smaller Life \& Biomass}};

\node[rectangle,  fill=atmDraw, rounded corners,
      minimum width=2.0cm, yshift=1.2cm, minimum height=1.2cm, text=atmText] 
(atmo) at ({90}:\radius) {\textbf{Atmosphere}\\[0.6mm] {\smaller Gases \& Winds}};

\node[rectangle, fill=oceDraw, rounded corners,
      minimum width=2.0cm, minimum height=1.2cm, xshift=-1.2cm, text=oceText] 
(ocean) at ({162}:\radius) {\textbf{Hydrosphere}\\[0.6mm] {\smaller Currents \& Salinity}};

\node[rectangle, fill=cryDraw, rounded corners,
      minimum width=2.0cm, minimum height=1.2cm, text=cryText] 
(cryo) at ({234}:\radius) {\textbf{Cryosphere}\\[0.6mm] {\smaller Ice \& Snow}};

\node[rectangle, fill=landDraw, rounded corners,
      minimum width=2.0cm, minimum height=1.2cm, xshift=0.7cm, text=landText] 
(land) at ({306}:\radius) {\textbf{Land}\\[0.6mm] {\smaller Soil \& Vegetation}};

\node[rectangle, fill=bioDraw, rounded corners,
      minimum width=2.0cm, minimum height=1.2cm, xshift=1.2cm, text=bioText] 
(bio) at ({18}:\radius) {\textbf{Biosphere}\\[0.6mm] {\smaller Life \& Biomass}};

\draw[-{Latex[length=2mm]}, thick, bend left=20] 
    (atmo) to node[midway, left, yshift = -0.7cm, xshift = 0.2cm,font=\scriptsize\itshape] {Heat \\ Flux} (ocean);
\draw[-{Latex[length=2mm]}, thick, bend left=40] 
    (atmo) to node[xshift=0.6cm, yshift = -1.8cm, midway, left, font=\scriptsize\itshape] {Radiation \\ Temp} (cryo);
  \draw[-{Latex[length=2mm]}, thick, bend right=20] 
    (land) to node[xshift=-1.05cm, yshift = 1.7cm, midway, right, font=\scriptsize\itshape] {Evapotransp.} (atmo);
  
\draw[-{Latex[length=2mm]}, thick, bend right=20] 
    (atmo) to node[midway, xshift=-2cm, yshift = -1.5cm, right, font=\scriptsize\itshape] {Precipitation \\ Temp} (land);
\draw[-{Latex[length=2mm]}, thick, bend left=40] 
    (ocean) to node[midway, left, yshift=-1cm, xshift = 2.5cm, font=\scriptsize\itshape] {Nutrients \\ Heat} (bio);
\draw[-{Latex[length=2mm]}, thick, bend right=20] 
    (land) to node[midway, right, yshift=-2mm, font=\scriptsize\itshape] {Water \\ Carbon} (bio);
\draw[-{Latex[length=2mm]}, thick, bend left=20] 
    (cryo) to node[midway, left, xshift=-1mm, font=\scriptsize\itshape] {Melting \\ Freshwater} (ocean);
\draw[-{Latex[length=2mm]}, thick, bend right=30] 
    (bio) to node[midway, right, xshift=1mm, font=\scriptsize\itshape] {CO$_2$ release\\  Evapotranspiration} (atmo);
\draw[-{Latex[length=2mm]}, thick, bend left=30] 
    (ocean) to node[midway, above, yshift=2mm, font=\scriptsize\itshape] {Heat \\ Flux} (atmo);
\end{tikzpicture}
	\caption{Climate system's main five components and selected interactions between them (illustrative, not exhaustive).}~\label{fig:ClimComp}
    \vspace{-0.4cm}
\end{figure}

\begin{figure*}[h!]
	\centering
	\includegraphics[width=0.94\textwidth]{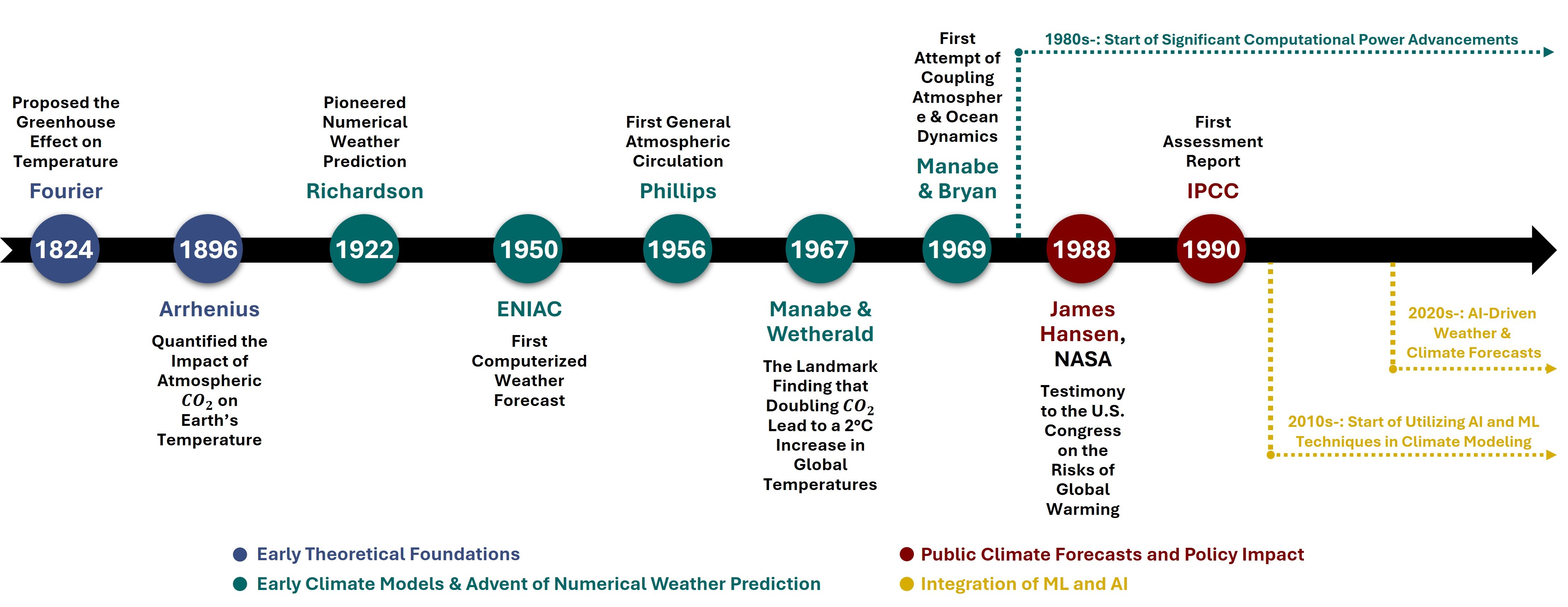}
	\caption{Key transitional milestones in the history of climate science.}~\label{fig:ClimateModHist} %\vspace{-0.5cm}
\end{figure*}

Modeling these components and their interactions is an intensively complex task. Yet, the small advances across many scientific areas have shaped the Earth science discipline and climate modeling. This progression dates back to at least the seventeenth century, beginning with the study of atmospheric science. The early efforts treat phenomena as stand-alone dynamics, translating them into mathematical formulations based on physical principles or observational data. Over time, researchers has begun studying interactions and coupling between these phenomena, which has led to the development of the early climate models. With advances in numerical simulations and the information revolution, modern climate models have become increasingly complex and arguably more accurate. 

A high-level overview of the key transitional milestones in the history of climate science is illustrated in Fig. \ref{fig:ClimateModHist}. These milestones have shaped climate and Earth science into one of the most widely discussed topics in theoretical research, academia, and policy. We present a brief history of climate science in what follows. 
\begin{enumerate}
	\item \textbf{Early Theoretical Foundations:}
	\begin{itemize}
		\item \textit{1824:} The French physicist \textit{Joseph Fourier} has proposed what we know now as greenhouse effect on the atmosphere's heat and accordingly temperature.
		\item \textit{1896:} The Swedish chemist \textit{Svante Arrhenius} has manually quantified the impact of atmospheric CO$_2$ on Earth's temperature, laying the groundwork for future climate studies \cite{arrheniusInfluenceCarbonicAcid1896}.
	\end{itemize}
	\item \textbf{Early Climate Models and Advent of Numerical Weather Prediction:}
		\begin{itemize}
			\item \textit{1922:} The British scientist \textit{Lewis Fry Richardson} has pioneered numerical weather prediction by attempting to calculate weather changes using mathematical equations, though his methods were limited by the computational tools of his time \cite{richardsonWeatherPredictionNumerical1922}.					
			\item \textit{1950:} The first successful computerized weather forecast is generated using the Electronic Numerical Integrator and Computer (ENIAC) computer, marking a significant leap in meteorological science \cite{brechtMENIACPhysicsInformedMachine2024}. 
			\item \textit{1956:} The Meteorologist \textit{Norman Phillips} has developed a two-layer atmospheric model capable of simulating general atmospheric circulation, representing a foundational climate model \cite{phillipsNumericalWeatherPrediction1960,edwardsHistoryClimateModeling2011}.
			\item \textit{1967:} \textit{Syukuro Manabe} and \textit{Richard Wetherald} have introduced a one-dimensional atmospheric radiative-convective model, demonstrating that doubling CO$_2$ levels could lead to a 2$^\circ$C increase in global temperatures. These results have been marked as a landmark finding in climate science, providing early evidence of human-induced climate change \cite{manabeThermalEquilibriumAtmosphere1967}.  
			\item \textit{1969:} \textit{Syukuro Manabe} and \textit{Kirk Bryan} have introduced the first attempt of coupling the atmosphere and ocean dynamics in one compact model \cite{manabeClimateCalculationsCombined1969}.
			\item \textit{1980s:} This decade marked the start of significant advancements in climate modeling, driven by increased computational power. These improvements have enabled more complex models that captured interactions between the different climate components such as atmosphere, oceans, and land surfaces, leading to the more accurate climate projections used today.
		\end{itemize}
	\item \textbf{Public Climate Forecasts and Policy Impact:}
	\begin{itemize}
		\item \textit{1988:} NASA scientist \textit{James Hansen} has testified before the US Congress, presenting model-based evidence of global warming, bringing climate modeling to the forefront of public policy discussions \cite{shabecoffGlobalWarmingHas1988}. 
		\item \textit{1990:} The Intergovernmental Panel on Climate Change (IPCC) has released its First Assessment Report, heavily relying on climate models to inform global policy on climate change \cite{changeIPCCFirstAssessment1990}. Since then, IPCC has published six more assessment reports in total, with the most recent being the Sixth Assessment Report (AR6) from 2021–2023. The IPCC is now in its seventh assessment cycle, which began in July 2023, with the Seventh Assessment Report (AR7) expected by late 2029.
	\end{itemize}
	\item \textbf{Integration of Machine Learning (ML) and Artificial Inelegance (AI):}
	\begin{itemize}
		\item \textit{2010s:} Researchers have begun exploring AI and ML to improve climate models, focusing on enhancing the simulations and predictions accuracy \cite{hauptHistoryPracticeAI2022}. 		
		\item \textit{2020s:} AI-driven models have been emerging, offering faster and potentially more accurate weather and climate forecasts by analyzing extensive datasets and identifying complex patterns \cite{niilerAILearningPredict2024}.
	\end{itemize}
\end{enumerate}

\begin{figure*}[h!]
	\centering
	\includegraphics[width=0.98\textwidth]{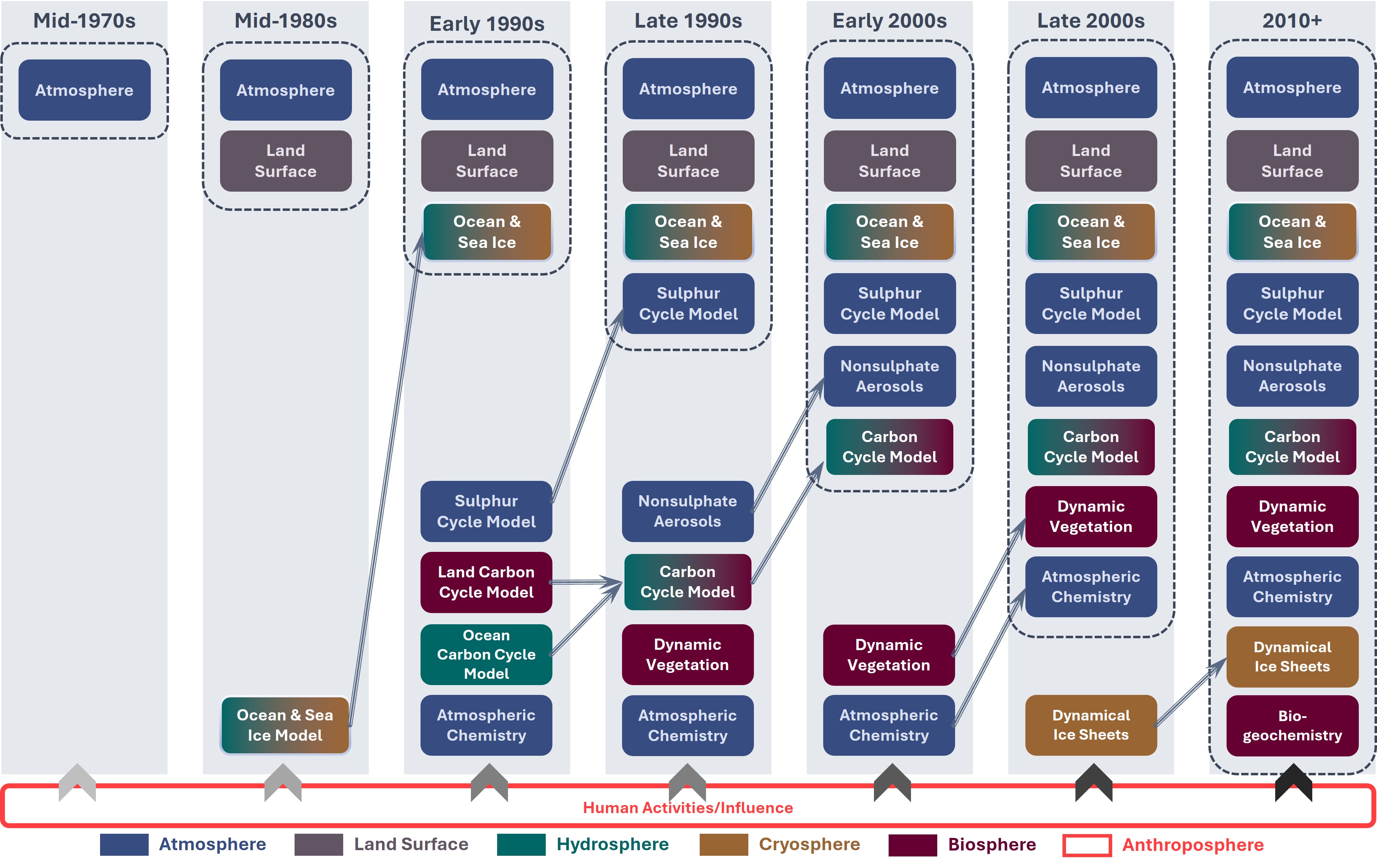}
	\caption{Replica of the chronological evolution of climate modeling from IPCC report summarized in \cite{TimelineClimateModels}, extended by explicitly including the anthroposphere which represents human activities and their influence on the climate system. Compared to~\cite{TimelineClimateModels}, this figure also maps processes to their respective climate system components and uses thicker arrows to highlight the increasing contribution of human activity to climate change, now explicitly incorporated into climate modeling frameworks.}~\label{fig:ClModEvo} \vspace{-0.5cm}
    % \caption{Replica of the chronological evolution of climate modeling from IPCC report summarized in \cite{TimelineClimateModels}. Compared to~\cite{TimelineClimateModels}, this figure illustrates which processes belong to which of the five main climate system components. \comment{add the anthroposhpere and define it within the figure and use thicker arrows to indicate increased human contribution to climate}}~\label{fig:ClModEvo}
\end{figure*}

% \comment{add more content here and consider giving this paragraph a header in bold} 
\vspace{0.1cm}
\noindent \textbf{The Anthroposphere: Modeling Our Mark on the Planet.} Another important element influencing the climate, beyond natural elements, is the \textit{anthroposphere}. This term refers to parts of the Earth's system that humans have created or modified for activities and habitats \cite{himiyamaHumanSphereEarth2020}. Such activities include urbanization, deforestation, and GHGs emissions, leading to land-use changes and alterations in the atmospheric radiation balance. In many climate models, these human-induced processes are incorporated as external forcings, treated as prescribed inputs to the system based on historical data or scenario-based projections. Fig.~\ref{fig:ClModEvo} and the extended schematic of climate modeling evolution emphasizes the growing influence of human activities on climate dynamics. As models become more complex, the anthroposphere is no longer treated as merely external, but as an active component interacting with other parts of the climate system, revealing the importance of incorporating societal drivers into Earth system modeling.

% \section{Climate Models: Past, Present, and Future}
\section{Climate Models: Evolution, Scales, and Applications}~\label{sec:ClimMdlEvol}
While early climate models evolve progressively, as explained in the previous section, the development of modern climate models has become a more complex process. However, it has also become more achievable with the computational tools available today. This section, we present what climate models are known for: their evolution, their relevance to the various climate components, and what kind of climate models are used for specific applications.

\subsection{Evolution and Classification of Climate Models}
In Fig. \ref{fig:ClModEvo}, we present the chronological evolution of climate models as documented in the IPCC report \cite{pachauriIPCCFourthAssessment2007} and summarized in \cite{TimelineClimateModels}, with an important addition: we highlight which processes belong to which of the five main climate system components. As illustrated, efforts have focused on individual processes till they are well-established before integrating them into unified models that incorporate multiple interacting components.

Climate models are classified based on several aspects, including but not limited to: \textit{(i)} dimensionality, \textit{(ii)} spatial and temporal resolution, and \textit{(iii)} the processes and dynamics included. These aspects are interdependent, meaning that the choice of one affects the others. For example, in a two-dimensional space model, the included processes may be limited to large-scale atmospheric or oceanic circulations, excluding finer-scale turbulence or convection that happens at a coarser spatial scale. The selection of processes, in turn, influences the required temporal resolution, as certain dynamics demand finer time steps to ensure stability of numerical solution to PDEs and ODEs. Additionally, spatial resolution affects both the level of detail captured and the computational feasibility. 

\begin{table*}[t!]
	\caption{Overview of Prominent Climate Models: Types, Dynamics, Applications, and Limitations.}~\label{tab:ClimModelsOvw}
	\centering
	\renewcommand{\arraystretch}{1.2} % Adjusts row height
	%	\rowcolors{2}{gray!20}{gray!10} % Start coloring from the first data row
	\begin{tabularx}{\textwidth}{>{\bfseries}l X X X}
		\toprule
		Model Type & Resolution \& Dynamics Included & Main Applications & Limitations \\
		\midrule
		Energy Balance Models (EBMs) & Simplified representations focusing on the balance between incoming solar radiation and outgoing terrestrial radiation & Estimating global mean temperature responses to changes in GHGs concentrations & Lack spatial resolution; fails to capture regional climate variations or complex feedback mechanisms \\
		General Circulation Models (GCMs) & Comprehensive models simulating atmospheric and oceanic circulation, along with their interactions with land and ice & Projecting global climate change scenarios; studying large-scale climate phenomena & Computationally intensive; diner-scale processes simplifications lead to predictions uncertainty \\
		Regional Climate Models (RCMs) & High-resolution models focusing on specific regions, incorporating local topography and land-use data, which are often not adequately represented in larger GCMs & Assessing regional climate impacts; informing local adaptation strategies; climate-aware regional urban planning & Dependent on boundary conditions from GCMs; access to detailed regional land use and topography data can be limited \\
		Earth System Models (ESMs) & Extend GCMs by including biogeochemical cycles, such as carbon and nitrogen cycles & Studying interactions between climate and ecosystems; assessing feedback in the Earth system; policy implications by assessing climate change impacts on ecosystems and human societies & Increased complexity adds to computational demands; uncertainties in biogeochemical processes and interactions \\
		Integrated Assessment Models (IAMs) & Combine climate models with socio-economic factors to assess policy impacts on climate change & Evaluating mitigation and adaptation strategies; informing policy decisions & Simplifications may overlook critical interactions; results can be sensitive to assumptions and value judgments of climate change impacts \\
		\bottomrule
	\end{tabularx}
\end{table*}

A generalized classification of climate models is introduced in Tab. \ref{tab:ClimModelsOvw}. This classification includes the most widely utilized and prominent models, including EBMs, GCMs, RCMs, ESMs, and IAMs (acronyms are defined in the table), and lists their applications and limitations. In addition to the models outlined in Tab. \ref{tab:ClimModelsOvw}, microclimate variations and localized climate impacts over specific areas can be assessed using Local Climate Models (LCMs). These models offer high-resolution simulations tailored to these areas and are developed by downscaling the RCMs surrounding them, thereby informing urban planning and localized risk assessments. On the other hand, an emerging expanded version of the ESMs is the Human Earth System Models (HESMs) that aim to include more detailed representations of human decision-making and social dynamics within the Earth system model. 

In connection with the climate system's main five components in Section~\ref{sec:ClmtModels}, we emphasize herein the different combination of these components in these models classified in Tab. \ref{tab:ClimModelsOvw} depending on their purpose and complexity. The EBMs typically represent the atmosphere and land surface in a highly simplified form, often through globally averaged equations, to capture large-scale temperature responses. GCMs explicitly simulate the atmosphere and ocean (hydrosphere), and often include simplified representations of the cryosphere, land surface, and biosphere. RCMs refine the spatial resolution of GCM outputs over specific regions, enabling more detailed representations of land surface heterogeneity and localized atmospheric processes. ESMs expand upon GCMs by incorporating interactive biosphere and biogeochemical cycles, such as carbon and nitrogen flows, which play an important role in representing long-term climate feedbacks. IAMs take a broader systems perspective by loosely coupling simplified representations of all five components with economic and policy modules to support decision-making. LCMs, derived from RCMs, focus on fine-scale atmospheric and land surface interactions relevant to urban microclimates. Finally, HESMs aim to capture the coupled dynamics between human systems and the climate, requiring deeper integration of land, biosphere, and atmospheric processes with socio-economic behavior. As shown in Tab. \ref{tab:ClimCompCharct}, the modeling of each component involves different characteristic time and spatial scales, numerical approaches, and governing equations, which influence model structure and selection. 

To that end, while this classification provides a structured overview of climate models levels and types, it is important to note that under each category, multiple models exist, each designed with different assumptions, computational frameworks, and objectives. These models differ in structural design, spatial resolution, \textit{parameterization} schemes, and numerical methods used to solve the complex equations governing climate processes. Parameterization refers to the process of approximating small-scale processes that cannot be directly resolved, such as cloud formation and convection. The choice of a particular model depends on the specific research question, the scale of analysis, and the computational resources available.

For instance, in their Sixth Assessment Report (AR6), IPCC utilized outputs from the Coupled Model Intercomparison Project Phase 6 (CMIP6). This project encompasses simulations from approximately 100 distinct climate models developed by 49 different modeling groups worldwide. These models provide a comprehensive foundation for assessing future climate scenarios and understanding potential impacts. To enhance accessibility and facilitate flexible spatial and temporal analysis of climate data, the IPCC's AR6 also introduced the \textit{Interactive Atlas} \cite{gutierrezIPCCInteractiveAtlas2021}. This interactive tool allows users to explore regional and global climate projections derived from CMIP6 models, making complex climate information more accessible. Readers can leverage this platform to analyze climate projections, assess regional impacts, and gain deeper insights into climate variability and change. %\comment{add a figure that you generate from the website...the figure would break the flow. }

\begin{figure}[t!]
	\centering
	\includegraphics[width=0.5\textwidth]{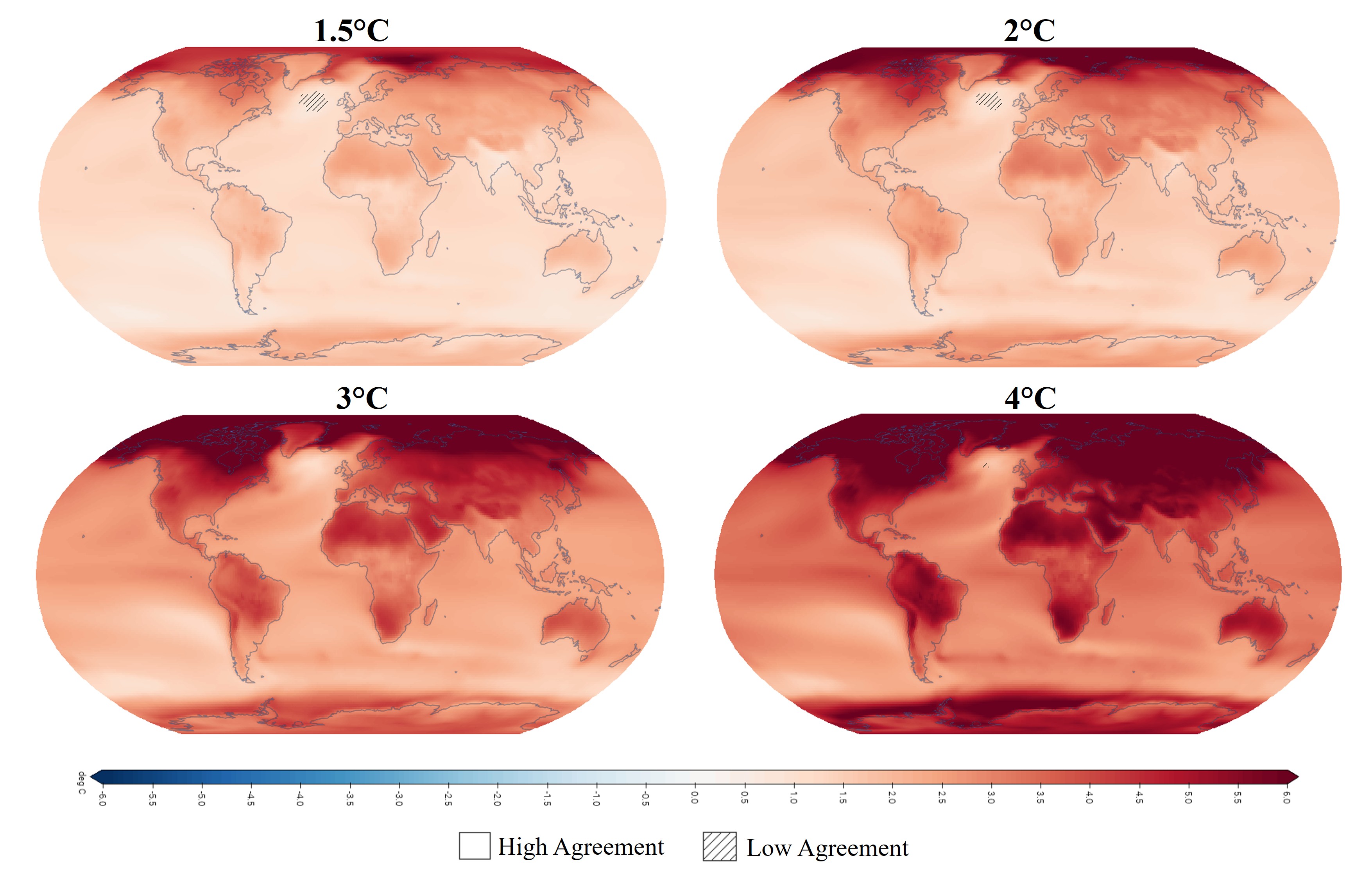}
	\caption{Figures generated by the IPCC AR6 Atlas~\cite{gutierrezIPCCInteractiveAtlas2021}, showing projected changes in mean annual temperature (°C) relative to the 1850–1900 baseline under different emission scenarios leading to average global warming of $1.5^\circ$C, $2^\circ$C, $3^\circ$C, and $4^\circ$C. Solid areas indicate high agreement among CMIP6 models (34 models for the first three scenarios, 20 for the last), while hatched areas indicate low agreement.}~\label{fig:IPCCAtlas}
    \vspace{-0.6cm}
\end{figure}

\vspace{0.1cm}
\noindent \textbf{An Example on Predicting and Computing Climate Projections.} Here, we illustrate how climate projections---that have become part of the climate discourse---are computed via dynamic climate models. For example, Fig.~\ref{fig:IPCCAtlas} presents visualizations generated by the IPCC's Interactive Atlas, displaying projected changes in mean annual temperature relative to the 1850–1900 baseline\footnote{The 1850–1900 baseline is used by the IPCC as a proxy for pre-industrial conditions, as it represents the average temperature before significant industrialization and the associated increase in GHGs emissions. It is computed by averaging observed or reanalyzed surface temperatures over that period. While direct instrumental coverage was sparse, especially in the Southern Hemisphere, recent studies \cite{hawkinsEstimatingChangesGlobal2017,rajamaniLegalCharacterOperational2018} show that the baseline is accurate within about $\pm0.1^\circ$C, making it suitable for long-term comparisons and global warming targets.} under four 
%\comment{why these scenarios or levels? because other scientists in ocean science, coral reefs, energy systems, ... predict or...}
global warming levels: $1.5^\circ$C, $2^\circ$C, $3^\circ$C, and $4^\circ$C. These specific levels represent policy-relevant thresholds and scientific milestones in the context of climate change mitigation and adaptation. The $1.5^\circ$C and $2^\circ$C targets are directly tied to the \textit{Paris Agreement}, \cite{ParisAgreementUNFCCC}---an international treaty adopted under the United Nations Framework Convention on Climate Change (UNFCCC) that aims to limit global warming to well below 2$^\circ$C, with efforts to pursue a 1.5$^\circ$C target above pre-industrial levels.  Each additional degree of warming brings increasingly severe consequences. That being said, the $3^\circ$C and $4^\circ$C levels serve as illustrative benchmarks for high-emission scenarios. These levels are also widely used in impact assessments across various domains, such as ocean science, water availability, energy systems, agriculture, and public health, where researchers estimate physical and societal impacts under different degrees of warming. This makes them a common language for linking model projections to real-world consequences.

% \comment{expand on this point...then mention that they have excluded some models because A,B,C...} 
These projections are derived from a multi-model ensemble of CMIP6 simulations, using 34 dynamic models for the first three scenarios and 20 for the fourth. The reduced number of models at the $4^\circ$C warming level is primarily due to the fact that not all modeling centers have submitted simulations for high-emission or extended scenarios, and some models were excluded due to missing output fields or inconsistencies that affect comparability. From a systems and dynamics perspective, each of these models solves a set of coupled nonlinear ODEs and PDEs that govern the evolution of climate variables such as temperature, wind, ocean currents, and moisture. These PDEs are discretized over spatial grids and time steps, often with horizontal resolutions ranging from tens to hundreds of kilometers, which means that generating a single simulation involves high computational cost and long runtimes. For example, simulating a century-long climate trajectory at high spatial and temporal resolution requires significant computational infrastructure, limiting how many runs some centers can produce, especially for extreme scenarios. The plots shown are constructed by spatially averaging the simulated surface temperature fields over time for each model grid, followed by computing the ensemble mean across all models included for a given warming level. 
%\comment{Add an example...relate it to grids discretization the PDEs i guess?} The plots are constructed by spatially averaging the simulated temperature outputs over time for each model, then computing the ensemble mean across all models used for each warming level. 
Agreement and confidence are visually encoded: solid-shaded regions indicate strong agreement among the models (i.e., consistent sign of change in at least 80\% of models), while hatched regions indicate areas of low confidence where models disagree on the direction of change or the signal-to-noise ratio is weak. 

The Interactive Atlas offers additional features for other climate variables such as surface wind and precipitation, and enables users to generate regional time series, assess seasonal variability, and explore different emissions scenarios through an interactive interface. Tools such as AR6 Atlas and the CMIP6 simulations are a rich source that can be used as testbed for generating data that can be used to solve climate-relevant control engineering problems as emphasized in Section \ref{sec:Clmt&CntrlEng}.

\subsection{How are Climate Models Useful?}
Beyond predicting the adverse effects of climate change, climate models can be useful for a plethora of applications that also involve dynamic system models. Indeed, climate modeling is exploited as a preliminary tool across various sectors to provide the needed insights for real-time decision-making in infrastructure planning, resource management, disaster preparedness, and more. These models help predict long-term trends and short-term variations, allowing industries and governments to make informed choices that are a function of physics-based climate models. 

One of the most significant drivers of climate modeling is that it serves as the foundation for understanding, predicting, and responding to climate change. These models help quantify the impact of GHGs emissions, assess long-term climate trends, and evaluate the potential consequences of different mitigation and adaptation strategies. The insights derived from climate models influence global policy frameworks, such as the Paris Agreement mentioned in the previous section, and guide regional and national climate action plans. %\textit{Paris Agreement} \cite{ParisAgreementUNFCCC}---an international treaty adopted under the United Nations Framework Convention on Climate Change (UNFCCC) that aims to limit global warming to well below 2$^\circ$C, with efforts to pursue a 1.5$^\circ$C target above pre-industrial levels---and guide regional and national climate action plans. 

That is, climate models have broad applications across various domains, where projects, trends, variations, shifts, and changes in climate variables and processes inform these domains' practices. The scope of these domains is wide, with varying levels of focus depending on the application and expected outcomes. Below, we outline some of these domains to give readers an overview and highlight the broader significance of these models beyond understanding the Earth and its climate. We also acknowledge the entities involved in investigating and addressing these areas, as well as those providing data and resources that support potential opportunities.
\begin{itemize}
	\item {\textit{Infrastructure and Urban Planning}:} Cities rely on climate projections to build resilient infrastructure, develop flood control strategies, and plan sustainable transportation systems in response to shifting climate patterns. NOAA's National Centers for Environmental Information (NCEI)~\cite{NationalCentersEnvironmental} provide climate data that assist in planning resilient infrastructure and urban development to withstand changing climate conditions.
	\item \textit{Power/Energy Systems and Integration of Renewables}: Climate models help optimize renewable energy deployment by forecasting long-term wind, solar, and hydrological trends, ensuring efficient energy grid management under varying climate conditions. These models provide the needed data for integrating wind and solar farms, optimizing hydropower operations, and planning energy storage solutions to balance supply and demand. Several organizations leverage climate modeling for energy planning. For instance, the International Energy Agency (IEA)~\cite{IEAInternationalEnergy} uses climate projections to assess renewable energy potential and develop policies for sustainable energy transitions. The National Renewable Energy Laboratory (NREL)~\cite{NationalRenewableEnergy} integrates climate models into its Renewable Energy Integration studies, evaluating how weather and climate variability impact grid stability and energy reliability. Additionally, the European Centre for Medium-Range Weather Forecasts (ECMWF) provides high-resolution climate data that support the optimization of wind and solar energy production~\cite{ForecastsECMWF}. The APEC Climate Center (APCC)~\cite{APECClimateCenter} conducts research on climate predictions, which supports the integration of renewable energy sources by forecasting climate variability. In short, dynamic climate models inform and occasionally impact other dynamic models in energy and power systems. 
	\item \textit{Agriculture and Food Security}: Farmers and policymakers use climate models to anticipate drought risks, changes in growing seasons, and extreme weather events, enabling adaptive agricultural practices and food security planning. Climate Information Services (CIS) \cite{ClimateInformationServices} offer forecasts and agrometeorological services, including tools that enable farmers to make informed decisions, enhancing food security.
	\item \textit{Water Resource Management}: Climate models inform reservoir operations, groundwater sustainability, and drought mitigation strategies, ensuring water availability for urban, industrial, and agricultural use. Through its Water Resources program, NOAA's (National Oceanic and Atmospheric Administration) Climate Program Office (CPO) advances approaches to prepare for and respond to climate-related impacts on water resources, supporting research and the development of strategies to address climate variability~\cite{WaterResourcesClimate}. Additionally, the International Water Management Institute (IWMI) integrates modeling, monitoring, and scenario planning for surface water and groundwater to help governments and partners plan and operate for climate change adaptation, further developing and deploying flood and drought monitoring and forecasting technologies~\cite{InternationalWaterManagement2025}. Similarly, RTI Center for Water Resources utilizes advanced, information-based technology and solutions to translate data into actionable information, strengthening water security and resilience against increasing demand, water scarcity, extreme weather events, and hydrologic uncertainty  \cite{WaterResourcesRTI2025}. In the United States, the Water Utility Climate Alliance (WUCA) emphasizes practical approaches to climate change adaptation, planning, and decision-making to ensure that water utilities and the communities they serve can thrive amidst emerging challenges. By leveraging collective utility experiences, WUCA develops leading practices in climate change adaptation and mitigation that are actionable, equitable, and serve as models for others \cite{WaterUtilityClimate}.
	\item \textit{Disaster Preparedness and Risk Management}: Governments and humanitarian organizations rely on climate models to develop early warning systems for hazards such as hurricanes, heatwaves, and wildfires, improving disaster response strategies. Initiatives like those led by the United Nations Office for Disaster Risk Reduction (UNDRR) \cite{ClimateActionDisaster2020} are taken to promote comprehensive disaster and climate risk management as central to development planning, emphasizing the integration of risk-centered approaches into national adaptation plans.
	%Advancements in AI have enhanced flood warnings, aiding in disaster preparedness, although challenges in risk communication and infrastructure resilience remain.
\end{itemize}

Other dynamic system application domains that climate models inform and impact are traffic networks, air quality, wildfires prediction, and spread of diseases.

\section{The Weather We Experience vs. The Climate We Predict}~\label{sec:ClmvsWth}
Having introduced the climate components and classified various climate models with their applications, we highlight in this section the fundamental difference between \textit{weather} and \textit{climate} before presenting some mathematical climate models in Section~\ref{sec:MathModels}. 

The concepts of weather and climate are sometimes confused and used interchangeably. Weather refers to the short-term atmospheric conditions in a specific place at a specific time. It encompasses daily variations in temperature, humidity, precipitation, wind, and visibility, along with severe conditions including droughts, tornadoes, and hurricanes. These conditions can change within minutes or hours. {In contrast, the atmosphere component in climate represents the long-term average of weather patterns over an extended period—months, years, or even centuries. As NOAA puts it: \textit{Climate is what you expect, weather is what you get.}}

From a control systems perspective, this distinction is analogous to analyzing a system’s short-term response (transients) versus its long-term behavior (average or steady-state response). While weather prediction involves tracking the immediate state trajectory of a highly nonlinear, chaotic system, climate modeling focuses on statistical or average behavior over time, making concepts like infinite-horizon or asymptotic stability, reachability, and observability relevant. This shift in focus from state evolution to long-term system characteristics is key for developing control-oriented strategies for climate adaptation and mitigation.

% \comment{more relevance to the controls community}

\subsection{Entities Providing Weather Forecasts and Climate Projections}
Several organizations worldwide are responsible for monitoring atmospheric conditions and providing weather forecasts, which influence daily activities, business operations, and decision-making. These organizations use various forecasting models---typically referred to as numerical weather prediction (NWP), differing in prediction timeframes and spatial resolutions. These agencies, among others, are part of the World Meteorological Organization (WMO), a specialized agency of the United Nations (UN). The WMO acts as a coordinating body for data sharing and best practices in global weather forecasting.
\begin{itemize}
	\item \textit{The National Weather Service} (NWS) in the US: Federal agency working under the umbrella of NOAA. The NWS operates two main forecasting models: the North American Mesoscale (NAM), which regional numerical weather prediction for the North American Continent, and the Global Forecast System (GFS), a global numerical weather prediction system.
	% that runs four times daily, providing forecasts up to 16 days in advance. 
	For more details on these models, refer to their official website \cite{usdepartmentofcommerceForecastModelsUtilizied}. %The GFS utilizes a finite volume cubed-sphere (FV3) dynamical core with an approximate horizontal resolution of 28 km, extending to 70 km for longer-range forecasts. 
	\item \textit{European Centre for Medium-Range Weather Forecasts} (ECMWF): An intergovernmental organization supported by European nations, providing global weather predictions and data. ECMWF operates the Integrated Forecasting System (IFS) \cite{ForecastsECMWF}.
	\item \textit{Met Office} in the UK: The UK's national weather service, delivering weather and climate-related services. The Met Office utilizes the Unified Model, a numerical weather prediction and climate modeling software suite \cite{UnifiedModelMet}. %This model operates across various spatial and temporal scales, providing forecasts from nowcasting to centennial climate projections. 
	\item \textit{Japan Meteorological Agency} (JMA): Japan's national weather service, which operates ensemble prediction systems for medium-range forecasts \cite{EnsembleModelPrediction}.
	%Responsible for gathering and reporting weather data in Japan and global forecasts, operating an ensemble prediction system \cite{EnsembleModelPrediction} %which provides medium-range forecasts up to 9 days ahead. Additionally, JMA runs the Meso-Scale Model (MSM) for short-range forecasts up to 33 hours ahead, focusing on severe weather phenomena. 
	\item \textit{Bureau of Meteorology} (BoM) in Australia: Uses the Australian Community Climate and Earth-System Simulator (ACCESS) for weather and climate modeling \cite{corporatenamenatWeatherForecastsBoM}.
\end{itemize}

In addition to the specialized models employed by various meteorological agencies, the Weather Research and Forecasting (WRF) Model stands out as a versatile, open-source numerical weather prediction system. Developed collaboratively by U.S. agencies and research institutions, WRF serves both atmospheric research and operational forecasting needs. WRF is distinguished for its adaptability across different scales, advanced physics and numerical methods, and data assimilation capabilities---for more detailed information visit \cite{WeatherResearchForecasting}.

\begin{shadedcvbox}
	{\textbf{\textit{Why does information about meteorological agencies and their forecasting models matter to the controls community?}}} \\
	Many of these agencies provide open access to full or partial datasets of historical and real-time weather data for academic research and scientific applications. Additionally, analyzing the differences in their model structures, time-frames, and methodologies and linking them to control engineering dynamics and applications discussed in this tutorial creates opportunities for further research and practical implementations.
\end{shadedcvbox}

\subsection{Relevance to Controls Community}
In this section, we pose an important question that pertains to the potentially important role the control engineering community can play in answering some climate-related questions. 

\begin{shadedcvbox}
\textit{\textbf{Are the control-theoretic insights and interpretations proposed in this tutorial applicable to weather forecast models?}} \\
The general answer is \textit{yes}, as these models are dynamical systems with close applications to the climate models. However, acknowledging the following key differences between climate and weather forecasting models to ensure that the application of control-theoretic insights is appropriately framed.
\begin{itemize}
	\item {\textit{Temporal Scales}:} Weather models focus on short-term forecasts (hours to weeks) whereas climate models simulate long-term trends over years to centuries.
	\item \textit{Required Conditions:} Weather models rely on initial conditions, which define the system's state at the start of a forecast. This accordingly highly impacts short-term accuracy. Climate models, in contrast, depend more on boundary conditions and external forcings, which set constraints like GHGs levels and ocean heat content to capture long-term trends.
	\item \textit{Resolution and Computational Complexity:} Weather models do not include processes that do not significantly impact short-term forecasts, such as atmospheric chemistry and the carbon cycle. On the other hand, climate models consider these processes to capture long-term interactions between the atmosphere, oceans, land, ice and biosphere. Additionally, they focus on long-term averages and trends, often smoothing out high-frequency fluctuations that influence weather forecast.
\end{itemize}
\end{shadedcvbox}

Thus, while control-theoretic approaches are applicable, they must be tailored to the specific constraints and objectives of each model type. {The next sections delve into actual dynamic models, highlighting the relationship between these models and the aforementioned forecasts and projections. }

\section{Simple Dynamic System Modeling of Climate Processes}\label{sec:MathModels}
Climate processes evolve over time, highlighting the dynamical nature of the system, which is driven by complex interactions between climate components. Climate models are built on fundamental physics-based principles. This includes conservation of energy, mass, and momentum.  After obtaining these models, the interactions between these components over space and time are studied, thereby capturing the variability and feedback mechanisms that shape Earth's climate. In this section, we showcase this dynamical nature through one of the simplest energy balance climate models (EBM). EBMs serve as a starting point and a pedagogical tool to motivate the more complex representations of climate dynamics discussed in Section~\ref{sec:MoreComplxModel}.

EBMs connect back to the broader modeling characteristics summarized in Tab.~\ref{tab:ClimCompCharct} and the classification outlined in Tab.~\ref{tab:ClimModelsOvw}, by exemplifying how even a minimal model can represent key climate behaviors in a control-oriented form. Historically, such simplified EBMs played a foundational role in early climate studies, offering insight into climate sensitivity, radiative forcing, and the concept of planetary energy balance. These early models have laid the groundwork for the development of modern, high-resolution climate models.

% \comment{relate back to the two tables and say CLEARLY that this IS THE SIMPLEST MODEL so it's just a motivation...}, followed by a discussion on more complex models and dynamics, and the associated complexities. \comment{add some historical background and say that this model formed some basis for climate studies.}

\subsection{Dynamic Climate Models 101: The Unforced EBM}
To introduce the control-oriented interpretation of climate models, we begin with a simple yet insightful example. The following EBM is a zero-dimensional model that simulates the change in the \textit{global mean surface temperature}\footnote{It may sound unintuitive, and frankly confusing, that the \textit{global} mean temperature is \textit{not} really an average or mean of anything. While the reader might think that this ODE models the average across different spatial $T_i(t)$ (each with different parameters $C_i, S_i, \alpha_i, \ldots$, in reality the model just considers the average parameters into a single ODE with one state.  In short, this simplification works by reducing all spatial variability into a single time-evolving state variable $T(t)$. This approach captures large-scale energy balance trends without needing spatial resolution.} over time. This model is expressed as a first-order ODE as in Eq. \eqref{eq:EBM}:
\begin{equation}~\label{eq:EBM}
	C \frac{dT(t)}{dt} = S(1 - \alpha(t)) - \epsilon \sigma T(t)^4.
\end{equation}

This EBM captures how the global mean temperature evolves over time in response to the imbalance between absorbed solar energy and emitted infrared radiation. In doing so, it implicitly represents the combined thermal response of the Earth's surface and atmosphere, abstracting away more detailed processes in the ocean, cryosphere, and biosphere.

% \comment{say here that the controls paper you cited in the intro uses this model.}
In~\eqref{eq:EBM}, $T(t)$ is the state variable, representing the global mean surface temperature in Kelvin (K) at time $t$. Parameter $C$ is the effective heat capacity of the Earth system, which determines how quickly the system responds to energy imbalances. Parameter $S$ is the average incoming solar energy per unit area (Watt/m$^2$, W$\cdot$m$^{-2}$), and $0\leq \alpha(t) \leq 1$ defines the planetary albedo which is typically a function of $T(t)$ that models the fraction of solar energy reflected by the Earth. 
%The average albedo for the Earth is about $0.3$. \comment{revisit} 
Parameter $0\leq \epsilon \leq 1$ is the emissivity of the Earth system, a coefficient that scales how effectively Earth emits energy back into space. This emissivity acts as a proxy for the greenhouse effect in this simple model. As GHGs concentrations increase, less energy escapes directly to space, which means the effective emissivity  decreases. Finally, $\sigma$ is the Stefan-Boltzmann constant, a physical constant that relates temperature to radiated energy ($5.67 \times 10^{-8}$  W$\cdot$m$^{-2}\cdot$K$^{-4}$). The term $\epsilon \sigma T^4$ represents the outgoing long wave radiation, which is considered a key feedback mechanism in the system.

This model describes how temperature $T(t)$ evolves continuously over time rather than giving a static or equilibrium value. The system responds to external conditions such as changes in solar radiation, albedo, or GHGs. The presence of $\frac{dT(t)}{dt}$ makes the model time-dependent, and the feedback term $\epsilon \sigma T^4$ introduces nonlinearity, which governs how the system stabilizes or amplifies---a more detailed discussion on this aspect is included later in this section.

From a control perspective, this model behaves as a nonlinear first-order system with a single state and no explicit control input, meaning that the system evolves on its own based on the physical parameters including solar radiation, albedo, and GHGs levels. The solution to this ODE gives the trajectory of global temperature over time, which can be analyzed numerically. 

% \subsection{Modeling Control Inputs and GHGs Emissions into~\eqref{eq:EBM}}
% \noindent \textbf{Modeling Control Inputs and GHGs Emissions into~\eqref{eq:EBM}.}
\subsection{Modeling Control Inputs and GHGs Emissions: Formulating a Forced EBM}

To influence the evolution of $T(t)$ in~\eqref{eq:EBM}, \textit{control inputs} can be appended to the dynamics. In the context of this simple EBM \eqref{eq:EBM}, these control inputs include geoengineering techniques and policy levers, as they can modify the model structure and/or parameters. As for GHGs emissions, they are typically \textit{not} included or modeled as control inputs in this framework. Instead of modeling GHGs directly via some other ODEs or even PDEs, we can alternatively model actions that influence emissions such as:
% \begin{custombox}{magenta}
\begin{custombox1}{white!60!blue}
\begin{itemize}[leftmargin=*, labelsep=1em, align=parleft, labelwidth=0pt]
    % \centering
    \item Carbon taxes or regulations $\rightarrow$ reduce emission rates.
    \item Renewable energy adoption $\rightarrow$ displaces fossil fuels.
    \item Carbon capture technology $\rightarrow$ removes CO$_2$ from the atmosphere.
\end{itemize}
\vspace{-0.1cm}
\end{custombox1}
% \end{custombox}

These policy or technological levers can be modeled as control inputs that affect the rate of emissions, which affects atmospheric CO$_2$ concentration, which affects emissivity $\epsilon$, which affects the feedback term $\epsilon \sigma T^4$, which changes the temperature evolution. 

Similarly, one of the geoengineering techniques that can be treated as control input is the stratospheric aerosol injection. Stratospheric aerosol injection is a geoengineering technique that involves releasing tiny reflective particles (usually sulfate aerosols) into the stratosphere (about 10--50 km above the Earth's surface) to reflect a portion of incoming solar radiation back into space. An interesting control engineering problem that can be posed here is \textit{how would introducing these aerosols injections impact the mean global temperature dynamics?} Conceptually, the below simple flow diagram illustrates the idea. 

% \begin{custombox}{magenta}
\begin{custombox1}{white!60!blue}
Stratospheric aerosol injection $\rightarrow$ reflects sunlight $\rightarrow$ increases albedo $\alpha$ in~\eqref{eq:EBM} $\rightarrow$ less energy absorbed $\rightarrow$ temperature $T(t)$ drops in~\eqref{eq:EBM}
\end{custombox1}
% \end{custombox}

In particular, we modify the EBM in \eqref{eq:EBM} into \eqref{eq:EBM_SAI} to include $\textcolor{blue}{u(t)}$ as the control input, representing the effective albedo increase from aerosol injection. $\textcolor{blue}{u(t)}$, bounded based on technical and environmental limits. By introducing this control input, the term $S(1-(\alpha+\textcolor{blue}{u(t)}))$ decreases, which means less energy absorbed, and thus cooling.

\begin{equation}~\label{eq:EBM_SAI}
	C \frac{dT(t)}{dt} = S(1 - (\alpha+\textcolor{blue}{u(t)})) - \epsilon \sigma T(t)^4.
\end{equation}

%Policy Levers as Control Input
%These are human actions that affect greenhouse gas emissions:
%
%Reducing CO$_2$ → slows the decrease in 
%$\sigma$ 
%
%Carbon pricing or emissions limits → indirectly control long-term temperature rise
%These are slower and indirect, but still count as inputs that influence system behavior over time.

So the broad idea here is as follows. By default, the model \eqref{eq:EBM} is an open-loop one but it can be extended to a closed-loop system \eqref{eq:EBM_SAI} by introducing such inputs. In addition to incorporating deliberate control inputs such as aerosol injection through $\textcolor{blue}{u(t)}$, external disturbances can also be introduced to capture the effect of GHG emissions. A common approach is to treat changes in emissivity $\epsilon$ as an uncertain and possibly time-varying quantity. Rising GHG concentrations reduce Earth's ability to emit energy back into space, which is effectively modeled by a \textit{decrease in} $\epsilon$. 

Instead of solving an additional ODE for atmospheric CO$_2$ concentration, we capture its impact on temperature dynamics by modifying $\epsilon$ as a time-varying function $\epsilon(t)$, or by modeling its deviation from a nominal value using an additive disturbance term $\textcolor{red}{w(t)}$. This leads to a disturbed version of the energy balance model:
\begin{equation}~\label{eq:EBM_disturbed}
	C \frac{dT(t)}{dt} = S(1 - (\alpha + \textcolor{blue}{u(t)})) - (\epsilon + \textcolor{red}{w(t)}) \sigma T(t)^4,
\end{equation}
where $\textcolor{blue}{u(t)}$ is the \textit{control input}, representing interventions such as stratospheric aerosol injection. However, variable $\textcolor{red}{w(t)}$ is an \textit{external disturbance} capturing the impact of increased GHG concentrations on emissivity.
    
If no explicit control is applied (i.e., $\textcolor{blue}{u(t)} = 0$) and emissions rise unchecked (i.e., $\textcolor{red}{w(t)} < 0$), the system experiences increased warming due to enhanced radiative trapping. Conversely, effective control ($\textcolor{blue}{u(t)} > 0$) and stable or reduced emissions ($\textcolor{red}{w(t)} \approx 0$) can lead to stabilization or even gradual cooling.

% \begin{figure}[t!] 
% \centering 
% \includegraphics[width=0.45\textwidth]{EBM_Control_Simulation.pdf} 
% \caption{Illustrative simulation of global mean temperature evolution under three scenarios using a zero-dimensional EBM: an unforced baseline, a scenario with control-based cooling intervention, and a scenario with external radiative forcing (disturbance), e.g., GHGs.} \label{fig:EBM_Control_Sim} 
% \end{figure}

 \begin{figure}[t!] 
\centering 
\includegraphics[width=0.45\textwidth]{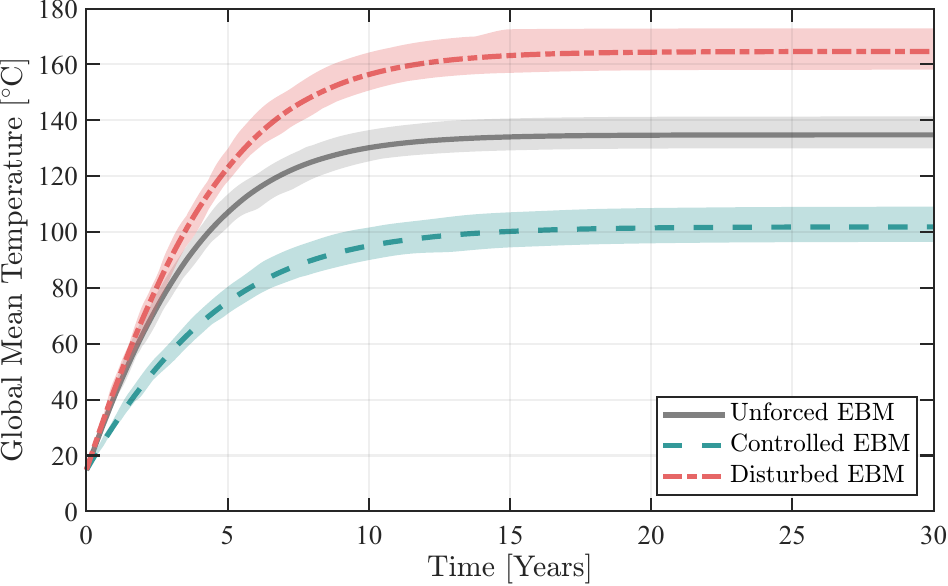} 
\caption{Illustrative simulation of global mean temperature evolution under three scenarios using a zero-dimensional EBM: an unforced baseline, a scenario with control-based cooling intervention, and a scenario with external radiative forcing (disturbance), e.g., GHGs. Each trajectory is simulated under process-level uncertainty modeled as multiplicative noise applied to the system dynamics. Shaded regions represent the 5th to 95th envelopes over 100 Monte Carlo simulations.} \label{fig:EBM_Control_Sim}
\end{figure}

To introduce the idea of framing the climate dynamics in \eqref{eq:EBM} in a control-theoretic framework, Fig.~\ref{fig:EBM_Control_Sim} presents an illustrative simulation that showcases how the global mean surface temperature evolves over time in three distinct scenarios: the unforced model representing the baseline climate system, a disturbed model where an external forcing (e.g., rising GHG emissions) reduces Earth's emissivity, and a controlled model where an artificial albedo increase is used to counteract warming. This example highlights a core idea from control theory: external disturbances such as increased emissions can push the system away from its nominal behavior, while appropriately designed control inputs can counterbalance these effects and reach desirable states. In this specific case, the control input mimics geoengineering strategies (e.g., stratospheric aerosol injection) that increase planetary albedo, leading to lower absorbed solar radiation and thus reduced surface temperatures. 

To reflect uncertainty in physical processes and imperfect model knowledge, we augment the system with two sources of uncertainty: \textit{(i)} process-level stochastic noise affecting the energy balance dynamics, and \textit{(ii)} random perturbations in key parameters, specifically the albedo $\alpha$ and the emissivity $\epsilon$. A total of 100 Monte Carlo simulations are performed for each scenario (unforced, controlled, and disturbed), with each realization using independently sampled noise trajectories and perturbed parameter values. The resulting temperature trajectories are used to construct full envelopes, showing the range within which the temperature may vary due to both internal stochastic variability and structural parameter uncertainty.

The parameters used in this simulation are: effective heat capacity $C = 8 \times 10^8\,\text{J/K/m}^2$, solar constant $S = 1367.6\,\text{W/m}^2$, base albedo $\alpha = 0.3$, and nominal emissivity $\epsilon = 0.61$. The control input corresponds to an increase in albedo of $\textcolor{blue}{u(t)} = 0.2$, while the disturbance is modeled as a decrease in emissivity of $\textcolor{red}{w(t)} = -0.15$. The initial temperature is set to $T_0 = 288\,\text{K}$ (equivalent to $15^\circ$C), and the simulation time span covers 30 years. Gaussian noise with a standard deviation of 10\% is applied to the model dynamics, while the parameters $\alpha$ and $\epsilon$ are randomly perturbed with a standard deviation of approximately 1\% around their nominal values to represent structural uncertainty. To maintain consistency with the rest of the tutorial, where temperature is presented in degrees Celsius, all plotted results are shifted from Kelvin to Celsius by subtracting 273.15. This makes the interpretation of the temperature trajectories more intuitive and aligned with conventional climate science practices.

It is important to emphasize that this example is conceptual in nature. The control input is not intended to represent a realistic or operational geoengineering scheme, nor does it account for complexities such as spatial heterogeneity, atmospheric chemistry, or long-term feedback mechanisms. Similarly, the temperature levels reached in the disturbed scenarios are unrealistically high relative to actual climate projections; they should not be interpreted literally but rather seen as illustrative. The objective here is not to simulate realistic climate outcomes, but to demonstrate how one might formally model, simulate, and reason about climate interventions using fundamental tools and concepts from control theory.

A more advanced EBM could also be derived considering the Earth as a grid with $m$ latitude and $n$ longitude points. Each cell in this grid of dimension $mn$ has its own dynamical model depicting surface temperature evolution $T_{ij}(t)$ in cell $(i,j)$ that basically extends~\eqref{eq:EBM}--\eqref{eq:EBM_disturbed} into the following:
\begin{align}
\small C_{ij} \frac{d {T}_{ij}(t)}{dt} = S_{ij}(t)\,\big(1 - \alpha_{ij} - \textcolor{blue}{u_{ij}(t)}\big) ~\label{eq:ebm2dx}
\end{align}
\vspace{-0.34cm}
$$\hspace{3cm} - \varepsilon_{ij}\,\sigma\,T_{ij}^{4}(t) +\mathcal{L}_{ij}(\{T_{kl}(t)\}) + \textcolor{red}{F_{ij}(t)},$$
where $i \in \{1,2,\ldots, m\}$ and $j \in \{1, 2, \ldots, n\}$ define the latitude and longitude indices. The other quantities in~\eqref{eq:ebm2dx} are defined similarly to \eqref{eq:EBM}--\eqref{eq:EBM_disturbed}. The major difference between the average EBM and the 2D model~\eqref{eq:ebm2dx} is the function $\mathcal{L}$ which represents the diffusive heat exchange with neighboring cells. The diffusive term \(\mathcal{L}_{ij}(\{T_{kl}\})\) accounts for horizontal heat transport between neighboring grid cells \((k,l)\in\mathcal{N}(i,j)\), with diffusion coefficients \(\kappa_{ij,kl}\) which is typically approximated with a linear model, where the values for $\kappa$'s vary according to the considered assumptions and to the specific application area of the climate model. The quantity \comment{\(F_{ij}(t)\)} represents external forcing or unresolved subgrid effects, akin to the $\textcolor{red}{w(t)}$ in~\eqref{eq:EBM_disturbed}.  This 2D EBM~\label{eq:ebm2d} presents itself as a networked system of temperature dynamics $T_{ij}(t)$ where the coupling appears in the $\mathcal{L}_{ij}(\cdot)$ function (albeit the coupling is very sparse as every local temperature $T_{ij}(t)$ only relates to its adjacent temperatures in the grid), in contrast to the simpler single-dimension EBM which averages individual $mn$ quantities $C_{ij}, S_{ij}, \alpha_{ij}, \epsilon_{ij}$ into single quantities. Later in this paper we show how more advanced models can be derived into controls-oriented state-space description.

Having looked at the basic EBM and its variants, some control-type questions can be posed such as: \textit{What input is needed to stabilize temperature at a target?} \textit{How sensitive is the system to delays or uncertainty in the control action?} These kinds of questions bring forward important concepts that connect climate modeling with insights from control engineering as discussed in Section \ref{sec:Clmt&CntrlEng}. 

\subsection{Interpreting Feedback in Climate Systems and Controls}

\begin{figure}[t!]
	\centering
	\includegraphics[width=0.5\textwidth]{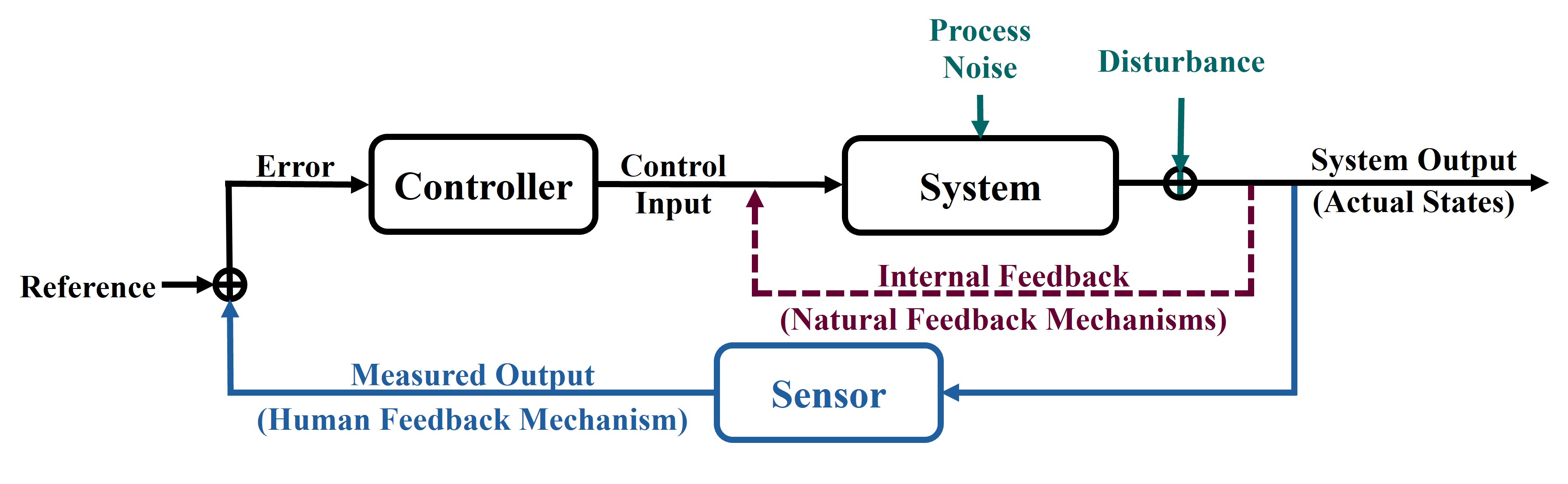}
	\caption{Generalized feedback structure distinguishing internal and external feedback mechanisms. Internal feedback represents natural dynamics updated over time within the system---feedback in climate science. External (human) feedback operates through measurement- and control-based approaches---feedback in control engineering.}~\label{fig:Feedback}
\end{figure}

One important concept here to discuss and be able to differentiate in climate modeling and control is \textit{feedback}. In climate systems, feedback refers to how a change in one part of the system influences other components, which then either amplify or dampen the original change. A common example of positive feedback is the ice-albedo effect: warming leads to ice melt, which lowers Earth's reflectivity (albedo), causing more solar absorption and further warming. On the other hand, negative feedback helps stabilize the system. For example, as temperature increases, the Earth emits more infrared radiation $\epsilon \sigma T(t)^4$, which counteracts the warming. In control engineering, feedback has a more formal definition: it refers to the process of measuring the output of a system and using that information to adjust the input. This can be either negative feedback, where the goal is to reduce the difference between a desired target and the actual output (stabilizing), or positive feedback, which reinforces changes and can lead to amplification or instability. In control terminology, feedback is part of the system architecture, often described using block diagrams or state-space representations, and is central to analyzing stability, robustness, and performance. What is referred to as {feedback} in climate modeling, such as ice-albedo or radiative cooling, corresponds in control terms to \textit{internal feedback loops} in the plant or process model. These are natural feedback mechanisms built into the system's dynamics, not part of a controller. Understanding this distinction is important when interpreting the behavior of the climate system from a control perspective---Fig. \ref{fig:Feedback} presents a generalized diagram with the two feedback mechanisms illustrated.

%\begin{itemize}
%	\item In both fields, negative feedback stabilizes, and positive feedback amplifies.
%	\item Control theory is more formal, with transfer functions, poles/zeros, and explicit stability analysis.
%	\item In climate, feedbacks are mostly physically emergent, but thinking of them using control theory helps understand the stability, sensitivity, and regulation potential of the climate system.
%\end{itemize}

\begin{shadedcvbox}
As a recap, we explain below how the dynamical nature of climate models connects to control concepts, using the EBM \eqref{eq:EBM} as an example:
\begin{itemize}
	\item The EBM is a nonlinear first-order dynamical system, where the state variable 
	$T(t)$ evolves over time based on the net energy balance.
	\item The term $\frac{dT(t)}{dt}$ represents the time evolution of the system, which is central to both climate modeling and control theory.
	\item In its default form, the model is open-loop: it evolves based on physical parameters (e.g., solar radiation, albedo, emissivity) without any external regulation.
	\item Introducing a control input, such as stratospheric aerosol injection, allows us to influence the system---this effectively makes the model a closed-loop system.
	\item In climate models, feedbacks such as ice-albedo (positive) or radiative cooling (negative) are natural feedback mechanisms within the system---what control theory refers to as internal feedback in the plant dynamics.
	\item In control terminology, feedback typically refers to measuring the system output and adjusting the input accordingly, which can be done in climate policy by adapting actions based on observed temperature trends.
	\item Framing climate dynamics of this EBM in control terms allows us to ask questions like:
		\begin{itemize}
			\item What input is needed to stabilize temperature at a certain level?
			\item How does the system respond to delays or uncertainty in the input?
			\item What is the role of feedback in achieving or maintaining stability?
		\end{itemize}
	\item This connection opens up new ways of analyzing and possibly influencing climate outcomes, using tools and insights from control engineering.
\end{itemize}
\end{shadedcvbox}

%\subsubsection*{Atmospheric Dynamics: Navier-Stokes Equations}
%The Navier-Stokes equations governing fluid motion in the atmosphere are:
%\begin{equation}
%	\rho \left( \frac{\partial \mathbf{u}}{\partial t} + (\mathbf{u} \cdot \nabla) \mathbf{u} \right) = -\nabla p + \mu \nabla^2 \mathbf{u} + \mathbf{f}
%\end{equation}
%where:
%\begin{align*}
%	\mathbf{u} & : \text{Velocity field} \\
%	\rho & : \text{Density} \\
%	p & : \text{Pressure} \\
%	\mu & : \text{Dynamic viscosity} \\
%	\mathbf{f} & : \text{External forces (e.g., gravity)}
%\end{align*}
%
%\section*{Ocean Circulation: Thermohaline Circulation Equations}
%The thermohaline circulation, driven by temperature and salinity differences, is described by:
%\begin{align}
%	\frac{\partial T}{\partial t} + \mathbf{u} \cdot \nabla T &= \kappa_T \nabla^2 T \\
%	\frac{\partial S}{\partial t} + \mathbf{u} \cdot \nabla S &= \kappa_S \nabla^2 S
%\end{align}
%where:
%\begin{align*}
%	T & : \text{Temperature} \\
%	S & : \text{Salinity} \\
%	\kappa_T, \kappa_S & : \text{Thermal and saline diffusivities}
%\end{align*}
%
%% Ice Sheet Dynamics: Continuity Equation
%\section*{Ice Sheet Dynamics: Continuity Equation}
%The evolution of ice sheet thickness is governed by:
%\begin{equation}
%	\frac{\partial H}{\partial t} + \nabla \cdot (H \mathbf{u}) = a - m
%\end{equation}
%where:
%\begin{align*}
%	H & : \text{Ice thickness} \\
%	\mathbf{u} & : \text{Ice velocity} \\
%	a & : \text{Accumulation rate} \\
%	m & : \text{Melt rate}
%\end{align*}

Beyond the simple EBM introduced earlier in this section, we present more complex models in the next section and discuss how they can be interpreted as control systems, along with the associated complexities. As examples, we focus on the dynamics of the atmospheric circulation, highlighting how this large-scale model can be viewed and analyzed within a control-oriented framework, providing mathematical formulations and discussing numerical solutions, feedback mechanisms, and parameterizations.

\section{More Complex Dynamic Climate Model}~\label{sec:MoreComplxModel}
As a more complex dynamic climate model, we consider a large-scale atmospheric motion model that is simulated using the \textit{primitive equations}, derived from the Navier-Stokes equations under hydrostatic and shallow atmosphere assumption. This model captures the large-scale circulation processes in the atmosphere; see Tabs.~\ref{tab:ClimRef} and \ref{tab:ClimCompCharct} for relevance with other climate components and processes. The primitive model presented here captures the core dynamical structure of large-scale atmospheric motion. However, compared to comprehensive weather prediction systems, it neglects important moist processes, radiation, surface interactions, and chemical feedbacks. These additional processes, while increasing model complexity and computational cost, capture the full range of phenomena observed in weather systems---from convective storms to boundary-layer dynamics. These simplifications help isolate the fundamental atmospheric dynamical interactions between pressure, temperature, and wind, making the model particularly useful for conceptual understanding, mathematical analysis, and control-oriented formulations. 

%The simplifications in this model help isolate the essential structure of atmospheric flow and are valuable for understanding, control analysis, and conceptual development.

To that end, we do not present all the processes involved; rather, we describe the primitive model as a set of nonlinear PDEs that encode the conservation of momentum, energy, and mass. Then, a nonlinear ODE-based state-space representation is produced given the PDEs. The conservation laws are written in continuous form, assuming smooth spatial and temporal domains. The key state variables are the wind velocity $\mathbf{v}(\m r,t)$, temperature $T(\m r,t)$, and density $\rho(\m r,t)$, each modeled as a time-dependent field distributed over space, with spatial position denoted by $\boldsymbol{r} = [x, y, z]^{\top}$ and time by $t$. These quantities evolve dynamically through advection, external forcing, and internal physical interactions. Together, they form a compact state representation of the atmosphere’s large-scale dynamics, governed by the coupled system of equations shown as follows:
\begin{subequations}~\label{eq:NS_Atm}
	\begin{align}
		\frac{\partial \mathbf{v}}{\partial t} + (\mathbf{v} \cdot \nabla) \mathbf{v} + f\mathbf{k} \times \mathbf{v} &= -\frac{1}{\rho} \nabla p + \mathbf{F}_\text{viscous},~\label{eq:NS_Atm1} \\
		\frac{\partial T}{\partial t} + \mathbf{v} \cdot \nabla T &= Q_T,~\label{eq:NS_Atm2} \\
		\frac{\partial \rho}{\partial t} + \nabla \cdot (\rho \mathbf{v}) &= 0.~\label{eq:NS_Atm3}
	\end{align}
\end{subequations}
The wind velocity ${\mathbf{v} = [u, v, w]^\top}$ is a three-dimensional vector, with components in the zonal, meridional, and vertical directions, respectively. The temperature $T$, pressure $p$, and density $\rho$ are scalars. The Coriolis parameter is given by $f$, which depends on the Earth's rotation rate and the latitude. The unit vector $\mathbf{k}$ points in the vertical direction. The term $\mathbf{F}_\text{viscous}$ represents viscous forces, and $Q_T$ denotes sources or sinks of temperature and heat (e.g., radiative or latent heating). 

As mentioned, each field of states in this system is a function of space and time. %We denote spatial position by $\boldsymbol{r} = (x, y, z)$ and time by $t$. 
Specifically, $\mathbf{v}(\boldsymbol{r}, t) = \big[u(\boldsymbol{r}, t), v(\boldsymbol{r}, t), w(\boldsymbol{r}, t)\big]^\top$, $T = T(\boldsymbol{r}, t)$, $p = p(\boldsymbol{r}, t)$, and $\rho = \rho(\boldsymbol{r}, t)$, but we omit this explicit dependence in the notation for clarity. Similarly, the Coriolis parameter $f = f(\boldsymbol{r})$, which typically varies with latitude but may be assumed constant in simplified settings (e.g., the $f$-plane approximation). The unit vector $\mathbf{k} = [0, 0, 1]^\top$ points in the vertical direction and is constant in space and time. The terms $\mathbf{F}_\text{viscous} = \mathbf{F}_\text{viscous}(\boldsymbol{r}, t)$ and $Q_T = Q_T(\boldsymbol{r}, t)$ can vary in both space and time depending on the processes being modeled and approximations being adopted. 

In \eqref{eq:NS_Atm}, the symbol $\times$ denotes the vector cross product operator. The operator $\nabla$ represents the spatial gradient. When applied to a scalar field $\phi$, it yields the vector of partial derivatives: $\nabla \phi = \left[ \frac{\partial \phi}{\partial x}, \frac{\partial \phi}{\partial y}, \frac{\partial \phi}{\partial z}\right]^\top$. When applied to a vector field, $\nabla$ returns the Jacobian matrix of partial derivatives. The divergence operator $\nabla \cdot$ maps a vector field to a scalar; for example, $\nabla \cdot \mathbf{a} = \frac{\partial a_x}{\partial x} + \frac{\partial a_y}{\partial y} + \frac{\partial a_z}{\partial z}$ returns the net scalar outflow of $\mathbf{a}$. 

% \noindent\textbf{Expanded Explanation of the Atmospheric Primitive Equation System \eqref{eq:NS_Atm}.} 
\subsection{Understanding the Dynamics Behind the Atmospheric Primitive Equations} 
Having introduced the PDE~\eqref{eq:NS_Atm}, here we break down these systems of equations and how they interact with each other:
\begin{enumerate}
    \item \textit{Momentum Equation \eqref{eq:NS_Atm1}}: 
 governs the evolution of wind velocity $\mathbf{v}$ due to  advection (nonlinear self-interaction of the wind velocity field $(\mathbf{v} \cdot \nabla) \mathbf{v}$), and the coriolis term $f \mathbf{k} \times \mathbf{v}$ accounts for Earth's rotation. The pressure gradient force $\frac{1}{\rho} \nabla p$ drives wind from high to low pressure and the viscous terms $\mathbf{F}_\text{viscous}$  depict sub-grid friction or diffusion.
         The momentum equation is affected by density ${\rho}$ and pressure {${p}$}, which are in turn influenced by temperature ${T}$ via the equation of state (ideal gas law) $p=\rho RT$, where $R$ is the specific gas constant.
    \item \textit{Thermodynamic (Energy) Equation \eqref{eq:NS_Atm2}:} describes how temperature evolves due to \textit{(i)} advection by wind $\mathbf{v} \cdot \nabla T$ (temperature is transported spatially) and \textit{(ii)} heating term $Q_T$ which includes radiative heating, latent heat release, and sensible heat flux.
         Similarly, $\rho$ is affected via the ideal gas law, thus coupling back into the momentum equation.
    \item \textit{Continuity Equation \eqref{eq:NS_Atm3}:} 
        enforces mass conservation, links the density field to divergence/convergence in the flow ($\nabla \cdot (\rho \mathbf{v})$), and demonstrates how the changes in $\rho$ impact pressure gradients and therefore motion.
\end{enumerate}

All three equations are nonlinearly coupled, meaning that a change in one variable affects the others. Additionally, uncertainties are inherently included in the parameters and unresolved terms. This is what gives atmospheric models their complex, often chaotic behavior. Notice that, these PDEs are based on the following assumptions:
\begin{enumerate}
    \item \textit{Hydrostatic Approximation:}
        \begin{itemize}
            \item Assumes vertical pressure gradients balance gravitational acceleration, which is justified for large-scale motions where vertical accelerations are small (i.e., not deep convection): $\frac{\partial p}{\partial z}=-\rho g.$
            \item Removes the vertical momentum equation from the full Navier-Stokes system.
            \item The two previous assumptions greatly reduce numerical stiffness and computation time.
        \end{itemize}
    \item \textit{Shallow Atmosphere Approximation:}
        \begin{itemize}
            \item Assumes the thickness of the atmosphere is small compared to Earth's radius.
            \item Allows simplification in the geometry (e.g., assuming constant radius for Coriolis terms).
            \item Horizontal metric terms (like curvature) are simplified.
        \end{itemize}
\end{enumerate}

\subsection{Mathematical Interpretation of the Atmospheric Primitive Model}
To facilitate interpretation for the reader and to support the later discussion on how they are solved, we break down each equation in \eqref{eq:NS_Atm} and explain the application of the relevant operators.

\begin{enumerate}
        \item \textit{The Math in the Momentum Equation \eqref{eq:NS_Atm1}}: 
            \begin{itemize}
    \item $\displaystyle \dfrac{\partial \mathbf{v}}{\partial t}$ represents the dynamic temporal rate of change of the wind velocity field $\mathbf{v}$ at a fixed point in space. It is the time derivative of the state variable or the wind velocity.
    
    \item $(\mathbf{v} \cdot \nabla) \mathbf{v}$ is the advection (nonlinear transport) term, which captures how a fluid parcel's velocity changes as it moves through a spatially varying velocity field. Expanded component-wise as:
    \[
    (\mathbf{v} \cdot \nabla)\mathbf{v} = 
    \begin{bmatrix}
    u \frac{\partial u}{\partial x} + v \frac{\partial u}{\partial y} + w \frac{\partial u}{\partial z} \\
    u \frac{\partial v}{\partial x} + v \frac{\partial v}{\partial y} + w \frac{\partial v}{\partial z} \\
    u \frac{\partial w}{\partial x} + v \frac{\partial w}{\partial y} + w \frac{\partial w}{\partial z}
    \end{bmatrix}.
    \]
    This term introduces nonlinear coupling among the velocity components.
    
    \item $f \mathbf{k} \times \mathbf{v}$ is the Coriolis force, which represents the apparent force due to Earth's rotation. The Coriolis parameter $f$ depends on latitude, and $\mathbf{k} = [0, 0, 1]^\top$ is the unit vector in the vertical direction. The cross product expands as $f \mathbf{k} \times \mathbf{v} =
    [
    -fv, \
    fu, \
    0
    ]^\top.
    $
    
    \item $-\frac{1}{\rho} \nabla p$ is the {pressure gradient force}. A body force per unit mass that drives flow from high to low pressure regions. The gradient operator acts only on $p$ via
    $\nabla p = 
    \begin{bmatrix}
    \frac{\partial p}{\partial x}, \
    \frac{\partial p}{\partial y}, \
    \frac{\partial p}{\partial z}
    \end{bmatrix}^\top$ 
    and is scaled by the inverse of the local density $\rho$.

    \item $\mathbf{F}_\text{viscous}$ is the viscous force term, representing internal friction in the fluid. It acts to diffuse momentum and slow down sharp velocity gradients. This force appears as a vector field with components in the $x$, $y$, and $z$ directions as follows $\mathbf{F}_\text{viscous} = 
\begin{bmatrix}
F_x, \
F_y, \
F_z 
\end{bmatrix}^\top.$ 
%In many models, it is approximated by the Laplacian of velocity: $\mathbf{F}_\text{viscous} = \nu \nabla^2 \mathbf{v}$, where $\nu$ is the kinematic viscosity.

\item \textit{Expanded Formulation:} Putting all terms together, the momentum equation \eqref{eq:NS_Atm1} can be written in fully expanded matrix form as:
\begin{equation}~\label{eq:NS_AtmEx1}
\begin{aligned}
\frac{\partial}{\partial t}
\begin{bmatrix}
u \\
v \\
w
\end{bmatrix}
&=
-
\begin{bmatrix}
u \frac{\partial u}{\partial x} + v \frac{\partial u}{\partial y} + w \frac{\partial u}{\partial z} \\
u \frac{\partial v}{\partial x} + v \frac{\partial v}{\partial y} + w \frac{\partial v}{\partial z} \\
u \frac{\partial w}{\partial x} + v \frac{\partial w}{\partial y} + w \frac{\partial w}{\partial z}
\end{bmatrix}
-
f
\begin{bmatrix}
-v \\
u \\
0
\end{bmatrix}
\\
&\quad
-
\frac{1}{\rho}
\begin{bmatrix}
\frac{\partial p}{\partial x} \\
\frac{\partial p}{\partial y} \\
\frac{\partial p}{\partial z}
\end{bmatrix}
+
\begin{bmatrix}
F_x \\
F_y \\
F_z
\end{bmatrix}.
\end{aligned}
\end{equation}

\end{itemize}

\item \textit{The Math in the Thermodynamic Equation \eqref{eq:NS_Atm2}}: 
    \begin{itemize}
        \item $\frac{\partial T}{\partial t}$ represents the local rate of change of temperature at a fixed point in space. It is the temporal derivative of the scalar state variable $T$.

        \item $\mathbf{v} \cdot \nabla T$ is the advection of temperature by the velocity field. It is a directional derivative, expressing how temperature changes as a fluid parcel moves with the flow. Expanded as:
        \[
        \mathbf{v} \cdot \nabla T = 
        u \frac{\partial T}{\partial x} 
        + v \frac{\partial T}{\partial y} 
        + w \frac{\partial T}{\partial z}.
        \]
        This term introduces coupling between the velocity field and the temperature field.

        \item $Q_T$ represents external heating or cooling processes. This may include radiative heating, latent heat release, or sensible heat fluxes. It acts as a forcing term in the energy balance.
        
        \item \textit{Expanded Formulation:} Putting the terms together, the thermodynamic equation \eqref{eq:NS_Atm2} becomes:
        \begin{equation}~\label{eq:NS_AtmEx2}
        \frac{\partial T}{\partial t}
        =
        - \left(
        u \frac{\partial T}{\partial x} 
        + v \frac{\partial T}{\partial y} 
        + w \frac{\partial T}{\partial z}
        \right)
        + Q_T.
        \end{equation}
    \end{itemize}

% The term $-\frac{1}{\rho} \nabla p$ represents the pressure gradient force per unit mass. Here, both $p = p(\boldsymbol{r}, t)$ and $\rho = \rho(\boldsymbol{r}, t)$ are time- and space-dependent fields. The gradient operator acts only on $p$, returning:
% \[
% \nabla p = \begin{bmatrix}
% \frac{\partial p}{\partial x} \\
% \frac{\partial p}{\partial y} \\
% \frac{\partial p}{\partial z}
% \end{bmatrix}.
% \]
% The scalar field $1/\rho$ then modulates the magnitude of the force in each direction, depending on the local fluid density. Unlike the divergence of a product, this term does not require a product rule expansion since the gradient is not applied to $\rho$.

% The divergence operator, denoted $\nabla \cdot \mathbf{v}$, returns a scalar and is defined as:
% \[
% \nabla \cdot \mathbf{v} = \frac{\partial u}{\partial x} + \frac{\partial v}{\partial y} + \frac{\partial w}{\partial z}.
% \]
% This measures the net outflow of the vector field at each point. To avoid ambiguity, we express all operator actions explicitly when discretizing or linearizing the system later in this work.

\item \textit{The Math in the Continuity Equation \eqref{eq:NS_Atm3}}:
    \begin{itemize}
        \item $\frac{\partial \rho}{\partial t}$ represents the local rate of change of density $\rho$ at a fixed point in space. This captures how density evolves over time due to fluid motion and compression or expansion.

        \item $\nabla \cdot (\rho \mathbf{v})$ is the divergence of the mass flux. It accounts for the spatial variation in the flow of mass and can be expanded using the product rule as:
        \[
        \nabla \cdot (\rho \mathbf{v}) 
        = 
        \mathbf{v} \cdot \nabla \rho 
        + \rho \nabla \cdot \mathbf{v},
        \]
        where:
        \[
        \nabla \cdot \mathbf{v} = 
        \frac{\partial u}{\partial x} 
        + \frac{\partial v}{\partial y} 
        + \frac{\partial w}{\partial z}.
        \]
        This term captures how variations in velocity and density contribute to local mass conservation.

        \item \textit{Expanded Formulation:} The continuity equation \eqref{eq:NS_Atm3} becomes:
        \begin{equation}~\label{eq:NS_AtmEx3}
        \begin{aligned}
        \frac{\partial \rho}{\partial t}
        =
        & - \Bigg(
        u \frac{\partial \rho}{\partial x}
        + v \frac{\partial \rho}{\partial y}
        + w \frac{\partial \rho}{\partial z}
        \\ &+ \rho \left(
        \frac{\partial u}{\partial x}
        + \frac{\partial v}{\partial y}
        + \frac{\partial w}{\partial z}
        \right) \Bigg).
        \end{aligned}
        \end{equation}
    \end{itemize}

% When applied to a product of a scalar and vector field, such as the mass flux $\rho \mathbf{v}$ in the continuity equation, the divergence operator follows a product rule:
% \[
% \nabla \cdot (\rho \mathbf{v}) = \mathbf{v} \cdot \nabla \rho + \rho \nabla \cdot \mathbf{v}.
% \]
% This captures both the local expansion/compression of the velocity field and spatial variations in density. The continuity equation therefore accounts for changes in mass due to both fluid motion and density fluctuations.

%The equations are written in continuous form, and the spatial and temporal domains are assumed to be smooth. 

\end{enumerate}

Details on solving these equations and their discrete representation are provided in the following section.

\subsection{Solving the Atmospheric Primitive Model }

In summary, these equations simulate the evolution of wind velocity, temperature, and density fields in the atmosphere, governed by internal physical dynamics (advection, pressure gradients, and gravity). They also allow the incorporation of external forcings such as incoming solar radiation or anthropogenic GHGs concentrations as discussed later. Once formulated as a continuous set of those nonlinear PDEs [\eqref{eq:NS_AtmEx1}, \eqref{eq:NS_AtmEx2}, and \eqref{eq:NS_AtmEx3}], the system can be discretized in space and time to be solvable numerically. We now discuss the spatio-temporal discretization resolution, initial and boundary conditions, integrating external forcing, parametrization, and feedback mechanisms in this model.

% \begin{enumerate}
%   \noindent \textbf{Discretization and Resolution.} The atmosphere dynamics are represented over a 3D computational grid, often structured in latitude-longitude-height (or pressure) coordinates, by discretizing the PDEs \eqref{eq:NS_Atm} using numerical techniques. Numerical discretization techniques utilized examples:
%     % \begin{itemize}
%     %     \item Finite-difference methods (FDM): approximate derivatives using neighboring grid values (common in legacy models such as NCAR's CAM~\cite{CommunityAtmosphereModel}).
%     %     \item Spectral methods: decompose variables into basis functions (e.g., spherical harmonics); efficient for global models.
%     %     \item Finite-volume methods (FVM): integrate fluxes across control volumes; more robust for mass/energy conservation and common in newer models (e.g., FV3 in NOAA's models~\cite{FV3FiniteVolumeCubedSphere}).
%     % \end{itemize}

%     \begin{subequations} \label{eq:Disc_NS}
% 	\begin{align}
% 		\dot{\mathbf{v}} &= - (\mathbf{v} \cdot \nabla) \mathbf{v} - f\mathbf{k} \times \mathbf{v} - \boldsymbol{\rho}^{-1} \nabla \mathbf{p} + \mathbf{F}_\text{viscous}, \label{eq:Disc_NS1} \\
% 		\dot{\mathbf{T}} &= - \mathbf{v} \cdot \nabla \mathbf{T} + \mathbf{Q}_T, \label{eq:Disc_NS2} \\
% 		\dot{\boldsymbol{\rho}} &= - \nabla \cdot (\boldsymbol{\rho} \mathbf{v}). \label{eq:Disc_NS3}
% 	\end{align}
% \end{subequations}

% Tab. \ref{tab:Atm_NS_Var} defines each of the system \eqref{eq:NS_Atm} variables, states, parameters, and operators.

\begin{table*}[t!]
	\centering
	\caption{Atmospheric Primitive Equation Terms and Operators: Units, Discrete Dimensions, and Physical Interpretation.}~\label{tab:Atm_NS_Var}
	\renewcommand{\arraystretch}{1.2}
	\begin{tabularx}{\textwidth}{>{\bfseries}l l p{1.3cm} l X}
		\toprule
		\textbf{Symbol} & \textbf{Units} & \textbf{Discrete Dimension} & \textbf{Represents} & \textbf{Notes} \\
		\midrule
		$\mathbf{v} = [\mathbf{v}_x, \mathbf{v}_y, \mathbf{v}_z]$ & [m/s] & $\mathbb{R}^{3N}$ & Wind velocity vector field & Zonal (along $x$-axis), meridional (along $y$-axis), and vertical (along $z$-axis) wind components, stacked across grid points \\
		$\mathbf{T}$ & [K] & $\mathbb{R}^{N}$ & Temperature field & Scalar field on the 3D spatial grid. Evolves due to advection and heating \\
		$\boldsymbol{\rho}$ & [kg/m$^3$] & $\mathbb{R}^{N}$ & Air density & Scalar field. Varies with altitude, pressure, and temperature. Appears in continuity and momentum equations \\
		$\mathbf{p}$ & [Pa] & $\mathbb{R}^{N}$ & Pressure field & Drives wind through horizontal and vertical pressure gradients\\
		$\nabla$ & [1/m] & $\mathbb{R}^{3N \times N}$ & Gradient operator & Maps scalar fields $\rightarrow$ vector fields. Constructed via finite-difference or finite-volume stencils \\
		$\nabla \cdot$ & [1/m] & $\mathbb{R}^{N \times 3N}$ & Divergence operator & Maps vector fields $\rightarrow$ scalar fields. Used in mass conservation (continuity) equation \\
		$\frac{\partial}{\partial t}$ & [1/s] & -- & Time derivative operator & Used in time integration. Realized through explicit or implicit stepping schemes \\
		$f \mathbf{k} \times \mathbf{v}$ & [m/s$^2$] & $\mathbb{R}^{3N}$ & Coriolis acceleration term & Represents Earth's rotation effects. $f$ is Coriolis parameter and varies with latitude. $\mathbf{k}$ is unit vector in vertical direction. Cross-product per grid point \\
		$\nabla \mathbf{p}$ & [Pa/m] = [N/m$^3$] & $\mathbb{R}^{3N}$ & Pressure gradient & Spatial derivatives of pressure; drive acceleration in the momentum equation. \\
		${\boldsymbol{\rho}}^{-1} \nabla \mathbf{p}$ & [m/s$^2$] & $\mathbb{R}^{3N}$ & Pressure gradient force & Acceleration due to pressure gradient; nonlinearly depends on density. \\
		$\mathbf{F}_\text{viscous}$ & [m/s$^2$] & $\mathbb{R}^{3N}$ & Sub-grid viscous force & Represents turbulent diffusion or vertical mixing. Often parameterized via Laplacians. \\
		$(\mathbf{v} \cdot \nabla) \mathbf{v}$ & [m/s$^2$] & $\mathbb{R}^{3N}$ & Advective acceleration & Nonlinear advection of momentum by the wind field. Each component convects all others. \\
		$\mathbf{v} \cdot \nabla T$ & [K/s] & $\mathbb{R}^{N}$ & Advective heat transport & Represents temperature change due to wind-driven advection. \\
		$\mathbf{Q}_T$ & [K/s] & $\mathbb{R}^{N}$ & Net heating source term & Includes radiative, latent, and sensible heating. Can include sub-grid parameterized sources. \\
		$\nabla \cdot (\boldsymbol{\rho} \mathbf{v})$ & [kg/(m$^3$·s)] & $\mathbb{R}^{N}$ & Mass divergence & Appears in the continuity equation; determines local change in air density. \\
		\bottomrule
	\end{tabularx}
\vspace{-0.2cm}
\end{table*}

\vspace{0.1cm}
\noindent\textbf{Discretization and Resolution.}
The atmospheric primitive equations \eqref{eq:NS_Atm} describe the evolution of velocity, temperature, and density fields over a continuous spatiotemporal domain. To solve them numerically, the spatial domain is discretized over a 3D grid, commonly defined in latitude-longitude-height or pressure coordinates, and the temporal dimension is discretized using time-stepping schemes.

\vspace{0.1cm}
\noindent\textit{Spatial Discretization (Semi-Discrete Form).} The continuous spatial fields are replaced with grid-defined fields over $N$ nodes. Derivative operators such as $\nabla$ and $\nabla \cdot$ are approximated numerically using one of the following methods:
\begin{itemize}
        \item Finite-difference methods (FDM): approximate derivatives using neighboring grid values (common in legacy models such as NCAR's CAM~\cite{CommunityAtmosphereModel}).
        \item Spectral methods: decompose variables into basis functions (e.g., spherical harmonics). These methods are efficient for global models.
        \item Finite-volume methods (FVM): integrate fluxes across control volumes; more robust for mass/energy conservation and common in newer models (e.g., FV3 in NOAA's models~\cite{FV3FiniteVolumeCubedSphere}).
    \end{itemize}

After discretizing the dynamics over the space domain, each field becomes a finite-dimensional vector:
\[
\mathbf{v} \in \mathbb{R}^{3N}, \quad \mathbf{T} \in \mathbb{R}^{N}, \quad \boldsymbol{\rho} \in \mathbb{R}^{N}, \quad \mathbf{p} \in \mathbb{R}^{N}.
\]
In the continuous setting, the velocity field is written as $\mathbf{v} = (u, v, w)^\top$, where each component represents the zonal, meridional, and vertical wind velocity. After spatial discretization, each of these components becomes a vector in $\mathbb{R}^N$ defined at every grid point. The full velocity field is thus represented as a stacked vector $\mathbf{v} = [\mathbf{v}_x^\top, \mathbf{v}_y^\top, \mathbf{v}_z^\top]^\top \in \mathbb{R}^{3N}$, where $\mathbf{v}_x$, $\mathbf{v}_y$, and $\mathbf{v}_z$ correspond to the discretized fields for $u$, $v$, and $w$, respectively. In addition, the differential operators become sparse matrices or stencil-based operators (e.g., $\nabla \in \mathbb{R}^{3N \times N}$, $\nabla \cdot \in \mathbb{R}^{N \times 3N}$). For completeness, Tab.~\ref{tab:Atm_NS_Var} summarizes the system variables, parameters, and operators along with their units, discrete dimensions, and physical interpretations.

After spatial discretization, the PDEs \eqref{eq:NS_Atm} are transformed into a system of ODEs in time (semi-discrete form):
\begin{subequations} \label{eq:Disc_NS}
	\begin{align}
	\dot{\mathbf{v}} &= - (\mathbf{v} \cdot \nabla) \mathbf{v} - f\mathbf{k} \times \mathbf{v} - \boldsymbol{\rho}^{-1} \nabla \mathbf{p} + \mathbf{F}_\text{viscous}, \label{eq:Disc_NS1} \\
		\dot{\mathbf{T}} &= - \mathbf{v} \cdot \nabla \mathbf{T} + \mathbf{Q}_T, \label{eq:Disc_NS2} \\
		\dot{\boldsymbol{\rho}} &= - \nabla \cdot (\boldsymbol{\rho} \mathbf{v}). \label{eq:Disc_NS3}
	\end{align}
\end{subequations}
These equations describe the evolution of state vectors $(\mathbf{v}, \mathbf{T}, \boldsymbol{\rho})$ governed by nonlinear and coupled dynamics. The nonlinear terms such as $(\mathbf{v} \cdot \nabla)\mathbf{v}$ and $\nabla \cdot (\boldsymbol{\rho} \mathbf{v})$ are evaluated using stencil operations or matrix–vector products. Similarly to \eqref{eq:NS_AtmEx1}, \eqref{eq:NS_AtmEx2}, and \eqref{eq:NS_AtmEx3} in the continuous domain, the semi-discrete form can be expanded to:
\begin{subequations} \label{eq:Disc_NSEx}
\begin{align}
\begin{aligned}
\frac{d}{dt}
\begin{bmatrix}
\mathbf{v}_x \\
\mathbf{v}_y \\
\mathbf{v}_z
\end{bmatrix}
&=
-
\begin{bmatrix}
\mathbf{v}_x \odot \frac{\partial \mathbf{v}_x}{\partial x} +
\mathbf{v}_y \odot \frac{\partial \mathbf{v}_x}{\partial y} +
\mathbf{v}_z \odot \frac{\partial \mathbf{v}_x}{\partial z} \\
\mathbf{v}_x \odot \frac{\partial \mathbf{v}_y}{\partial x} +
\mathbf{v}_y \odot \frac{\partial \mathbf{v}_y}{\partial y} +
\mathbf{v}_z \odot \frac{\partial \mathbf{v}_y}{\partial z} \\
\mathbf{v}_x \odot \frac{\partial \mathbf{v}_z}{\partial x} +
\mathbf{v}_y \odot \frac{\partial \mathbf{v}_z}{\partial y} +
\mathbf{v}_z \odot \frac{\partial \mathbf{v}_z}{\partial z}
\end{bmatrix}
\\ &\quad
-
f
\begin{bmatrix}
-\mathbf{v}_y \\
\mathbf{v}_x \\
\mathbf{0}
\end{bmatrix}
-
\boldsymbol{\rho}^{-1} \odot
\begin{bmatrix}
\frac{\partial \mathbf{p}}{\partial x} \\
\frac{\partial \mathbf{p}}{\partial y} \\
\frac{\partial \mathbf{p}}{\partial z}
\end{bmatrix}
+
\begin{bmatrix}
\mathbf{F}_x \\
\mathbf{F}_y \\
\mathbf{F}_z
\end{bmatrix}
\end{aligned}
\label{eq:Disc_NSEx1}
\\
\begin{aligned}
\frac{d\mathbf{T}}{dt}
&=
-
\left(
\mathbf{v}_x \odot \frac{\partial \mathbf{T}}{\partial x} +
\mathbf{v}_y \odot \frac{\partial \mathbf{T}}{\partial y} +
\mathbf{v}_z \odot \frac{\partial \mathbf{T}}{\partial z}
\right)
+
\mathbf{Q}_T
\end{aligned}
\label{eq:Disc_NSEx2}
\\
\begin{aligned}
\frac{d\boldsymbol{\rho}}{dt}
&=
-
\left(
\mathbf{v}_x \odot \frac{\partial \boldsymbol{\rho}}{\partial x} +
\mathbf{v}_y \odot \frac{\partial \boldsymbol{\rho}}{\partial y} +
\mathbf{v}_z \odot \frac{\partial \boldsymbol{\rho}}{\partial z}
\right)
\\ &\quad
-
\boldsymbol{\rho} \odot
\left(
\frac{\partial \mathbf{v}_x}{\partial x} +
\frac{\partial \mathbf{v}_y}{\partial y} +
\frac{\partial \mathbf{v}_z}{\partial z}
\right) 
\end{aligned}
\label{eq:Disc_NSEx3}
\end{align}
\end{subequations}
where $\odot$ denotes element-wise multiplication across the $N$ grid points and the term $\boldsymbol{\rho}^{-1}$ represents the pointwise inverse of the discretized density field. That is, if $\boldsymbol{\rho} = [\rho_1, \rho_2, \dots, \rho_N]^\top$, then $\boldsymbol{\rho}^{-1} = [1/\rho_1, 1/\rho_2, \dots, 1/\rho_N]^\top$, and the expression $\boldsymbol{\rho}^{-1} \odot \nabla \mathbf{p}$ is computed element-wise across the grid.

\vspace{0.1cm}
\noindent\textit{Temporal Discretization (Fully Discrete Form).} 
Once the spatial discretization yields a system of ODEs (as in Eq.~\eqref{eq:Disc_NS}), temporal integration is applied to evolve the system forward in time. This results in a fully discrete representation of the system dynamics, or a set of nonlinear difference equations (NDE).  

The choice of time-stepping scheme influences numerical stability, accuracy, and computational cost:
\begin{itemize}
    \item {Explicit methods} (e.g., forward Euler, Runge-Kutta) evaluate the right-hand side of the ODE using known values at time step \( n \). These methods are simple and computationally cheap but subject to stability constraints that may require very small time steps.
    
    \item {Implicit methods} (e.g., backward Euler, Crank-Nicolson) involve evaluating the right-hand side at the unknown future time \( n+1 \), which generally requires solving a system of equations at each time step. They are more stable, especially for stiff systems, but computationally more intensive.
\end{itemize}

Let \( \Delta t \) denote the time step, and define the time-discretized state variables as
\begin{equation*}
\begin{aligned}
  &  \mathbf{v}^n := \mathbf{v}(t_n), \quad  \mathbf{T}^n := \mathbf{T}(t_n), \quad \boldsymbol{\rho}^n := \boldsymbol{\rho}(t_n), \\  
    & \text{where } t_n = n \Delta t \text{ and } n = 0, 1, \dots, N_t.
\end{aligned}
\end{equation*}
Meaning that, the simulation is carried out over a time span \( [0, t_{\text{final}}] \), where \( t_{\text{final}} = N_t \Delta t \), and \( N_t \) is the total number of time steps.

Using a forward Euler scheme as an illustrative example (first-order explicit method), the time integration of Eq.~\eqref{eq:Disc_NS} gives the fully discrete update equations:
\begin{subequations} \label{eq:Disc_NS_Discrete}
\begin{align}
\mathbf{v}^{n+1} 
&= \mathbf{v}^n + \Delta t \cdot \Big( 
- (\mathbf{v}^n \cdot \nabla) \mathbf{v}^n 
- f \mathbf{k} \times \mathbf{v}^n \notag \\
&\quad 
- \boldsymbol{\rho}^{-1,n} \odot \nabla \mathbf{p}^n 
+ \mathbf{F}_\text{viscous}^n 
\Big) \label{eq:Disc_NS_Discrete1} \\[1ex]
\mathbf{T}^{n+1} 
&= \mathbf{T}^n + \Delta t \cdot \Big( 
- \mathbf{v}^n \cdot \nabla \mathbf{T}^n 
+ \mathbf{Q}_T^n 
\Big) \label{eq:Disc_NS_Discrete2} \\[1ex]
\boldsymbol{\rho}^{n+1} 
&= \boldsymbol{\rho}^n + \Delta t \cdot \Big(
- \nabla \cdot (\boldsymbol{\rho}^n \mathbf{v}^n)
\Big) \label{eq:Disc_NS_Discrete3}
\end{align}
\end{subequations}
where the superscript \( n \) indicates the current time step, and \( n+1 \) the next; all nonlinear terms are evaluated at time step \( n \), consistent with the explicit Euler scheme; the inverse density term \( \boldsymbol{\rho}^{-1,n} \) is computed element-wise as in the semi-discrete form: if \( \boldsymbol{\rho}^n = [\rho_1^n, \dots, \rho_N^n]^\top \), then \( \boldsymbol{\rho}^{-1,n} = [1/\rho_1^n, \dots, 1/\rho_N^n]^\top \). Finally, the pressure gradient \( \nabla \mathbf{p}^n \) and all spatial derivatives are approximated using the same stencil or matrix-based discretization discussed earlier.
    
This formulation yields a first-order, fully discrete system that updates the state at each time step based solely on known quantities. Higher-order or implicit schemes can be applied in similar fashion, but may require iterative solvers or linearizations.

\begin{table*}[t!]
		\centering
		\caption{Typical Spatial and Temporal Resolution for Global Atmosphere and Weather Models Built Based on Systems such as \eqref{eq:Disc_NS}.}~\label{tab:AtmResolution}
		\renewcommand{\arraystretch}{1.2}
		\begin{tabularx}{\textwidth}{>{\bfseries}l l l l X}
			\toprule
			\textbf{Model Type} & \textbf{Horizontal Resolution} & \textbf{Vertical Layers} & \textbf{Time Step} & \textbf{Notes} \\
			\midrule
			Global GCM (coarse) & $\sim$100--250 km & $\sim$20--60 layers & 10--60 min & Suitable for long-term atmospheric climate projections. Lower computational demand. \\
			Global GCM (fine)   & $\sim$25--50 km   & $\sim$100+ layers & 1--10 min  & Higher spatial fidelity allows better representation of changes in parameters and external forcings. \\
			RCM / Weather Model & $\sim$1--25 km    & $\sim$40--100+ layers & seconds-minutes & High-resolution simulations for localized weather or regional climate. RCM requires external boundary conditions (from GCM or reanalysis). \\
			\bottomrule
		\end{tabularx}
	\end{table*}

\vspace{0.1cm}
\noindent \textit{Discretized System's Resolution.}
The spatial and temporal resolutions (e.g., the grid size and time step) are determined based on the type of model as shown in Tab.~\ref{tab:AtmResolution} and the discretization scheme used, with stability and accuracy constraints taken into account. The reader is also referred to Tab. \ref{tab:ClimModelsOvw} for more details on those climate models types. The choice of resolution reflects a tradeoff between computational tractability and the ability to represent fine-scale processes, which is also influenced by the model's scientific objectives (e.g., climate trends vs. short-term forecasting) and the availability of computational resources. As such, the structure and interpretation of the discrete equations~\eqref{eq:Disc_NS_Discrete} presented here remain applicable across a wide range of model configurations, from coarse-resolution climate simulations to high-resolution, event-driven forecasts.

\vspace{0.1cm}
    \noindent \textbf{Initial vs. Boundary Conditions (Weather vs. Regional Climate Models).}
        Global weather models obtain their forecasts by solving an atmospheric dynamical system (e.g., \eqref{eq:Disc_NS_Discrete}) forward from a detailed \textit{initial} atmospheric state (at $t=0$). On the other hand, regional weather/climate models (based on these same dynamics) solve them with boundary conditions applied at the edges of the domain: $\mathbf{v}, \mathbf{T}, \boldsymbol{\rho}$ prescribed at domain boundaries every 3--6 hours, to ensure consistency with large-scale atmospheric patterns. These boundary values are interpolated from a GCM or reanalysis dataset, and they feed into the right-hand sides of \eqref{eq:Disc_NS_Discrete} via advection or pressure gradient terms.

\vspace{0.1cm}
    \noindent \textbf{Parameterization and External Forcing.} In Eq. \eqref{eq:Disc_NS_Discrete}, the term $\mathbf{F}_\text{viscous}$ represents unresolved turbulent diffusion and frictional effects. Since many atmospheric motions (e.g., small eddies and turbulence) occur at spatial and temporal scales smaller than the grid resolution, they cannot be explicitly modeled and must be parameterized. Therefore, it is counted as an unresolved processes and its effect is approximated based on a resolved state. For example, turbulent mixing in the planetary boundary layer is often modeled using eddy diffusivity schemes, which in return can be expressed as function of the wind velocity $\mathbf{v}(t)$. 

   External forcing refers to influences and external inputs to that affect the atmospheric state but are not determined by the internal dynamics of the system. Common examples for the systems expressed in \eqref{eq:NS_Atm} and \eqref{eq:Disc_NS_Discrete} include solar radiation (shortwave), GHGs  concentrations, aerosols, land use changes, and sea surface temperature boundary conditions. These forcings typically enter the system through the heating term ($Q_T$ in \eqref{eq:NS_Atm} and $\mathbf{Q}_T$ \eqref{eq:Disc_NS_Discrete})   and/or through modified boundary conditions for regional models. 

    \vspace{0.1cm}
    \noindent \textbf{Feedback Mechanisms.} Even in this core dynamic system, internal feedback mechanisms naturally emerge from the coupling of variables:
	\begin{itemize}
		\item \textit{Thermodynamic–momentum coupling:} A change in temperature \( T \) modifies air density \( \rho \) via the ideal gas law, which in turn alters pressure gradients \( \nabla p \) and affects wind velocity \( \mathbf{v} \). The wind field then advects temperature, feeding it back into the temperature field and completing the loop as shown below:
\vspace{-0.2cm}
% \begin{custombox}{magenta}
% \centering
% $T \rightarrow \rho \rightarrow \nabla p \rightarrow \mathbf{v} \rightarrow \nabla T \rightarrow T$
% \end{custombox}
%     \vspace{-0.2cm}
\begin{shadedcvbox}
% \begin{custombox}{magenta}
\centering
\begin{tikzpicture}[baseline=(current bounding box.center), every node/.style={font=\normalsize}]
    \node (T1) at (0,0) {\( T \)};
    \node (rho) at (0.8,0) {\( \rho \)};
    \node (gradp) at (1.8,0) {\( \nabla p \)};
    \node (v) at (2.8,0) {\( \mathbf{v} \)};
    \node (gradT) at (3.8,0) {\( \nabla T \)};
    \node (T2) at (4.8,0) {\( T \)};

    \draw[->] (T1) -- (rho);
    \draw[->] (rho) -- (gradp);
    \draw[->] (gradp) -- (v);
    \draw[->] (v) -- (gradT);
    \draw[->] (gradT) -- (T2); 
    % Straight arrow under with corners
    \draw[->] (T2.south) -- ++(0,-0.25) -- ++(-4.8,0) -- (T1.south);
\end{tikzpicture}
\end{shadedcvbox}
% \end{custombox}

	\item \textit{A heating anomaly }(\(Q_T > 0\)) increases the air temperature \(T\), which reduces density \(\rho\) through the ideal gas law and alters pressure gradients \(\nabla p\). This accelerates the wind field \(\mathbf{v}\), which transports heat and moisture, further modifying the temperature field. The increase in \(T\) also raises water vapor concentration, a GHG, enhancing greenhouse trapping and amplifying the initial warming via radiative feedback. These couplings form a closed internal feedback loop in the atmospheric system:
%     \vspace{-0.2cm}
% 	\begin{custombox}{magenta}
% 	$Q_T > 0 \rightarrow T \rightarrow \rho \rightarrow \nabla p \rightarrow \mathbf{v} \rightarrow \nabla T \rightarrow \text{GHG forcing} \rightarrow Q_T$
% \end{custombox}\vspace{-0.2cm}
\vspace{-0.2cm}
\begin{shadedcvbox}
% \begin{custombox}{magenta}
\begin{tikzpicture}[remember picture, baseline=(box.center)]
    \node[inner sep=0pt] (box) {
        \parbox{0.92\columnwidth}{\centering
        $Q_T > 0 \rightarrow T \rightarrow \rho \rightarrow \nabla p \rightarrow \mathbf{v} \rightarrow \nabla T \rightarrow$\\
        GHG forcing $\rightarrow Q_T$
        }
    };

    % Exact arrow shape under wrapping
    \coordinate (start) at ([xshift=-6.9em,yshift=-0.1em]box.south east);
    \coordinate (benddown) at ($(start)+(0,-0.6em)$);
    \coordinate (middle) at ($(benddown)+(-0.6\columnwidth,0)$);
    \coordinate (endup) at ($(middle)+(0,2.25em)$);
    \coordinate (end) at ([xshift=1.3em,yshift=1.55em]box.south west);

    \draw[->] (start) -- (benddown) -- (middle) -- (endup) -- (end);
\end{tikzpicture}
% \end{custombox}
\end{shadedcvbox}
\vspace{-0.2cm}
\end{itemize}
% \end{enumerate}

% \noindent \textbf{Interpreting the Atmospheric Model \eqref{eq:NS_Atm} Through a Control-Theoretic Lens. \comment{CHANGE}}
\vspace{-0.3cm}
\subsection{State-Space Outlook on the Atmospheric Model}
Having introduced the structure, physical, and mathematical interpretation of the atmospheric primitive equations, we now shift toward viewing this system through the lens of control engineering. While originally formulated to simulate large-scale atmospheric dynamics, this model can also be understood as a high-dimensional, nonlinear dynamical system with coupled state variables, external inputs, and internal feedback loops. By identifying the system's states, inputs, outputs, and feedback structure, we lay the groundwork for applying tools from control engineering such as observability, controllability, stability, state estimation, and feedback control design. This enables potentially improved understanding, analysis, and prediction of the behavior of the atmospheric system.

% \noindent The mapping from PDEs $\rightarrow$ ODEs $\rightarrow$ difference equations is summarized as:
% \[
% \text{Continuous PDE} \quad \rightarrow \quad \text{Semi-Discrete ODE: } \frac{d\mathbf{x}}{dt} = \mathbf{f}(\mathbf{x}, t) \quad \rightarrow \quad \text{Discrete Update: } \mathbf{x}^{n+1} = \mathbf{x}^n + \Delta t \cdot \mathbf{f}(\mathbf{x}^n)
% \]
% where $\mathbf{x} = [\mathbf{v}^\top, \mathbf{T}^\top, \boldsymbol{\rho}^\top]^\top \in \mathbb{R}^{5N}$ is the full state vector of the system.

% $$ x_{t+1} = f(x_t, \underbrace{u_t}_{\text{cont. forces}}, \overbrace{w_t}^{\text{uncont. forces}}, \underbrace{p_t}_{\text{time-varying params}}; \overbrace{c}^{\text{time invariant params.}}) $$
% $$ 0 = g(x_t, \underbrace{u_t}_{\text{cont. forces}}, \overbrace{w_t}^{\text{uncont. forces}}, \underbrace{p_t}_{\text{time-varying params}}; \overbrace{c}^{\text{time invariant params.}})$$
% $$ y_t = C_x x_t + C_u u_t + C_w w_t + C_p p_t$$

% $x_t$ is a vector with large scale dimension that concatenates $v(t,x), T(t,x), \rho(t,x)$ 

The handling of the atmospheric primitive system pipeline progresses through three stages:
\[
\text{PDEs \eqref{eq:NS_Atm}} 
\rightarrow 
\text{Semi-discrete ODEs \eqref{eq:Disc_NS}: } \mathbf{\dot{x}}(t) = \mathbf{f}(\mathbf{x}, t) 
\]
\[
\rightarrow 
\text{Discrete-time NDEs  \eqref{eq:Disc_NS_Discrete}: } 
\mathbf{x}^{n+1} = \mathbf{x}^n + \Delta t \cdot \mathbf{f}(\mathbf{x}^n)
\]
where $\mathbf{x}$ is the full state vector gathering the wind velocity $\mathbf{v}, \mathbf{T},$ and $\boldsymbol{\rho}$.
% in the discrete form is
% \[
% \mathbf{x}(t) = \begin{bmatrix} \mathbf{v}_x^\top(t) & \mathbf{v}_y^\top(t) & \mathbf{v}_z^\top(t) & \mathbf{T}^\top(t) & \boldsymbol{\rho}^\top(t) \end{bmatrix}^\top \in \mathbb{R}^{5N}.
% \]
% Here, \( N \) is the number of grid points in the spatial domain, and each subfield (velocity components, temperature, and density) is a length-\( N \) vector.

We can now reinterpret the atmospheric model using a state-space framework commonly adopted in control engineering. This provides a compact and general way to represent system dynamics, inputs, outputs, and parameters.

\noindent\textit{State-Space Form of the Semi-Discrete ODE System.}
The system of equations in \eqref{eq:Disc_NS} describes the evolution of the atmospheric state variables over time following spatial discretization. This system can be expressed in a general nonlinear state-space form:
% \begin{equation} \label{eq:StateSpace_ODE}
% \frac{d\mathbf{x}}{dt} = \mathbf{f}(\mathbf{x}(t),\, \underbrace{\mathbf{u}(t)}_{\text{controlled inputs}},\, \overbrace{\mathbf{w}(t)}^{\text{uncontrolled forcings}},\, \underbrace{\boldsymbol{\theta}(t)}_{\text{time-varying parameters}};\, \overbrace{c}^{\text{fixed parameters}})
% \end{equation}
\begin{subequations} \label{eq:StateSpace_ODE}
\begin{align}
 &   \frac{d\mathbf{x}}{dt} = \mathbf{f}(\mathbf{x}(t),\, {\mathbf{u}(t)},\, {\mathbf{w}(t)},\, {\boldsymbol{\theta}(t)};\, \mathbf{c}),\\
  &  \mathbf{y}(t) = \mathbf{h}(\mathbf{x}(t),\, \mathbf{u}(t),\, \mathbf{w}(t),\, \boldsymbol{\theta}(t)).
\end{align}
\end{subequations}

\noindent In this representation, \( \mathbf{x}(t) = [\mathbf{v}_x(t)^\top,\, \mathbf{v}_y(t)^\top,\, \mathbf{v}_z(t)^\top,\, \mathbf{T}(t)^\top,\, \boldsymbol{\rho}(t)^\top]^\top \in \mathbb{R}^{5N} \) is the state vector, composed of the time-varying discretized fields of wind, temperature, and density. The input \( \mathbf{u}(t) \) denotes potential control actions, such as prescribed heating, anthropogenic forcing, or modified boundary fluxes. The term \( \mathbf{w}(t) \) represents uncontrolled environmental inputs like volcanic eruptions or natural solar variability. The time-dependent parameter vector \( \boldsymbol{\theta}(t) \) includes evolving but externally provided fields such as surface albedo, land cover, or radiative coefficients. Lastly, \( \mathbf{c} \) represents time-invariant physical parameters (e.g., gas constant, gravity), and the output vector \( \mathbf{y}(t) \) captures observable quantities such as surface temperature, pressure, or wind components at selected locations or aggregated regions.

\noindent\textit{Fully Discrete Nonlinear Dynamical System.}
By further discretizing the temporal domain using a numerical integration scheme such as forward Euler (as in \eqref{eq:Disc_NS_Discrete}), the system becomes a NDEs:
\begin{equation} \label{eq:StateSpace_NDE}
\mathbf{x}^{n+1} = \mathbf{f}_d(\mathbf{x}^n,\, \mathbf{u}^n,\, \mathbf{w}^n,\, \mathbf{p}^n;\, c)
\end{equation}
\begin{equation} \label{eq:Output_NDE}
\mathbf{y}^n = \mathbf{h}_d(\mathbf{x}^n,\, \mathbf{u}^n,\, \mathbf{w}^n,\, \mathbf{p}^n)
\end{equation}

\noindent where \( \mathbf{x}^n \in \mathbb{R}^{5N} \) is the state at time \( t_n = n \Delta t \), all variables are evaluated at discrete time steps indexed by \( n \), and the function \( \mathbf{f}_d \) results from applying the chosen time-stepping scheme to \eqref{eq:StateSpace_ODE}.

Having reformulated the atmospheric dynamics in both continuous and discrete state-space form, we are now equipped to draw explicit parallels with the standard language and structure of control systems. This interpretation is not only useful for re-expressing the model mathematically, but it also enables the application of a wide array of techniques from estimation, feedback control, and system analysis. {The following summary boxes highlight how the atmospheric primitive equations, originally derived for physical simulation, can be naturally framed as a large-scale nonlinear dynamical system amenable to control-theoretic reasoning.}

\begin{theo}[Model Representation]
	\vspace{-0.25cm}
	The system is originally described by nonlinear PDEs governing momentum, thermodynamics, and mass conservation. Upon discretization in space and time (i.e., Eq. \eqref{eq:Disc_NS_Discrete}), the system becomes a high-dimensional, nonlinear discrete-time dynamical system. The dimensionality of the resulting system is directly tied to the spatial resolution of the grid and the number of physical variables tracked at each grid point. \\
         In GCMs, which typically operate with horizontal resolutions between 100 and 250 kilometers (or down to 25-50 km in higher-resolution configurations as shown in Tab. \ref{tab:AtmResolution}), the total number of state variables can range from tens to hundreds of millions, depending on the number of vertical levels and included fields. For example, a typical GCM might use a horizontal resolution of 100 km with 60 vertical levels on a grid covering the entire globe (\(3.6 \times 10^6\) horizontal grid points). Tracking three primary fields of states: wind velocity \((\mathbf{v}_x, \mathbf{v}_y, \mathbf{v}_z)\), temperature \(\mathbf{T}\), and density \(\boldsymbol{\rho}\) at each of the ~\(2.2 \times 10^8\) total grid points (horizontal $\times$ vertical) results in over one billion state variables.\\         
         RCMs and NWP systems, while applied over smaller domains, use much finer resolution, often between 1 and 25 kilometers, and thus also produce systems with millions of coupled state variables. Another example here is a RCM applied over a smaller domain (e.g., \(1000 \times 1000\) km) at 10 km resolution with 50 vertical levels would yield \(10^6\) horizontal points and about \(5 \times 10^7\) total variables for the same set of fields. Reducing the horizontal resolution to 5 km would quadruple the number of horizontal points and increase the system size accordingly.\\
         These discretized systems retain the nonlinear and tightly coupled structure of the original equations, making them ideal candidates for control-theoretic interpretation and analysis, especially in the context of large-scale estimation, uncertainty quantification, and feedback structure. 
    		% \vspace{-0.1cm}
\end{theo}

\begin{theo}[States and Measurable Outputs]
	\vspace{-0.25cm}
	The state vector includes grid-point values of wind velocity \(\mathbf{v}\), temperature \(\mathbf{T}\), and density \(\boldsymbol{\rho}\) across the 3D model domain. Measurable outputs typically include surface or upper-air observations of temperature, wind speed, and pressure, obtained from satellites, radiosondes, or ground stations. In addition to states, some external forcing terms or boundary inputs, such as radiative heating, surface fluxes, or pressure gradients, may be estimated or prescribed from measurements or reanalysis data. Since the full state and all inputs are not directly observable, estimation and inference play a central role. These challenges are well-studied in the control engineering field, and their relevance to atmospheric modeling is further discussed in Section~\ref{sec:Clmt&CntrlEng}.
    		% \vspace{-0.1cm}
\end{theo}

\begin{theo}[Control Inputs]
	\vspace{-0.25cm}
	 Similarly to the EBM model \eqref{eq:EBM_SAI}, control inputs can be conceptually introduced to the atmospheric system to represent hypothetical or indirect interventions. In physical terms, these inputs would correspond to modifications of the external forcing terms (e.g., heat sources \(Q_T\)), boundary conditions, or parameterized sub-grid processes. Unlike in the EBM, where the input can directly modify surface albedo or emissivity in a scalar ODE, here the influence would be distributed in space and time across the 3D model grid. For example, injecting stratospheric aerosols would alter the radiative balance in the energy equation by increasing reflectivity and thus reducing incoming solar energy---a localized change in the heating term. Similarly, implementing large-scale land-use changes or reflective surfaces could alter surface properties and boundary conditions. Though these inputs are not part of standard weather or climate prediction workflows, they are useful constructs in a control-theoretic interpretation of the system: they represent degrees of influence over the model's evolution, enabling the posing of questions such as: what form, location, or magnitude of intervention is needed to influence atmospheric temperature distributions or circulation patterns? These inputs can be thought of as distributed, constrained actuators that interface with the physical system, similar in spirit to how geoengineering or emission policies have been framed as control inputs in the EBM case.
    		% \vspace{-0.1cm}
\end{theo}

 \vspace{0.1cm}
\noindent \textbf{Raising Control Engineering Questions.} The models presented in \eqref{eq:NS_Atm} and {\eqref{eq:Disc_NS_Discrete}} open up classical and modern control questions that are related to core control concepts, such as the following, pertaining to specifically observability and controllability. 
% \begin{itemize}
For {observability-related analyses,} the following questions can be asked: Which parts of the atmospheric state $\m x(t)$ can be reconstructed from the available measurements $\m y(t)$? What and how many sensors are required to ensure desirable observability metrics for any of the derived climate models? These questions are often posed in dynamic system sciences but are hardly navigated in climate research.

As for controllability analysises, the following questions can be posed. If hypothetical actuators are introduced to imitate controlled practices (e.g., geoengineering techniques) and policy-driven mitigation schemes, then mapped to present control vectors in \eqref{eq:NS_Atm} and \eqref{eq:Disc_NS_Discrete}, then how effective are these control measures at steering the system toward a more desirable climate trajectory? What are the resulting controllable subspaces and how can reachability analysis improve our understanding of the decade-long evolution of climate states? 
% Instead of just simply running extended period simulations of climate models, can reachability analysis inform or predict reachable spaces given improved mitigation strategies and specific control actions? 
What is the impact of placing control nodes as actuators in specific geographic locations to thereby improve the trajectory of climate states? Can this be posed as an actuator placement problem? 
% \end{itemize}

While these questions are intuitive to the controls community, they may appear unfamiliar or even abstract to climate scientists. However, many of these concepts have meaningful counterparts in climate modeling, even if they are approached from different directions or embedded implicitly in other tasks. This connection is further elaborated and emphasized in Section~\ref{sec:Clmt&CntrlEng}. Furthermore, one example of such a connection is highlighted in the next section. 

{Note that, while the above-stated questions were directed toward a simple EBM and an atmospheric process model, the same questions, and the connections drawn in Section~\ref{sec:Clmt&CntrlEng}, are both applicable and extendable to more complex, coupled climate models that represent the Earth system as a whole. The aim of this section is not to reduce climate to a single component, nor to equate it with abstract notions of weather prediction, but rather to develop conceptual and illustrative examples that generalize beyond the specific models presented here.}

Section~\ref{sec:GeoEngCntProb} focuses on geoengineering, which has received increased attention in recent years as a proposed climate intervention strategy. Several core challenges in geoengineering design and---such as tuning intervention strength, timing, and spatial deployment---are, in fact, control problems when viewed through a control engineering lens.

\section{Geoengineering Strategies as Control Problems}~\label{sec:GeoEngCntProb}
% \textcolor{blue}{ive rewritten this section and i need your help with (a) re-reading everything and fixing any errors/typos; (b) cleaning the table of any inaccuracy; (c) fixing the last 3 paragraphs of subsection A where i have the red comment. the rest should be fine but please read carefully}
In this section we examine geoengineering\footnote{We emphasize here that this topic of geoengineering (solutions devised to improve climate dynamics or counteract climate change) is not free of controversy as many studies have discussed their potential adverse and unintended effects. We briefly discuss these effects in this section, although the objective herein is to summarize the literature on control methods for geoengineering.} as a class of proposed climate interventions and frame specific design tasks as control problems. Most of the geoengineering studies utilize the vast array of climate models derived in the earlier sections and beyond. To provide value for the reader, we focus on a specific set of representative problems drawn from the literature: stratospheric aerosol injection (SAI) designs, control of spatially patterned solar forcing, and multi-objective SAI under actuator and feasibility limits. Note that, SAI belongs to the broader class of solar radiation management (SRM) strategies that are counted as geoengineering interventions---and SRM is just one geoengineering strategy (SAI~$\subset$~SRM~$\subset$~Climate~Intervention\,/\,Geoengineering). 

At a high level, two complementary formulations are common in the literature:
\textit{(i)} \emph{open-loop optimization} where forcing patterns are pre-computed to minimize deviations from a desired reference climate state such as global mean temperature; 
\textit{(ii)} \emph{feedback control} where interventions are adjusted dynamically based on observed climate variables to correct deviations as they arise. SAI provides an illustrative example of both: the latitudinal distribution or injection rate of aerosols acts as the actuator; the climate response (e.g., global or regional temperature) serves as the measured output; and the objective is to minimize the deviation from a target climate state. The following section uses this structure to interpret several geoengineering designs; the reproduced literature list is illustrative, not exhaustive. The goal is to make the control structure explicit and connect it to the broader discussion in Section~\ref{sec:Clmt&CntrlEng}.

\subsection{Control- and Optimization-Based Geoengineering Formulations}

Early studies asked a simple question: with a small number of latitude \textit{knobs}, how close can we get to a reference climate condition or state? In this context, the control is a latitude‐dependent distribution of imposed shortwave reduction (for idealized SRM) or the latitude pattern of stratospheric aerosol optical depth (AOD)\footnote{AOD is a key metric in climate research. It quantifies how much sunlight aerosols scatter and absorb in the atmosphere. This indicates the extent to which aerosols reduce solar radiation transmission, thereby impacting Earth's energy balance. Higher AOD values correspond to more aerosols and greater radiative impact.} in SAI regimes. A small set of basis patterns (for example: uniform, hemispheric dipole, and a high-vs-low latitude tilt) provides knobs that can be mixed to place cooling where it is most useful. The knobs are virtually the controls. We briefly summarize some of the SRM/SAI-focused studies that use system optimization and feedback control in what follows. The studies are summarized in Tab.~\ref{tab:geo}.

\begin{table*}[t]
\centering
\scriptsize
\renewcommand{\arraystretch}{1.25}
\setlength{\tabcolsep}{3pt}
\caption{Comparison of Some SRM/SAI Geoengineering Studies: Problem, Controls, Objectives, Dynamics, and Outcomes}
\label{tab:geo}
\begin{tabular}{p{0.5cm} p{3.1cm} p{2.8cm} p{3.2cm} p{3.0cm} p{3.5cm}}
\hline
\textbf{Paper} & \textbf{Problem Studied} & \textbf{Control Variables (knobs)} & \textbf{Objective Functions} & \textbf{Considered Dynamics} & \textbf{Outcomes} \\
\hline
\cite{brody2025using} &
Plan SAI strategy across seven latitude/season shares to better meet multiple climate goals using optimization. &
Nonnegative shares over seven SO$_2$ injection policies (EQ, $\pm15^\circ$, $\pm30^\circ$ year–round; $\pm60^\circ$ spring–only), constrained to sum to the total planned cooling. &
Convex nonnegative least–squares to reduce departures in $T_0$, $P_0$, $T_1$–$T_2$, ITCZ latitude, and September Arctic sea–ice (SSI). &
Small-signal, quasi-static linear surrogate calibrated from CESM2/WACCM6-MA ensembles; linearity validated near $\sim\!1^\circ$C cooling; notes microphysics/circulation limits at larger amplitudes. &
Maps feasible target sets (e.g., $T_0$ with $P_0$ co-manageable); gains at $\sim\!1^\circ$C often comparable to internal variability; temperature-focused mixes emphasize EQ and/or high latitudes. \\
\cite{kravitz2016geoengineering} &
Shift from evaluation to design: choose SRM patterns to meet multiple climate goals. &
Latitude-dependent insolation forcing pattern $F(\phi)$. &
Multi-objective design: minimize deviations in $\{T_{\text{Arctic}}, P_{\text{tropics}}, T_{\text{global}}\}$. &
Community Earth system coupled atmosphere–ocean (AOGCM/ESM class); linearized local response for design. &
Demonstrates simultaneous achievement of regional targets; introduces four design principles (objectives, control inputs, uncertainty, validation). \\

\cite{ban2010geoengineering} &
 SRM via optimization: how and where much solar reduction to offset CO$_2$ warming while limiting side effects. &
Fractional solar insolation reduction through variable $u$. &
Minimize deviation from target climate: $J=\int (T-T_{\text{ref}})^2\,dt + \lambda C(u)$. &
Full equilibrium simulations from the nonlinear global climate model NCAR CAM3.1. &
Shows near-linearity of climate response to moderate SRM forcing; highlights limitations (nonlinearities, ocean acidification). \\

\cite{macmartin2017climate} &
Design injection latitude strategies to tailor temperature patterns. &
SO$_2$ injection rates at multiple latitudes (e.g., 30°S, 15°S, 15°N, 30°N). &
Minimize mean-square error in temperature: $\sum (T_{ij}(u)-T_{ij}^{\text{target}})^2$. &
Approx. linear superposition of responses; fully coupled whole-atmosphere chemistry climate dynamics CESM1(WACCM). &
Multi-latitude control improves temperature compensation of CO$_2$ forcing by $\sim$30\%; adds 3 effective control DoFs. \\

\cite{macmartin2014dynamics} &
Model SRM under feedback (closed-loop) control with delays. &
SRM radiative forcing $u(t)$ is a PI control accounting for delays. &
Maintain $T(t)\!\approx\!T_{\text{target}}$ while avoiding oscillations; stability analysis. &
Low-order box-diffusion model + time-delay feedback. &
Closed-loop feedback can amplify variability and oscillations for high gains; identifies human–climate as coupled dynamical system. \\
\cite{bonetti2018multiple} &
Develop robust/adaptive SRM control for latitudinal zones. &
SRM forcing in three latitude boxes: $u_N, u_C, u_S$. &
Minimize weighted temperature error with adaptive/PI control under uncertainty. &
Three-box energy-balance model with heat transport and control feedback. &
Adaptive controller mitigates hemispheric temperature asymmetry; identifies controllability limits; neglects full climate feedback.  \\
% \cite{rasch2008overview} &
% Survey of SAI impacts on temperature, precipitation, ozone. &
% Aerosol injection rate, altitude, latitude. &
% Qualitative: reduce warming while minimizing side effects. &
% Chemistry–climate and GCM simulations of sulfate aerosol–radiation coupling. &
% Confirms SAI potential to offset global mean warming but regional disparities and ozone impacts expected; major uncertainties noted. \\
\hline
\end{tabular}
\end{table*}

The study \cite{ban2010geoengineering} formulates SRM as an open-loop optimization problem over a geographical area, rather than a simple one-variable tuning which was typical of earlier SRM studies. Specifically, this study presents an optimization architecture to determine where and how much solar radiation to reflect and control. The study considers idealized latitude-based solar dimming \textit{patterns} as control knobs to optimize climate outcomes.
% , followed by \textit{(ii)} designing interventions that minimize regional climate side effects while attempting to ensure desired global outcomes. 
The formulated mathematical objective focuses on minimizing the global mean temperature anomalies or regional deviations from a target climate state; the associated control variables are the aforementioned knobs. A dynamic climate model is used---as a proxy for a plant in our language---to inform the constraints and the objectives of this optimization through sensitivity matrices. The optimization then yields \textit{spatially non-uniform} SRM interventions that \textit{could} reduce undesirable regional effects compared to \textit{uniform} global shading, thereby showing the value of the optimization.  In contrast, the study \cite{macmartin2017climate} advances this line of work by considering physical injection locations and rates while considering a more realistic stratospheric transport and radiative effects in a climate model.  In particular, this paper showcases how injecting spatially-different amounts of $\text{SO}_2$, the spatial pattern of
AOD can be partially controlled.  Compared to earlier studies, this work demonstrates how to inject specific amounts of aerosols different latitudes, and how SAIs can serve as actual control inputs that approximate the optimized patterns/knobs in earlier work. 

Both studies~\cite{ban2010geoengineering,macmartin2017climate} and their variants in the literature produce open-loop optimization that should not be conflated with truly a closed-loop one or a model predictive controller (MPC) that explicitly accounts for high-level system dynamics in the constraints. This is understandable as integrating complex climate dynamics in a controller design would require expensive simulations-optimization architecture, whereas open-loop architectures build explicit (yet not real-time or \textit{predictive}) relationships between the controls and their effects on climate states. The following discussion shows how simple models can be integrated to perform feedback control.

% applies the same idea with explicit SAI physics in a fully coupled whole-atmosphere chemistry climate model CESM1 (WACCM). Instead of prescribing abstract patterns, sulfur dioxide SO\(_2\) is injected at selected latitudes (30\(^\circ\)S, 15\(^\circ\)S, 15\(^\circ\)N, 30\(^\circ\)N). Single-latitude experiments provide empirical \textit{gain} columns linking injections to AOD and then to temperature moments. Linear combinations of these injections synthesize three useful AOD modes: a nearly uniform mode that controls \(T_0\) (global mean), an interhemispheric difference that controls \(T_1\), and a high- vs low-latitude tilt that controls \(T_2\). Planning multi-site injections becomes a constrained least-squares fit with nonnegativity. The study also shows where additivity weakens at higher loads because of aerosol microphysics and transport, \textit{which motivates closing the loop rather than relying only on a one-time open-loop plan.}

% \subsection{Feedback Operation of SRM/SAI}
Although many studies focus on open-loop control due to its simplicity, some studies in geoengineering research investigates closed-loop control via simple control techniques and linear(ized) climate models. The study~\cite{macmartin2014dynamics} presents the simple mathematical dynamics, through transfer functions, of implementing solar geoengineering through proportional integral (PI) control. A reduced box-diffusion model is used to simulate the climate system's response to SRM policies, resulting in an explicit relationship between the policies and radiative forces $F(t)$ and the mean temperatures $T(t)$ in frequency domain $s$ via transfer function $G(s)=T(s)/F(s)$. In particular, this pioneering work at the intersection of linear control and climate dynamics treats the radiative forces $F(t)$ as the control input to be adjusted via the pre-computed PI gains (this is still high-level compared to other SRM/SAI strategies that design injection schedules). In contrast, the study \cite{kravitz2016geoengineering} poses a direct tracking problem: choose a small set of large-scale diagnostics---such as $T_0$ (global-mean surface temperature), $T_1$ and $T_2$ (interhemispheric and equator-to-pole temperature gradients), and ITCZ latitude (the mean position of the tropical rain belt)---and estimate channel gains and time lags from targeted model runs, and update actuator amplitudes each year with PI rules tuned for slow bandwidth and adequate phase margin. The climate model is comprehensive atmosphere-ocean general circulation model (AOGCM/ESM class), but the control layer remains low-order by design. The logic is standard: feedback corrects model error, slow drift, and modest nonlinearitieswhile avoiding responses to short-term internal variability. The same structure carries to other SRM actuators and is summarized in \cite{soldatenkoDesigningPrioriScenarios2025}. 

% \comment{it was not clear in these two papers what the control variables are and what are we trying to track down or regulate.}

% \textcolor{red}{this is a chapter in a book named "Geoengineering and Climate Change: Methods, Risks, and Governance," should we acknowledge that?}

% Feedback makes operation adaptive to changing conditions (e.g., evolving CO$_2$ or aerosol efficiency), and the few-mode basis keeps MIMO interactions simple enough for transparent oversight: which signals are tracked, which actuators move, and how fast loops run are all explicit.

% \subsection{Robust and Adaptive Control in Low-Order Models}
Low-order climate models provide a reasonable testbed for robustness and adaptation. 
% \textcolor{red}{explain. do not reproduce text that is not defined properly} 
The work in \cite{bonetti2018multiple} uses a three-box (and extended five-box) energy-balance model with multiple SRM actuators that affect regional radiative forcing. The goal is to track chosen targets while keeping meridional gradients in check under parameter uncertainty and disturbances. Controllers range from carefully tuned PI to model reference adaptive control. The study makes standard robust control points in a climate context: low-order controllers are enough if gains are conservative; adaptation helps when the system drifts from the models; and decoupling by design (aligning actuators with dominant modes) prevents cascading effects. The box model clarifies controllability and observability trade-offs that are harder to see cleanly in full GCMs. The study produces interesting ideas that have unfortunately not become mainstream in climate science, so further development vis-\'{a}-vis the intersection of climate science and robust control could lead to more fruitful contributions in this area. 

% \comment{i could not figure out yet again: what the model was, what the control objective is, what the optimization variable or control input is. the discussion here is so general. also, this seems like feedback control so perhaps we should keep it in the above section (please remove the subsection label)...also, remarkably, this paper has a discussion on observability and controllability of this simple 3-box climate model, so maybe we should ACK it later in the paper as I didn't see anything similar. of course the issue is that the paper seems not to be cited enough, which is interesting to observe. i dont know what to make of it. }

% \subsection{Multi-Objective, Constrained Planning at System Scale}
Recent work casts SAI strategy design as an explicit planning problem \cite{brody2025using}. The control variables are nonnegative shares over seven SAI injection policies, year-round at the Equator, $\pm15^\circ$, $\pm30^\circ$, and spring-only at $\pm60^\circ$. Those shares sum to the total planned cooling. Using ensembles from the fully coupled Earth system CESM2/WACCM6-MA model, the study identifies a small-signal linear surrogate that maps these shares to a set of large-scale diagnostics: $T_0$, $P_0$ (global-mean precipitation), $T_1$, $T_2$, ITCZ latitude, and September Arctic sea-ice extent (SSI). Strategy selection is then posed as a convex least-squares fit that chooses the shares to reduce departures from chosen targets. The surrogate is validated for about $1^\circ$C of cooling, where mixtures behave approximately as weighted sums of single-policy responses, while noting that aerosol microphysics and circulation limit linearity at larger amplitudes. The simulations indicate clear degrees-of-freedom limits, some goals such as holding $T_0$ with $P_0$ can be co-managed, whereas jointly satisfying $T_0$, $T_1$, $T_2$, ITCZ, and SSI leaves residual pattern errors, and gains at $\sim\!1^\circ$C are often comparable to internal variability, with temperature, focused mixes tending toward equatorial and/or high-latitude injections. In control terms this is multi-objective target matching under input constraints, calibrated from CESM2/WACCM6-MA; reliable operation would still benefit from feedback to handle drift and uncertainty.

\subsection{Second-Order Effects of Geoengineering}
Geoengineering sits inside a coupled Earth-human system. Even when large-scale metrics (e.g., global-mean temperature) are well managed, second-order effects can emerge in chemistry, hydrology, ecosystems, and governance. In this section, we report both sides as they appear in the literature, where there is broad agreement and where doubts and uncertainty remain, and note what they imply for design and operation whenever is possible.\footnote{We note here, at the time of the writing and to the best of the authors' knowledge, SRM and SAI strategies have \textit{not} been implemented at large-scale although some small-scale, controlled experiments have taken place with little generalizations. }

% \paragraph{Climate Responses under SRM}
There is broad model and observational support that stratospheric SAI would cool the planet and could slow ice loss and sea-level rise---that is the core benefit motivating SRM proposals \cite{robock2009benefits,rasch2008overview}. At the same time, responses are not uniform: patterns of precipitation, monsoons, and circulation can shift even if the global mean is held fixed. Ozone chemistry on sulfate surfaces and aerosol microphysics (size control) add additional constraints and uncertainties \cite{rasch2008overview}. Recent design studies suggest that using multiple injection latitudes can supply a few independent degrees of freedom to better manage meridional temperature gradients alongside the global mean, improving fit relative to equatorial-only injection \cite{macmartin2017climate}. These are promising levers, but they do not eliminate all tradeoffs. The recent study~\cite{haywood2022assessing} goes into great detail describing the unintended second-order effects of SAI strategies by showing that using small amount of sunlight-absorbing aerosols can make solar geoengineering less effective and more risky with various unintended effects such as heating the stratosphere, shifting rainfall, and disturbing ozone and wind patterns.

% \paragraph{Ocean Chemistry and Unresolved Drivers}
% SRM does not address ocean acidification because atmospheric \(\mathrm{CO_2}\) remains high even if temperature is stabilized; this point is common ground across reviews \cite{robock2009benefits,rasch2008overview}. Thus, \emph{SRM is at most a complement to emissions cuts and carbon removal, not a substitute.}

% \paragraph{Operational Hazards and Longer-Term Risks}
Another major concern is \textit{termination shock}: if an SRM program is abruptly stopped while $\mathrm{CO}_2$ remains elevated, global temperatures and precipitation patterns would rebound rapidly, potentially faster than in a no-SRM scenario; multi-model studies confirm this swift climate response upon sudden suspension \cite{parker2018risk}. This argues for conservative bandwidth (slow control) and robust contingency planning in any operational design. Other geoengineering strategies such as carbon dioxide removal (CDR), which targets the root cause (atmospheric \(\mathrm{CO_2}\)), have also received scrutiny. At large scales, CDR can compete with land and water uses or perturb biogeochemistry. These risks, benefits, and feasibility limits are treatment-specific and not uniform; policy assessments emphasize careful, method-by-method evaluation and governance \cite{RoyalSoc2009,NAS2021}.

% \paragraph{Governance, Ethics, and Research Norms}
% Major reports argue simultaneously for caution and for structured, transparent research governance: develop capability to answer policy-relevant questions \emph{and} ensure oversight, public engagement, and international coordination before any outdoor testing or deployment \cite{NAS2021,RoyalSoc2009}. Others warn about moral hazard (reduced mitigation effort if SRM appears to "work") and about geopolitical inequities if regional outcomes differ; proposals to break the governance deadlock focus on clear rules for small-scale research and international norms before scaling \cite{parson2013end}. For control design, this translates to explicit operating envelopes, clear observables, human-in-the-loop guardrails, and conservative gain and phase margins that prioritize stability and reversibility.

% \vspace{0.2cm}
% \noindent \textbf{Takeaways for Control Community Readers.} 
\subsection{Constrained Geoengineering and Control Problems}

Across the literature, three patterns recur: \textit{(i)} a small, well-chosen actuator set can manage a small set of large-scale targets; \textit{(ii)} feedback is essential for reliability under model error, time-delays and slow drift; \textit{(iii)} some outputs (chemistry, ecosystems, equity) are only indirectly controllable and therefore must be treated as monitored constraints with explicit guardrails. Those points motivate conservative loop design, robust identification, and multi-objective formulations that surface tradeoffs rather than hide them. Furthermore, very few studies focused on modeling large-scale climate dynamics through signals of dynamic models that would then be utilized in an MPC framework to perform constrained and responsible geoengineering. In particular, a pertinent study could find ways to generate lower order models of coupled Earth-human systems, that could then be used in a predictive way with sensor observations to perform shorter-term SAI or SRM regulation under constraints from other geological systems. As in other application areas of MPC, the model accuracy is \textit{not} as critical when coupled with a substantial amount of observations. This entails that such reduced order model for complex climate models can be especially useful in constrained geoengineering formulations.

% \newpage
\section{Analogies Between Control Engineering and Climate Science Problems}~\label{sec:Clmt&CntrlEng}

In this section, we draw and observe structured connections between challenges (and research problems) in climate science and foundational ideas in control engineering. The goal is to suggest that some known climate problems can be potentially cast as control engineering ones. By using the conceptual lexicon of control engineering (states, inputs, outputs, observability, controllability, stability feedback, model order reduction, system identification, and uncertainty propagation), we build a shared framework for interdisciplinary thinking. We argue here that control systems researchers can benefit from the produced analogies in this section by hopefully applying some control engineering algorithms for large-scale climate models. We admit that this might be very ambitious, but the point of this tutorial was to initiate a conversation and hopefully motivate some folks to delve into climate science and seek collaborators from this crucial field of science. 

The previous sections have laid the groundwork for this discussion by emphasizing the dynamic and time-evolving nature of climate models. We have introduced a simple EBM, a first-order nonlinear dynamical system representing the evolution of global mean temperature. This model serves as a starting point to illustrate how climate behavior can be framed in terms of input-output relationships, feedback mechanisms, and externally driven disturbances. We have introduced hypothetical control inputs, such as geoengineering strategies to increase planetary albedo, and external disturbances, i.e., radiative forcing from greenhouse gas emissions, to demonstrate how climate models can be extended into open- or closed-loop control systems. These formulations naturally raise control-oriented questions regarding stabilization, responsiveness, sensitivity, and robustness, even in their simplest form.

To go beyond the conceptual ODE-level intuition, we then have introduced a more complex and physically realistic climate model: a large-scale atmospheric system based on the primitive equations. These nonlinear PDEs govern the evolution of wind, temperature, and density fields under the conservation of momentum, mass, and energy. After spatial and temporal discretization, the resulting system becomes a high-dimensional nonlinear dynamical model with millions of coupled variables. This transformation highlights the rich dynamical structure of climate systems and presents a direct opportunity to apply control-theoretic reasoning. The primitive model exposes the system’s internal feedbacks, distributed inputs and outputs, external forcing, and sensitivity to uncertainty---key features that align closely with control formulations.

Building on these conceptual extensions, this section formalizes and expands the parallels between climate modeling and control system analysis. Control engineering provides a rich set of tools for modeling, analyzing, and designing dynamical systems under uncertainty, partial observability, and feedback. These are exactly the features that characterize large-scale climate systems. While traditional climate models are built from first principles, they often face challenges that align closely with those addressed by control theory: high dimensionality, sparse data, parameter uncertainty, delayed responses, and the need for optimal intervention under constraints.

Rather than treating climate models as black boxes or fixed simulation tools, the control perspective views them as dynamical systems whose behavior can be shaped, predicted, and interpreted through formal system-theoretic properties. In doing so, it becomes possible to translate familiar engineering tasks, such as state estimation, feedback design, reachability, or robustness analysis, into meaningful questions within climate science. We do not imply here that control theory will change the discourse of climate science; we argue that control engineers and researchers can perhaps learn from climate problems and perhaps be more involved in the most timely research topic of our time. 

This section presents a structured exploration of these connections, with the aim of fostering cross-disciplinary insight and enabling the use of control engineering frameworks to deepen our understanding of climate dynamics and intervention planning. That is, the remainder of this section highlights several core concepts in control theory and their relevance to climate modeling. Tab. \ref{tab:ParalCont&Clm} summarizes these parallels.

\begin{table*}[t!]
	\centering
    \caption{Parallels Between Climate Science and Control Engineering, Emphasizing Shared Challenges and Transferable Tools Across Disciplines.}~\label{tab:ParalCont&Clm}
\begin{tikzpicture}
	
	% Define colors
	\definecolor{pastelblue}{RGB}{70,130,180}
	\definecolor{pastelgreen}{RGB}{0,40,92}
	% \definecolor{pastelgreen}{RGB}{0,128,128}
	\definecolor{slategray}{RGB}{70, 80, 90}
	
	% Column widths and height
	\def\slim{3.5cm}
	\def\wide{5.4cm}
	\def\height{16.6cm}
	\def\headerheight{0.6cm}
	\def\gap{0.05cm}
	
	% Draw column boxes
	% Climate Science - Problem
	\draw[fill=pastelblue!50, draw=pastelblue!50] (0,0) rectangle ++(\slim,\height);
	% Climate Science - Definition
	\draw[fill=pastelblue!20, draw=pastelblue!20] (\slim+\gap,0) rectangle ++(\wide,\height);
	
	% Control Engineering - Problem
	\draw[fill=pastelgreen!30, draw=pastelgreen!30] (\slim+\wide+2*\gap,0) rectangle ++(\slim,\height);
	% Control Engineering - Definition
	\draw[fill=pastelgreen!15, draw=pastelgreen!15] (\slim+\wide+\slim+3*\gap,0) rectangle ++(\wide,\height);
	
	% Group headers (bottom line only, color matched, centered text)
	% Define colors

% Group header background rectangles
% Climate Science header background
\draw[fill=pastelblue!90, draw=none]
  (0, \height) rectangle (\slim+\gap+\wide, \height + \headerheight);

% Control Engineering header background
\draw[fill=pastelgreen!80, draw=none]
  (\slim+\wide+2*\gap, \height) rectangle (\slim+\wide+2*\gap+\slim+\gap+\wide, \height + \headerheight);

% Group header text (on top of dark background)
\node[align=center, font=\bfseries, text=pastelblue!5] 
  at ({(\slim+\gap+\wide)/2}, \height + \headerheight/2) {Climate Science};

\node[align=center, font=\bfseries, text=pastelgreen!5] 
  at ({\slim+\wide+2*\gap + (\slim+\gap+\wide)/2}, \height + \headerheight/2) {Control Engineering};

    % Bidirectional arrow between Climate Science and Control Engineering
\draw[<->, line width=2pt, color=white] 
  (\slim+\gap+\wide-0.4cm, \height + \headerheight/2) 
  -- (\slim+\wide+2*\gap-0.5*\gap+0.4cm, \height + \headerheight/2);

	% Sub-headers (pushed clearly into the top of each rectangle)
	\node[align=center, font=\bfseries,  text=pastelblue!5] 
	at (\slim/2, \height - 8) {Problem};
	
	\node[align=center, font=\bfseries, text=pastelblue!80] 
	at (\slim+\gap+\wide/2, \height - 8) {Definition};
	
	\node[align=center, font=\bfseries, , text=pastelgreen!5] 
	at (\slim+\wide+2*\gap+\slim/2, \height - 8) {Problem};
	
	\node[align=center, font=\bfseries, text=pastelgreen!80] 
	at (\slim+\wide+2*\gap+\slim+\gap+\wide/2, \height - 8) {Definition};
	
	% Line under Climate Science subheaders
	\draw[line width=1.0pt, color=pastelblue] 
	(0, \height - 16.0) -- (\slim+\gap+\wide, \height - 16.0);
	
	% Line under Control Engineering subheaders
	\draw[line width=1pt, color=pastelgreen] 
	(\slim+\wide+2*\gap, \height - 16.0) -- (\slim+\wide+2*\gap+\slim+\gap+\wide, \height - 16.0);
	
	\node[align=left, font=\bfseries, anchor=north west, text width=\slim, text=slategray]
	at (0, \height - 18)
	{Missing Data \& Data Assimilation};
	
	\node[align=left, text width=\wide, anchor=north west, text=slategray] 
	at (\slim, \height - 18) 
        {Data assimilation merges model forecasts with sparse or noisy observations to estimate the full system state. This enables better initialization of forecasts, especially when key variables are unobserved or partially measured.};
	
	\node[align=left, font=\bfseries, anchor=north west, text width=\slim, text=slategray] 
	at (\slim+\wide+2*\gap, \height - 18) 
	{Observability \& State Estimation};

	\node[align=left, text width=\wide-0.6, anchor=north west, text=slategray] 
	at (\slim+\wide+2*\gap+\slim, \height - 18) 
	{Control systems use observer design and estimation filters (e.g., Kalman filters) to reconstruct internal states from outputs. These techniques parallel climate data assimilation, offering formal tools for state reconstruction and uncertainty handling.};
	
		% Line under Climate Science
	\draw[line width=0.8pt, color=pastelblue] 
	(0, \height - 75.0) -- (\slim+\gap+\wide, \height - 75.0);
	
	% Line under Control Engineering 
	\draw[line width=0.8pt, color=pastelgreen] 
	(\slim+\wide+2*\gap, \height - 75.0) -- (\slim+\wide+2*\gap+\slim+\gap+\wide, \height - 75.0);

    \node[align=left, font=\bfseries, text width=\slim, anchor=north west, text=slategray]
    at (0, \height - 77) % adjust Y position
    {Climate Intervention Planning};

    \node[align=left, text width=\wide, anchor=north west, text=slategray]
    at (\slim, \height - 77)
    {Climate interventions aim to alter system trajectories, e.g., reducing global temperature or emissions. The effectiveness of these actions depends on how responsive the climate system is to applied forcing, much like input response in control systems.};

    \node[align=left, font=\bfseries, text width=\slim, anchor=north west, text=slategray]
    at (\slim+\wide+2*\gap, \height - 77)
    {Controllability of \\ Dynamical Systems };

    \node[align=left, text width=\wide-0.6, anchor=north west, text=slategray]
    at (\slim+\wide+2*\gap+\slim, \height - 77)
    {Controllability characterizes whether and how a system can be steered to desired states using inputs. Tools such as input Gramians, metrics, and actuator design offer formal ways to plan effective interventions: concepts increasingly relevant in climate control strategies.};

    	% Line under Climate Science
	\draw[line width=0.8pt, color=pastelblue] 
	(0, \height - 135.0) -- (\slim+\gap+\wide, \height - 135.0);
	
	% Line under Control Engineering 
	\draw[line width=0.8pt, color=pastelgreen] 
	(\slim+\wide+2*\gap, \height - 135.0) -- (\slim+\wide+2*\gap+\slim+\gap+\wide, \height - 135.0);

    \node[align=left, font=\bfseries, text width=\slim, anchor=north west, text=slategray]
    at (0, \height - 137) % adjust Y position
    {Policy Levers \& \\ Climate Goals};

    \node[align=left, text width=\wide, anchor=north west, text=slategray]
    at (\slim, \height - 137)
    {Climate actions like carbon pricing, energy transition, or geoengineering serve as levers to steer the system toward goals (e.g., net-zero emissions, temperature caps). Planning these involves choosing when and how intensely to act, under uncertainty and constraints.};

    \node[align=left, font=\bfseries, text width=\slim, anchor=north west, text=slategray]
    at (\slim+\wide+2*\gap, \height - 137)
    {Optimal Control \& \\ Trajectory Design};

    \node[align=left, text width=\wide-0.6, anchor=north west, text=slategray]
    at (\slim+\wide+2*\gap+\slim, \height - 137)
    {Optimal control computes input trajectories to meet system objectives while respecting dynamics and constraints. This maps to climate policy design, where actions are optimized over time to achieve climate targets under physical and socioeconomic limits.};

% Line under Climate Science
	\draw[line width=0.8pt, color=pastelblue] 
	(0, \height - 195.0) -- (\slim+\gap+\wide, \height - 195.0);
	
	% Line under Control Engineering 
	\draw[line width=0.8pt, color=pastelgreen] 
	(\slim+\wide+2*\gap, \height - 195.0) -- (\slim+\wide+2*\gap+\slim+\gap+\wide, \height - 195.0);

    \node[align=left, font=\bfseries, text width=\slim, anchor=north west, text=slategray]
    at (0, \height - 197) % adjust Y position
    {Forecasting \& \\ Climate Projections};

    \node[align=left, text width=\wide, anchor=north west, text=slategray]
    at (\slim, \height - 197)
    {Forecasts estimate short-term evolution (e.g., seasonal rain patterns), while projections map long-term plausible futures under various scenarios. Both explore how the system might evolve from today’s conditions under different influences.};

    \node[align=left, font=\bfseries, text width=\slim, anchor=north west, text=slategray]
    at (\slim+\wide+2*\gap, \height - 197)
    {Reachability Analysis};

    \node[align=left, text width=\wide-0.6, anchor=north west, text=slategray]
    at (\slim+\wide+2*\gap+\slim, \height - 197)
    {Reachability defines the set of states a system can reach given initial conditions and inputs. Like climate projections, it maps possible futures in forms of envelopes rather than certainties, helping guide planning under uncertainty.};

    % Line under Climate Science
	\draw[line width=0.8pt, color=pastelblue] 
	(0, \height - 254.0) -- (\slim+\gap+\wide, \height - 254.0);
	
	% Line under Control Engineering 
	\draw[line width=0.8pt, color=pastelgreen] 
	(\slim+\wide+2*\gap, \height - 254.0) -- (\slim+\wide+2*\gap+\slim+\gap+\wide, \height - 254.0);

            \node[align=left, font=\bfseries, text width=\slim, anchor=north west, text=slategray]
    at (0, \height - 256) % adjust Y position
    {Uncertainty Quantification \\ \& Climate Risk};

    \node[align=left, text width=\wide, anchor=north west, text=slategray]
    at (\slim, \height - 256)
    {Uncertainty is central to climate modeling, arising from internal variability, incomplete data, and model limitations. Climate scientists use ensemble projections, confidence intervals, and scenario spreads to assess how robustly future states can be anticipated.};

    \node[align=left, font=\bfseries, text width=\slim, anchor=north west, text=slategray]
    at (\slim+\wide+2*\gap, \height - 256)
    {Robustness \& \\ Uncertainty Handling};

    \node[align=left, text width=\wide-0.6, anchor=north west, text=slategray]
    at (\slim+\wide+2*\gap+\slim, \height - 256)
    {In control engineering, robustness ensures that systems perform reliably under uncertainty in dynamics or disturbances. Techniques like $H_\infty$ control, tube MPC, and bounded-error design echo ensemble-based and worst-case strategies in climate risk management.};

     % Line under Climate Science
	\draw[line width=0.8pt, color=pastelblue] 
	(0, \height - 315.0) -- (\slim+\gap+\wide, \height - 315.0);
	
	% Line under Control Engineering 
	\draw[line width=0.8pt, color=pastelgreen] 
	(\slim+\wide+2*\gap, \height - 315.0) -- (\slim+\wide+2*\gap+\slim+\gap+\wide, \height - 315.0);

            % New line under Uncertainty
    \node[align=left, font=\bfseries, text width=\slim, anchor=north west, text=slategray]
    at (0, \height - 317) % adjust Y position
    {Tipping Points \\ and System Resilience};

    \node[align=left, text width=\wide, anchor=north west, text=slategray]
    at (\slim, \height - 317)
    {Climate systems may experience abrupt, irreversible shifts if critical thresholds are crossed. Rather than returning to a pre-disturbance state, resilience analysis focuses on maintaining the system within safe functional regimes and avoiding destabilizing feedback loops.};

    \node[align=left, font=\bfseries, text width=\slim, anchor=north west, text=slategray]
    at (\slim+\wide+2*\gap, \height - 317)
    {Stability and \\ Dynamical Behavior};

    \node[align=left, text width=\wide-0.6, anchor=north west, text=slategray]
    at (\slim+\wide+2*\gap+\slim, \height - 317)
    {In control theory, stability ensures that a system returns to equilibrium after perturbations. Techniques such as Lyapunov analysis, eigenvalue tests, and bifurcation theory characterize how systems behave near equilibria, concepts that help frame climate tipping dynamics and resilience thresholds.};

    % Line under this new row
	\draw[line width=0.8pt, color=pastelblue] 
	(0, \height - 385.0) -- (\slim+\gap+\wide, \height - 385.0);
	
	\draw[line width=0.8pt, color=pastelgreen] 
	(\slim+\wide+2*\gap, \height - 385.0) -- (\slim+\wide+2*\gap+\slim+\gap+\wide, \height - 385.0);

        % New line for Model Calibration and System Identification
    \node[align=left, font=\bfseries, text width=\slim, anchor=north west, text=slategray]
    at (0, \height - 387) % adjust Y position
    {Model Calibration \\ \& Learning};

    \node[align=left, text width=\wide, anchor=north west, text=slategray]
    at (\slim, \height - 387)
    {Model calibration tunes parameters of dynamical models (e.g., cloud, land, ocean processes) to better fit observations. Learning techniques like ML and ensemble methods have been used to infer subgrid processes, design surrogate models, and replace or enhance traditional physical parameterizations, while aiming to maintain physical consistency.};

    \node[align=left, font=\bfseries, text width=\slim, anchor=north west, text=slategray]
    at (\slim+\wide+2*\gap, \height - 387)
    {System Identification \\ \& Model Discovery};

    \node[align=left, text width=\wide-0.6, anchor=north west, text=slategray]
    at (\slim+\wide+2*\gap+\slim, \height - 387)
    {System identification builds mathematical models from input-output data. Techniques range from ARX models and subspace identification for linear systems to modern approaches like sparse identification (SINDy) and physics-informed ML for nonlinear systems. Hybrid modeling blends physical knowledge with learned components to improve prediction and control.};

 %    % Line after this new row
 %    \draw[line width=0.8pt, color=pastelblue] 
	% (0, \height - 455.0) -- (\slim+\gap+\wide, \height - 455.0);

 %    \draw[line width=0.8pt, color=pastelgreen] 
	% (\slim+\wide+2*\gap, \height - 455.0) -- (\slim+\wide+2*\gap+\slim+\gap+\wide, \height - 455.0);

\end{tikzpicture}
\end{table*}

% Define colors
	\definecolor{pastelblue}{RGB}{70,130,180}
	\definecolor{pastelgreen}{RGB}{0,128,128}
	\definecolor{pastelpink}{RGB}{255, 182, 193}
	\definecolor{pastelorange}{RGB}{255, 218, 185}
	\definecolor{slategray}{RGB}{70, 80, 90}

\subsection{Missing Data Points \& Data Assimilation vs. Observability \& State Estimation}
% \noindent \textbf{\textcolor{pastelblue}{Missing Data Points, Data Assimilation} \& \textcolor{pastelgreen}{Observability, State Estimation}.} 
We start out the technical details in this section by building parallels between the climate problems of \textit{missing data points} and data assimilation versus the control problems of observability and state estimation.  

The challenge of reconstructing the system parameters and state from partial or missing data arises in both climate science and control engineering, though each field has developed its own terminology and tools. In control engineering, \textit{observability} refers to the ability to infer the internal state of a dynamical system using only output measurements. Building on this, \textit{state estimation} techniques, such as Kalman filtering or observer design, are used to reconstruct those internal states over time, even in the presence of noise or data loss \cite{kirk2004optimal,trentelman2002control,simon2006optimal,roinodottenberg1970observability,zabczyk2020mathematical,besanccon2007nonlinear,gill2000state,makhlouf2023state,kulikov2024state}. In climate science, similar problems are addressed under the umbrella of \textit{data assimilation} \cite{talagrand2010data,kalnay2024earth,asch2016data,latini2003handling,park2013data,nakagawa2015missing,afrifa2020missing,schneider2001analysis}. Here, incomplete or uncertain observational datasets (e.g., from satellites or ground sensors) are merged with model predictions to yield improved estimates of climate variables such as temperature, pressure, or humidity. This is especially critical when direct observations are sparse or missing over space or time. The parallels between these two domains enable a cross-disciplinary perspective. We produce particular examples and discussions.

\subsubsection{Kalman Filtering and Data Assimilation}

The Kalman filter, a foundational tool in control engineering for linear Gaussian systems, has been adapted in climate science as the basis for various data assimilation schemes \cite{houtekamer1998data,koshin2020ensemble,houtekamer2005atmospheric,zheng2010ensemble,pedatella2014ensemble,bach2023multi,wang2023kalman}. Ensemble Kalman Filters and extended Kalman filters are direct generalizations that handle nonlinear climate models and ensemble-based uncertainty quantification. An example of this adaptation is the assimilation system developed by Koshin \textit{et al.}~\cite{koshin2020ensemble}, where a four-dimensional local ensemble transform Kalman filter (4D-LETKF) is used with a high-top general circulation model to reconstruct the atmospheric state up to the lower thermosphere. Observations assimilated include conventional ground-based measurements from available datasets and satellite temperature profiles from the Aura Microwave Limb Sounder (MLS), covering heights from the troposphere through the mesosphere. This has enabled state estimation across scales, from synoptic dynamics to mesospheric processes, critical for capturing phenomena such as sudden stratospheric warmings (SSWs) and inter-hemispheric coupling.

    Another example is the work by Pedatella \textit{et al.}~\cite{pedatella2014ensemble}, who have applied an ensemble adjustment Kalman filter (EAKF) within the whole atmosphere community climate model (WACCM) framework. Their system assimilates satellite observations to improve simulations of large-scale events like the 2009 sudden stratospheric warming. Observations include satellite temperature and ozone data, and the model has spanned the entire atmosphere up to 140 km altitude, demonstrating how Kalman-based assimilation could improve the consistency and predictability of climate variables even under extreme dynamical disturbances.

    The utilization of basic Kalman filtering techniques illustrates that there is an opportunity for control engineers to: \textit{(i)} test out advanced filtering and state estimation algorithms (such as advanced observers, moving horizon estimators, learning-based predictors/estimations for large-scale nonlinear climate models); \textit{(ii)} develop new state estimation algorithms that perform well for climate models with real data. This is because it indeed is known that the basic Kalman filter---and its many variants---suffer in handling unmodeled uncertainty, complex nonlinearities, and unknown initial conditions. 

\subsubsection{Handling Missing Data}    

In control theory, missing or unmeasured outputs are often managed through robust estimation algorithms and strategic sensor placement to maximize system observability. Observers are carefully designed to compensate for the lack of complete output information, ensuring reliable state recovery even with incomplete sensing. In climate science, missing observations can arise due to sensor malfunctions, satellite coverage gaps, or inaccessible regions (e.g., polar areas, deep oceans). These gaps are typically addressed by interpolation, reanalysis products, or machine-learning-based gap-filling techniques. As an example, reanalysis datasets like ERA5 \cite{hersbach2020era5} merge sparse, noisy measurements with physical models to reconstruct missing atmospheric and oceanic fields over decades. The missing data problems in climate science are important because they allow: \textit{(i)} reconstructing or estimating past climate trajectories,  \textit{(ii)} validating physics-based climate models (such as the ones outlined earlier in the paper), and \textit{(iii)} accurately predicting with some certainty future climate. All of these climate problems and outcomes \textit{(i)}--\textit{(iii)} are common in control systems expositions and hence pursuing them in a control lens can be fundamentally new---and potentially useful.

\subsubsection{Observer Design and Climate States Reconstruction}

Observer theory provides formal designs for estimating the hidden state of a system using known inputs and outputs. Analogously, climate models can integrate these principles to estimate unobserved fields (e.g., subsurface ocean temperature) using observable proxies (e.g., sea surface temperature), thereby improving model fidelity and forecast accuracy.

These connections observed in this section underscore a key opportunity: methods from control engineering, grounded in estimation theory, observability analysis, and feedback structure, can be tailored and extended to support climate science needs in data-limited regimes. By bridging these frameworks, we enable more robust and theoretically sound strategies for climate parameters and state reconstruction and prediction.

\subsection{Climate Interventions, Policy Levers, \& Climate Goals vs. Controllability \& Optimal Control} 
% \textbf{\textcolor{pastelblue}{Climate Interventions} \& \textcolor{pastelgreen}{Controllability in Dynamical Systems}.} 
Having thoroughly explained connections between state estimation and data assimilation in climate science, it is only natural to pose questions pertaining to controllability analysis and its corresponding relevance in climate studies. The question of how much influence we can exert on a system, whether to guide it toward a desired outcome or away from a harmful trajectory, is a shared challenge in both climate science and control engineering. While control theory provides formal tools to assess this influence through controllability analysis, climate science deals with it more implicitly, through studies on the effectiveness and consequences of human interventions such as emissions reductions or geoengineering. 

In control engineering, \textit{controllability} refers to the ability to drive the system from any initial state to any final state using admissible inputs. This is analyzed using mathematical tools such as the controllability matrix, Gramian analysis, metrics, and actuator placement strategies. Inputs are well-defined and can be applied selectively in space and time to shift the system in desired directions \cite{trentelman2002control,kirk2004optimal,zabczyk2020mathematical}. In climate science, \textit{interventions} represent policy or engineering efforts to alter the trajectory of climate variables---for example, by reducing CO$_2$ emissions, increasing surface albedo, or deploying stratospheric aerosols \cite{sabel2022fixing,hawken2017drawdown,gates2021avoid}. These are analogous to control inputs, as introduced in Section~\ref{sec:MathModels} and reflected by introducing aerosol injections as control input in the simple EBM model \eqref{eq:EBM_SAI}. However, unlike classical control systems, the input-output pathways in climate models are uncertain, delayed, and often spatially heterogeneous.

Despite these differences, several concepts in control engineering can inform and enhance how we model, evaluate, and plan climate interventions. Framing climate action as a control problem allows us to systematically assess not just what interventions are possible, but how effective and efficient they are likely to be, and where control energy may be wasted. 

\begin{figure*}[t]
    \centering
    \includegraphics[width=0.85\linewidth]{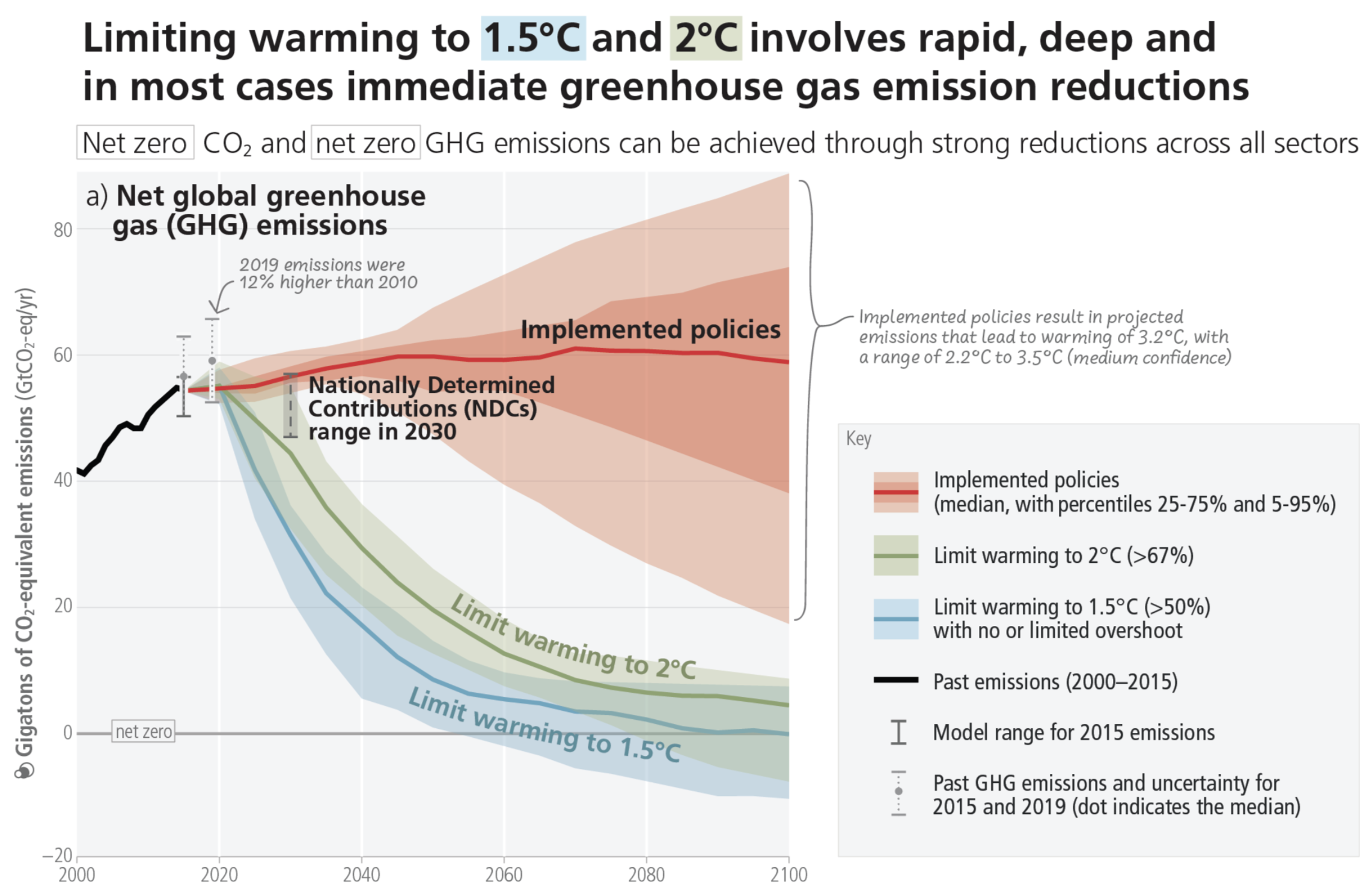}
    \caption{Projected net global GHG emissions under various policy scenarios. The red shading shows expected GHG emissions based on currently implemented policies, while the green and blue pathways represent the more aggressive reductions needed to limit global warming to 2$^\circ$C and 1.5$^\circ$C, respectively. Shaded regions represent uncertainties in model projections (5--95\% confidence intervals). The data illustrate how different levels of intervention can significantly alter emissions trajectories. Figure adapted from the IPCC Sixth Assessment Report (AR6)~\cite{calvinIPCC2023Climate2023}. The figure sheds light into the potential role of control in climate science. Current mitigation trajectories are not designed through formal optimal control formulations where climate dynamics are used directly to compute minimum-control strategies. In contrast, optimal control theory offers a systematic framework for designing interventions: defining explicit objectives, modeling system dynamics, and optimizing trajectories.  By framing climate action through the lens of control, we can move beyond ad-hoc aggregate targets and toward deeply planned, dynamically feasible pathways that respect both physical and socioeconomic constraints. A pertinent investigation is hence whether plots such as this one can be generated via specific control actions, and/or ranges for states can be computed via reachability analysis considering a fixed set of mitigation strategies and control signals.}
    \label{fig:GHGPolicies}
\end{figure*}
As an illustrative example, Fig.~\ref{fig:GHGPolicies} shows another important dimension: the projected pathways for global GHG emissions under different levels of policy action. Current policies, as shown by the red band, are not sufficient to meet the targets associated with limiting warming to 1.5$^\circ$C or 2$^\circ$C. Achieving those targets would require far more aggressive and coordinated emissions reductions, as depicted by the green and blue trajectories. Importantly, these emissions scenarios are typically constructed and simulated by aggregating policy pledges and expert assumptions, rather than by solving \textit{climate dynamics-constrained} optimization or optimal control problems. That is, current mitigation trajectories are not designed through formal optimal control formulations where climate dynamics (governed by ODEs/PDEs such as the models presented in Sections \ref{sec:MathModels} and \ref{sec:MoreComplxModel}) are used directly to compute minimum-control strategies to stabilize the system at target states (e.g., CO$_2$ concentration or surface temperature). In contrast, optimal control theory offers a systematic framework for designing interventions: defining explicit objectives (e.g., limiting temperature rise), modeling system dynamics, applying control inputs (e.g., emissions reductions), and optimizing trajectories over time while satisfying system constraints. This perspective not only enables finding feasible mitigation pathways but also identifies trade-offs, energy costs, and the sensitivity of outcomes to uncertainties. By framing climate action through the lens of dynamic systems and control theory, we can move beyond ad-hoc aggregate targets and toward deeply planned, dynamically feasible pathways that respect both physical and socioeconomic constraints.

% \vspace{0.1cm}
% \noindent \textbf{\textcolor{pastelblue}{Policy Levers and Climate Goals} \& \textcolor{pastelgreen}{Optimal Control and Trajectory Design}.} 

That is, the challenge in climate science involving determining what actions should be taken, when, and at what intensity is a structure that directly parallels optimal control problems in engineering. In both domains, the objective is to influence the evolution of a dynamical system in order to achieve certain outcomes, such as minimizing a cost or reaching a target state.

% \url{https://www.ipcc.ch/report/ar6/syr/downloads/report/IPCC_AR6_SYR_LongerReport.pdf}

% \comment{add two figures from the IPCC report: (A) one of them should go to the first section of the paper (basically a teaser) that shows basically the global average temperatures under different policies (BAA, paris agreement, ....) and how in this tutorial we will talk about a) such predictions/figures are generated via what kind of models; b) how control engineering can solve problems related to predicting better pathways for mitigation...or more theoretical ones that lead to improved outcomes... (B) then another figure here in this section which shows the GHG under different policies, then relate it to the math in previous sections, and the idea that currently these policies and their implications are kinda ad-hoc and not designed via climate-dynamics-constrained ODEs/PDEs, but rather using the aggregate policies of countries that decide to (for example) reduce GHG by 50\%...contrast it with optimal control theory where we design control actions to meet a certain target, which in this case are GHG reductions, maintaining 1.5-2 C temperatures, etc...}

In control engineering, \textit{optimal control} provides a framework to compute input trajectories that minimize an objective function while satisfying system dynamics and constraints. Classical examples include minimizing energy use, time to reach a target, or deviation from a reference trajectory. In climate science, this maps to the design of long-term \textit{policy pathways}, where actions such as emissions reduction, energy transition, or geoengineering are applied over time to meet targets like temperature stabilization or CO$_2$ budgets. These policies act as control inputs, while climate models represent the system dynamics.

Several concepts in optimal control have natural analogues in climate decision-making:

\begin{itemize}
    \item \textit{Objective Functions:} Control systems minimize energy, cost, or error. Climate policy optimizes for carbon budgets, economic cost, risk, or social impact which often use IAMs---refer to Tab. \ref{tab:ClimModelsOvw}.
    
    \item \textit{Constraints:} Just as control problems enforce bounds on states and inputs, climate policies must respect physical, political, and technological limits (e.g., renewable deployment rates, economic costs, fairness constraints).
    
    \item \textit{Control Horizon and Receding Horizon Control:} Optimal control often looks ahead over a fixed time window. Climate policy planning does the same, with tools like MPC increasingly adapted to accommodate feedback, uncertainty, and mid-course correction.
    
    \item \textit{Trade-offs and Multi-Objective Control:} In both domains, competing goals (e.g., economy vs. environment, short-term vs. long-term) are common. Analyses such as Pareto front and weighted cost formulations help frame these trade-offs mathematically.
\end{itemize}

Other problems and analogies are still relevant.

\subsection{Forecasting \& Prediction vs. Reachability Analysis}
% \noindent\textbf{\textcolor{pastelblue}{Forecasting and Prediction} \& \textcolor{pastelgreen}{Reachability Analysis}.}
Forecasting in climate science and reachability analysis in control engineering address a common question: \textit{What future states can the system reach from its current condition?} In climate science, a distinction is made between \textit{forecasts} and \textit{projections} \cite{kovacs2020weather,national1999making,national2006completing,robertson2018sub}. Forecasts are short- to medium-term weather predictions (typically days to seasons) of climate variables like temperature or precipitation, based on current initial conditions and dynamically evolving physical models, much like state prediction in control systems. These forecasts aim to be precise and are evaluated against reality. In contrast, \textit{projections} refer to long-term simulations that estimate how the climate system might evolve under different assumed external forcings or socio-economic pathways, such as varying greenhouse gas emission scenarios. These are often simulated as not direct predictions but conditional \textit{if-then} explorations of plausible futures.

In control engineering, \textit{reachability analysis} plays a parallel role by determining all states that a system can attain over time given an initial condition and a set of admissible inputs \cite{bujorianu2012stochastic,meyer2021interval,kvasnica2015reachability}. It provides bounds on what is possible, rather than what is certain; much like how climate projections outline the range of possible future climates under specific pathways. The ensemble of climate model outputs from scenarios like those in CMIP6 (e.g., Shared Socio-economic Pathways SSP1-2.6, SSP5-8.5) creates a \textit{projection envelope} that is conceptually similar to a reachable set in control theory. Both approaches aim to map out a future space that informs planning and decision-making under uncertainty. Importantly, while traditional climate projections require computationally intensive ensemble simulations across numerous scenarios, reachability analysis offers a promising alternative. By characterizing the entire set of possible future states using system dynamics and input bounds, it allows us to infer whether critical or \textit{dangerous} states, such as crossing a temperature threshold or triggering a tipping point, could be reached. This approach can potentially, in principle, circumvent the need to re-run large-scale models for every new scenario, providing guarantees over the space of uncertainty without exhaustive simulation. To that end, key parallels between the domains include: 
\begin{itemize} 
\item Both rely on system dynamics and are sensitive to initial state uncertainty and input variability. 
\item Reachable sets in control match the role of scenario-based projections in climate modeling. 
\item Techniques used in MPC for safety-constrained forecasting have conceptual overlap with constrained climate mitigation planning. 
\end{itemize}

These parallels can be further pursued and rigorously investigated. 

\subsection{Uncertainty Quantification \& Climate Risk vs. Robustness \& Decision Confidence}
% \noindent \textbf{\textcolor{pastelblue}{Uncertainty Quantification and Climate Risk} \& \textcolor{pastelgreen}{Robustness and Decision Confidence}.}
Uncertainty is a central theme in both climate science and control engineering. In each field, the presence of imperfect knowledge, noisy observations, and model approximations demands tools and frameworks that explicitly account for uncertainty in both analysis and decision-making. Although the terminology and contexts differ---climate scientists speak of ensemble projections and confidence intervals, while control engineers use concepts like robustness and bounded uncertainty---the underlying goal is the same: to understand the limits of what can be known and to make decisions that remain valid across a range of possible conditions.

In control engineering, robustness refers to the ability of a system or controller to maintain acceptable performance despite uncertainties in model parameters, external disturbances, or sensor noise. This includes classical approaches like $\mathcal{H}_\infty$ control, min-max optimization, and tube-based MPC, all of which aim to ensure reliable operation under worst-case or bounded uncertainty~\cite{zhou1998essentials,khalil1996robust,liu2016robust,mackenroth2013robust}. Meanwhile, climate science approaches uncertainty through formal uncertainty quantification (UQ), using statistical, stochastical, and ensemble-based methods to capture the range of possible future outcomes. This includes multi-model ensembles (e.g., CMIP6), Monte Carlo sampling, and Bayesian frameworks to characterize and communicate confidence in projections~\cite{smith2024uncertainty,curry2023climate,palmer2006primacy,franzke2015stochastic,qian2016uncertainty}.

Despite disciplinary differences, the treatment of uncertainty in both fields reveals strong structural parallels:
\begin{itemize} 
\item \textit{Uncertainty Sets and Ensembles:} In robust control, uncertainty is represented by structured sets (e.g., intervals, polytopes, ellipsoids). In climate modeling, ensembles of models or parameter sets perform a similar role, each representing a plausible system realization.

\item \textit{Disturbance Rejection vs. Scenario Spread:} Robust controllers are designed to reject or attenuate unknown inputs. Climate projection ensembles under varying scenarios (e.g., SSP pathways) represent a spread in potential external forcings, highlighting sensitivity to input uncertainty.

\item \textit{Safety Margins and Exceedance Thresholds:} Control systems include margins to ensure safety under worst-case disturbances. Similarly, climate policy aims to avoid crossing critical thresholds (2°C warming), expressed in terms of exceedance probabilities or confidence bounds.

\item \textit{Decision-Making under Uncertainty:} Both fields rely on optimization techniques that explicitly consider uncertainty, such as stochastic MPC in control or robust emissions pathways in climate IAMs.

\end{itemize}

By recognizing uncertainty as a shared, intrinsic feature of dynamical systems, we can better appreciate the synergy between climate modeling and control theory. Tools from both domains can inform one another, not through transplantation but through analogy. 

% \vspace{0.1cm}
% \noindent \textbf{\textcolor{pastelblue}{Tipping Points and System Resilience} & \textcolor{pastelgreen}{Stability and Dynamical Behavior}.} 
\subsection{Tipping Points \& System Resilience vs. Stability \& Dynamical Behavior}
Both climate science and control engineering are concerned with how systems respond to perturbations—but the questions they ask are shaped by the nature of their systems. In control engineering, stability analysis assesses whether a system returns to an equilibrium state after a disturbance. Concepts like asymptotic stability, bounded-input bounded-output stability, and Lyapunov functions are used to evaluate how reliably systems settle back into desired behavior \cite{martynyuk2022stability,khalil2002nonlinear,la1976stability,bacciotti2005liapunov,zabczyk2020mathematical,haddad2008nonlinear}.

In contrast, climate science deals with systems that may not have a single stable equilibrium—or whose equilibria may themselves evolve over time due to external forcing. Rather than expecting the climate system to return automatically to a pre-disturbance state, the focus is on \textit{resilience}: the ability of the system to absorb shocks without undergoing a qualitative, often irreversible shift. \textit{Tipping points}, such as the collapse of the Atlantic Meridional Overturning Circulation or the rapid loss of the Greenland Ice Sheet, represent loss of resilience. After crossing such thresholds, returning to the original state may become physically impossible or require disproportionately large interventions \cite{wunderling2024climate,lenton2019climate,lenton2011early,national2024tipping,wunderling2020basin,lynas2008six}.

Several core differences emerge between the two fields:

\begin{itemize} 
\item \textit{Equilibria vs. Regimes:} Control theory often studies stability around specific equilibria or trajectories. Climate science, instead, focuses on maintaining the system within broad functional regimes, without assuming an easy return to a specific baseline.

\item \textit{Perturbation Size:} In engineering, stability typically concerns relatively small perturbations. In climate science, perturbations (e.g., anthropogenic CO$_2$) can be large, pushing the system into fundamentally different basins of attraction.

\item \textit{Stability Tools:} While both fields use phase space methods, climate science relies heavily on bifurcation diagrams, resilience metrics, early warning indicators (e.g., critical slowing down), and potential well analyses to understand the proximity to tipping points.

\item \textit{Control Objective:} In control, stability ensures that desired performance is robust against disturbances. In climate science, interventions (e.g., emissions reductions, geoengineering) aim to maintain the climate system within acceptable safe boundaries, recognizing that recovery to the original state may not be possible.
\end{itemize}

Despite these differences, shared mathematical tools enable fruitful cross-fertilization. Concepts such as basins of attraction, stability margins, and feedback sensitivity provide a bridge to reason about climate resilience with a control-theoretic mindset. Moreover, understanding the climate system as a high-dimensional, feedback-rich dynamical system invites deeper adoption of dynamical systems theory methods, such as identifying early warning signals for critical transitions, optimizing interventions to maximize system resilience, and framing climate stability problems as control challenges under extreme uncertainty.

\subsection{Model Calibration \& Learning vs. System Identification}
% \vspace{0.1cm}
% \noindent\textbf{\textcolor{pastelblue}{Model Calibration and Learning} \& \textcolor{pastelgreen}{System Identification}.} 
% \textcolor{red}{Ahmad: I believe this section here needs more expansion. Maybe we should move it to the start of the main section and have a subsection on AI and ML? do you think you can work on it while I am working on the earlier comment? Also, can you add one paragraph or two on autoregressive models here and relate it to Sys ID? }

The problem of learning dynamic models from data and calibrating them to better reflect reality arises in both climate science and control engineering, though with different traditions and terminologies. However, the underlying ideas, identifying model parameters, uncovering dynamics, and improving predictive skill, are deeply paralleled across the two fields.

In control engineering, \textit{system identification} refers to the process of building mathematical models of dynamic systems from observed input-output data~\cite{eykhoff1974system,ljung1998system,ljung1999system,goodwin1977dynamic,keesman2011system,ljung2010perspectives}. Techniques range from linear regression models for simple dynamics to nonlinear black-box models like neural networks, depending on the system complexity and the quality of available data. The primary objective is to either identify the full system dynamics (structure and parameters) or estimate unknown parameters within a known structure. Classical techniques include prediction error methods (PEM), subspace identification, and more recently ML-based methods for nonlinear systems. Rich toolboxes exist for these tasks, emphasizing guarantees like consistency, efficiency, and convergence.

Similarly, in climate science \textit{model calibration} and \textit{learning} are critical tasks. Climate models---whether EBMs, intermediate-complexity models, or full Earth system models (such as in Tab. \ref{tab:ClimModelsOvw})---rely on parameterizations of unresolved or poorly understood processes (e.g., cloud dynamics, ocean eddies). Calibration involves tuning these parameters based on observational data, reanalysis products, or high-resolution simulations. Traditionally, this was done manually by expert judgment, but more systematic approaches like ensemble Kalman inversion, Bayesian calibration, and ML surrogates have become prominent \cite{schneider2017earth,mcneall2020bias,couvreux2021process,williams2021model}. A key parallel between the two fields is that both deal with models that are incomplete, noisy, and computationally expensive. Learning hence must balance fidelity to data with physical realism. Several important connections and trends are worth highlighting:

\begin{itemize} 

\item \textit{Parameter Estimation:} In control, unknown parameters are estimated through system identification. In climate modeling, parameters governing cloud microphysics, land-surface interactions, and biogeochemical cycles are similarly estimated using data assimilation, optimization, or statistical learning techniques.

\item \textit{Learning Full Models:} Control engineering has long explored data-driven model discovery, from AutoRegressive with eXogenous inputs (ARX) models to modern neural network-based nonlinear models. In climate science, recent efforts have extended beyond tuning parameters to learning full sub-grid scale parameterizations (e.g., for convection, ocean eddies) using deep learning \cite{eyring2024pushing,bracco2025machine}. Offline and online learning paradigms, as detailed in \cite{bracco2025machine}, now play a major role.

\item \textit{Hybrid Modeling:} A growing trend in both fields is hybrid modeling, where physics-based models are augmented by data-driven components. In control, this parallels gray-box identification \cite{youssry2024experimental,sievers2023jax,ghosh2021model}; in climate studies, it includes neural-network parameterizations embedded within Earth system models (e.g., NeuralGCM \cite{kochkov2024neural}).

\item \textit{Challenges of Generalization and Stability:} In both fields, learned models often face challenges when applied beyond the training regime (out-of-sample climates, unseen operating conditions). Issues such as physical constraint violations, instabilities, and error accumulation have motivated the development of physics-informed learning, regularized identification, and hybrid methods.

\end{itemize}

Recent landmark contributions in the climate community highlight the increasing adoption of machine learning:
\begin{itemize} 
\item ML techniques have been used for \textit{parameter tuning} (e.g., Gaussian processes, ensemble Kalman inversions) \cite{couvreux2021process}.

\item \textit{Offline learning of parameterizations} from high-resolution simulations to substitute for handcrafted physical models \cite{rasp2018deep,huntingford2025potential}.

\item \textit{Online learning} of parameterizations during coupled model simulation to improve system statistics \cite{rasp2020coupled,eyring2024pushing}.

\item \textit{Equation discovery} using symbolic regression and sparse identification methods to find interpretable models for processes like ocean mesoscale eddies \cite{zanna2020data,huntingford2025potential}.

\end{itemize}

Similarly, in dynamic systems, advances in learning dynamics and system identification under uncertainty, high-dimensionality, and partial observability mirror these developments. Techniques like Koopman operator theory, dynamic mode decomposition, and hybrid neural-physics models are increasingly central to the literature.

% \noindent \textbf{Example Applications:} 
% \begin{itemize} 
% \item \textit{Climate:} Neural networks trained on high-resolution cloud-resolving models to replace traditional cloud parameterizations (e.g., using SPCAM5 as training data) \cite{gentine2018mlconvection,bracco2025mlphysics}.

% \item \textit{Control:} Learning aerodynamic models for drones from sparse flight data using Gaussian process regression or deep neural networks, ensuring stability margins under uncertainty.

% \end{itemize}

In short, the rich traditions of model learning and system identification in control engineering offer valuable insights for modern climate modeling. Conversely, the massive scale and complexity of climate applications are pushing the boundaries of what learning-based modeling needs to achieve---generalizability, physical consistency, and uncertainty-aware predictions---which can in turn inspire new developments in control and system identification methodologies.

\section{Tutorial Summary and Concrete Control for Climate Science Problems}~\label{sec:Conc}
In this tutorial, we have produced the very first control systems-focused tutorial on climate science models and contemporary research problems in this rapidly evolving field. We hope that this tutorial bridges the disciplines of control engineering and climate science by exploring how control-theoretic principles can be applied to the behavior of climate systems. By viewing climate dynamics through a control systems lens, we highlight potential benefits for modeling, prediction, and strategic intervention. This perspective offers an alternative way to think about climate behavior and supports the idea of developing robust, physics-informed approaches for climate mitigation and adaptation. Integrating insights from both fields holds promise for better decision-making and more effective policy in response to climate change.

Building on this tutorial, we include ideas for future research papers that are well-contained in their scope. 

\begin{itemize}
    \item \textit{Reachability Analysis:} Given the following: \textit{(i)} an ensemble of extremely complex, nonlinear, and large-scale climate models, \textit{(ii)} some measurable initial conditions of climate states (or at least a region of initial climate state space), \textit{(iii)} a long-term time-horizon of decades, and \textit{(iv)} specific climate change mitigation strategies and various policy scenarios (as ones outlined in Fig.~\ref{fig:GHGPolicies}), one very relevant control engineering study could focus on investigating reachability analysis that considers \textit{(i)}--\textit{(iv)}. This results in realizing whether certain state space regions of climate spaces are indeed possible, under the various sources of uncertainty, without requiring running the large-scale models for different realizations of uncertainty. This problem of reachability is indeed well-contained if studied for one specific climate model as~\eqref{eq:Disc_NS} for example. This problem is still theoretically challenging as reachability analysis for nonlinear systems is still an active field of research in systems science. Applying common reachability analysis in this context is of interest, but also designing specific climate science-based reachability analysis tools could be more pertinent. 
    
    \item \textit{Nonlinear Model Order Reduction:} This augments or replaces the need to run expensive simulations that (ironically) consume large amount of electric power and hence further contribute to GHG emissions.  As has been demonstrated in some recent work, the carbon footprint~\cite{acosta2024computational} of running climate models has exponentially increased recently. There has also been some proposed work to reduce the computational burden and the corresponding carbon footprint~\cite{palomas2024reducing}. As a control theoretic alternative, model order reduction (MOR) for nonlinear dynamic systems can potentially be proposed to preserve the system properties (controllability, observability, stability) for the reduce order models. MOR algorithms basically reduce the state-space dimension, thereby resulting in faster forward simulations of the state-space dynamics. There has been recent significant advances in MOR theory for nonlinear systems~\cite{peherstorfer2022breaking,touze2021model}, and their investigation and application for large-scale complex climate climate models is a promising area of research. 
    \item \textit{Comparative analysis of system identification algorithms for climate models:} There are virtually hundreds of control theoretic studies on the topic of system identification and learning dynamic models. We have avoided thoroughly discussing or citing any of this literature for brevity. Relevant to this paper, we note that it is still virtually unclear what is the overwhelmingly \textit{better} method to perform large-scale system identification in complex nonlinear dynamic networks. In short, there is also an absence of computational studies that focus on comparing various identification and model learning approaches. Unfortunately, the abundant control engineering literature focuses on testing theoretical algorithms on toy examples that are not always meaningful or scalable. Dynamic climate models could hence be good benchmarks for testing the performance of system identification algorithms seeing that these models are riddled with uncertainty, extremely large-scale, often under-sensed, and have various forms of complex nonlinear relationships present in their dynamics. 

    % \color{blue}
    
    \item \textit{Sensor Placement, Satellite Data, Orbital Designs, and Observability Analysis:} A key problem in dynamic system theory is the sensor placement problem (where to geographically install sensors and design their sampling rates), its relationship with observability analysis (how different sensor placements result in varying observability metrics), and their joint optimization. There indeed is a broad literature on this topic with applications in virtually every large-scale dynamic system application domain (in gas, power, robotic, transportation, water, and hydrological systems, to mention a few). 
    
    In climate science, however, this problem has been studied through the lens of Kalman filtering and how varying number of climate data sensors results in different state estimation and model calibration errors for some specific climate phenomena and models. However, the notion of system-theoretic observability analysis for linearized or nonlinear climate models (through the Kalman rank condition, Lie derivatives, or empirical observability Gramians) is not investigated or presented in climate research, to the best of our knowledge. The survey paper~\cite{carrassi2018data} on climate data assimilation sheds some light on this point by: \textit{(i)} discussing how data assimilation and state estimation are used by different communities to mean the same thing and \textit{(ii)} presenting the computational challenges of estimation problems for large-scale climate models. While subtle, the paper also  includes an initial-condition-dependent definition of observability, similar to the observability condition in linear systems~\cite[Appendix A, \textit{Dependence on the Initial Condition}]{carrassi2018data}. We see that system-theoretic observability analysis---and how it informs data assimilation for nonlinear climate models through constructing unobservable subspaces of climate states, given available sensor data---is an interesting future research direction that is largely underexplored. This analysis should be tailored to climate models, as generic observability analysis of nonlinear dynamic models is not scalable (which is, remarkably, an issue with many control-theoretic algorithms in nonlinear systems). There is then an opportunity of delivering theoretical methods that scale well for climate model observability analyses. 

\vspace{0.15cm}

    Relevant to this topic, is the timely research problem of satellite orbital designs and maneuvers---the former can be perceived as an offline design problem of determining orbital parameters whereas the latter determines adjustments in response to uncertainty and uncontrollable conditions---to maximize the \textit{information usefulness} or observability of satellite data.  Satellites play an important role in both climate modeling and climate prediction by providing high-resolution observations that are otherwise impossible to obtain from ground-based sensors. Satellite data is widely used for model calibration and data assimilation through the aforementioned orbital designs and maneuvers. Some climate science efforts~\cite{stroud2010ensemble,shprits2025observing,balthazor2015methodology,nicklas2025efficient,hoffman2016future} have tackled these problems thereby maximizing system \textit{observability} (emphasis here to distinguish observability from the control-theoretic one, which has not been pursued in the literature) through Kalman filter surrogates or occasionally through exhaustively testing different orbital scenarios. These all fall under the broad umbrella of Observing System Simulation Experiments (OSSE), which is a framework to evaluate and predict the impact of new sensors (e.g., new satellite instruments, orbit configurations, or other climate sensor networks) on the performance of weather or climate prediction systems prior to installing these sensors.

    \normalcolor
    
    % \item \textit{Applications to power and water infrastructure:} \comment{can you summarize that section  you had earlier in the paper here? just a few lines. basically list one specific research problem each for drinking water or power grids here? be specific. no need for a length paragraph here. something short and sweet}
    \item \textit{Applications to Infrastructure Systems:} Climate models enable predictive control opportunities in infrastructure systems affected by environmental variability. For instance in the water sector, projected shifts in precipitation, snowmelt, and drought severity can inform dynamic scheduling of water delivery, pump control, and reservoir operations to ensure supply reliability and minimize energy use---these have all been formulated and solved as optimal control problems in the literature. Similarly, in power systems, long-range and short-term climate forecasts can be integrated into real-time dispatch and control of renewable resources where temperature, wind, and solar radiation forecasts affect both load demand and generation profiles. In particular, the recent work~\cite{prudhvi2025vulnerability} explains how climate change-induced, changing temperature patterns could lead to increased stress on power distribution networks, potentially causing more blackouts.  These applications highlight the potential for closed-loop climate-informed control frameworks to enhance the resilience of critical infrastructure.
\end{itemize}

In the past two years, this tutorial presented itself as an opportunity for us to study and learn about a new problem, outside of the scope of our research interests. We were careful in our discussions and presented expositions. We used descriptive and welcoming language that can be understood by all controls and climate research readerships. We tried to not overhype the promise of control engineering in potentially addressing some climate science related problems, but also showcased promise of controls-guided climate science. We welcome suggestions, corrections, and clarifications from the curious and interested readers. 

\bibliographystyle{IEEEtran}
\bibliography{ClimateCT}

\end{document}